\pdfoutput=1

\newcommand{\forpreamble}{ 
  }
\providecommand{\ifmain}[1]{#1}

\ifmain{
  \documentclass[12pt]{amsart} %
  \setlength{\topmargin}{0in}\setlength{\textheight}{8.6in}
  \setlength{\oddsidemargin}{.25in}\setlength{\evensidemargin}{.25in}\setlength{\textwidth}{6in}
  \newcommand{\tpath}{/Users/speter18/4t/4geek/tex} 

\listfiles %
\usepackage{ 
  amssymb, %
  amsmath, %
  latexsym, %
  graphpap, %
  graphics, %
  verbatim, %
  color, %
  natbib, 
  har2nat, %
  layout, %
  amsbsy, %
  calc, %
  pstricks, %
  pst-plot, %
  pst-pdf, %
  pstricks-add, %
  xr, %
  mathdots, %
  enotez, %
  lipsum, 
  stmaryrd, %
  textcomp %
  }

\usepackage[bottom]{footmisc} %
 
\usepackage[inline]{enumitem} %

\newcommand{\ifdraft}[1]{} %
\newcommand{\ifdrafttext}[1]{} %
\newcommand{\iffinal}[1]{#1} %
\newcommand{\ifallkeys}[1]{} %
\newcommand{\ifbeamer}[1]{} %
  \newcommand{\eput}{ \ifmain{\showbib \end{document}}  } 

\newcommand{\sella}[1]{} 
\newcommand{\sellb}[1]{} 
\newcommand{\sellc}[1]{} 
\newcommand{\selld}[1]{} 
\newcommand{\selle}[1]{} 
\newcommand{\sellf}[1]{} 
\newcommand{\sellg}[1]{} 
\newcommand{\sellh}[1]{} 
\newcommand{\selli}[1]{} 
\newcommand{\sellj}[1]{} 

\newtheorem{lemma}{\indent Lemma} 
\newtheorem{nthm}[lemma]{\indent Theorem} 
\newtheorem{ndef}[lemma]{\indent Definition} 
\newtheorem{ncrly}[lemma]{\indent Corollary} 
\newtheorem{prop}[lemma]{\indent Proposition}

\theoremstyle{definition} %
\newtheorem{pfa}[lemma]{\indent Proof} 
\newenvironment{npf} %
  { \begin{pfa} } 
  { \hfill $\Box$ \end{pfa} }

\theoremstyle{remark} %

\newtheorem{absatz}[lemma]{\hspace{-1ex}\tt} %
\newcommand{\alabel}{} 
\newcommand{\amark}{} 
\newcommand{\ai}[2][]{\setcounter{equation}{\thelemma} 
  \begin{absatz}\label{#2}\amark \renewcommand{\alabel}{#2} 
  \begin{subequations} {\em#1}} %
\newcommand{\ao}{\end{subequations} \end{absatz}}

\newtheorem*{pfb}{\indent Proof} 
\newenvironment{pf} 
  { \begin{pfb} } 
  { \hfill $\Box$ \end{pfb}  } 

\newtheorem{exa}[lemma]{\indent Question} %
\newtheorem{soln}[lemma]{\indent Answer to Question} 
\newcommand{\exercomment}{} %
 
\newcommand{\ifbothexeransr}[1]{} 
  { \begin{exa} \label{#1} \exercomment} %
  { \end{exa} } 
  { \end{soln} } %

\newcounter{blistno} %
  { \begin{list} %
      { (\arabic{blistno}) } %
      { \setlength{\leftmargin}{0mm} %
        \setlength{\rightmargin}{0mm} %
        \setlength{\itemindent}{-1mm} %
        \setlength{\labelwidth}{0mm} %
        \setlength{\labelsep}{0mm} %
        \setlength{\itemsep}{1mm} %
        \setlength{\topsep}{1mm} %
        \usecounter{blistno} 
        \nopagebreak } } %
  { \end{list} }

\newcounter{clistno} %
  { \begin{list} %
      {  \arabic{clistno}. } %
      {  \setlength{\leftmargin}{0mm} %
        \setlength{\rightmargin}{0mm} %
        \setlength{\itemindent}{3mm} %
        \setlength{\labelwidth}{0mm} %
        \setlength{\labelsep}{0mm} %
        \setlength{\itemsep}{2mm} %
        \setlength{\topsep}{2mm} %
        \usecounter{clistno} 
        \nopagebreak } }%
  { \end{list} } 

\newcommand{\tlistindent}{0mm} 
\newcounter{tlistno} %
\newenvironment{tlist}[1][] %
  { \begin{list} %
      {\indent(\alph{tlistno}#1) } %
      { \setlength{\leftmargin}{0mm} %
        \setlength{\rightmargin}{0mm} %
        \setlength{\itemindent}{\tlistindent} 
        \setlength{\labelwidth}{0mm} %
        \setlength{\labelsep}{0mm} %
        \setlength{\itemsep}{0mm} %
        \setlength{\topsep}{0mm} %
        \usecounter{tlistno} } } %
  { \end{list} } 
\newcommand{\hglistindent}{0mm} 
\newcounter{hglistno} %
  { \begin{list} %
      {\indent\fbox{\q}\hspace{1ex} } %
      { \setlength{\leftmargin}{0mm} %
        \setlength{\rightmargin}{0mm} %
        \setlength{\itemindent}{\hglistindent} 
        \setlength{\labelwidth}{0mm} %
        \setlength{\labelsep}{0mm} %
        \setlength{\itemsep}{0mm} %
        \setlength{\topsep}{0mm} %
        \usecounter{hglistno} } } %
  { \end{list} } 

\newcounter{cllistno} %
\newenvironment{cllist} %
  { \begin{list} %
      {\emph{Claim \arabic{cllistno}:} } 
      { \setlength{\leftmargin}{0mm} %
        \setlength{\rightmargin}{0mm} %
        \setlength{\itemindent}{\parindent} %
        \setlength{\labelwidth}{0mm} %
        \setlength{\labelsep}{0mm} %
        \setlength{\itemsep}{1.2mm}%
        \setlength{\topsep}{1.2mm}%
        \usecounter{cllistno} } } %
  { \end{list} } 
 
\newcommand{\unskipcl}{\vspace{-5.2mm}} %

\newcommand{\itemi}{ \begin{itemize} } 
\newcommand{\itemo}{ \end{itemize} } 
\newcommand{\yl}[1]{\item\label{#1}} 

\newcommand{\ili}{\begin{enumerate*}[label=$\lbrack$\alph*$\rbrack$]} 
\newcommand{\ilr}{\begin{enumerate*}[resume*]} 
\newcommand{\ilo}{\end{enumerate*}} 
\newcommand{\ilitem}[1]{\item\label{#1}\ilo\hspace{-1mm}\ifallkeys{\hspace{10mm}}} 
\newcommand{\ilc}[1]{\ili\ilitem{#1}} 
\newcommand{\il}[1]{\ilr\ilitem{#1}} 
 
\newcommand{\lii}{\begin{enumerate*}[label=$\lbrack$\arabic*$\rbrack$]} 
\newcommand{\lir}{\begin{enumerate*}[resume*]} 
\newcommand{\lio}{\end{enumerate*}} 
\newcommand{\liitem}[1]{\item\label{#1}\lio\hspace{-1mm}\ifallkeys{\hspace{10mm}}} 
\newcommand{\lic}[1]{\lii\liitem{#1}} 
\newcommand{\li}[1]{\lir\liitem{#1}}

\newcommand{\ttc}[2]{\begin{enumerate*}[label={#1}]\item\label{#2} 
  \end{enumerate*}\hspace{-0.8ex}\ifallkeys{\hspace{10mm}}}

\newcommand{\ttr}[2]{{(\ttc{#1}{#2})}} 
\newcommand{\tts}[2]{{\ttc{[#1]}{#2}}} 
\newcommand{\ttz}[2]{{\ttc{#1}{#2}}}

\newcommand{\zz}{} 
\newcommand{\subi}{\begin{subequations}} 
\newcommand{\subo}{\end{subequations}} 
\newcommand{\nt}{\notag \\ }

\newenvironment{maneq}{%
  \setlength{\arraycolsep}{.1ex} 
   
  \begin{array}{lc} \rule{5ex}{0ex} & \rule{79ex}{0ex} \\[-3ex] }{%
  \end{array}} 
\newcommand{\mqi}{\begin{maneq}} 
\newcommand{\mqo}{\end{maneq}} 
 
\newcommand{\ttt}[2]{\tag*{#1}\label{#2}} %
 
\newcommand{\ci}{\par\begin{center}} 
\newcommand{\co}{\end{center}\par\noindent}

\newcommand{\mi}{ \left[ \begin{matrix} } 
\newcommand{\mo}{ \end{matrix} \right] } 
\newcommand{\miz}{ \left. \begin{matrix} } 
\newcommand{\moz}{ \end{matrix} \right. } 
\newcommand{\mip}{ \left( { \ } \begin{matrix} } 
\newcommand{\mop}{ \end{matrix} { \ } \right) } 
\newcommand{\mib}{ \left[ { \ } \begin{matrix} } 
\newcommand{\mob}{ \end{matrix} { \ } \right] } 
\newcommand{\smi}{ \left[ \begin{smallmatrix} } %
\newcommand{\smo}{ \end{smallmatrix} \right] } 
\newcommand{\casei}{ \left\{ \begin{matrix} } %
\newcommand{\caseo}{ \end{matrix} \right. } 
 
\newcommand{\mpi}{ \begin{minipage} } 
\newcommand{\mpo}{ \end{minipage} } 
\newcommand{\mpic}{\begin{minipage}{4cm}\begin{center}} 
\newcommand{\mpoc}{\end{center}\end{minipage}}

\newcommand{\datestmp}{%
  $ \the\year \backslash \the\month \backslash \the\day \backslash 
    \the\time $ } 
 
\newcommand{\filestamp}[1]{} %

\newcommand{\sectitle}{}
\newcommand{\markb}[1]{\markboth{#1}{#1}} %
\newcommand{\showit}{\ifdraft{\vspace{-.7cm}{⋅}\vspace{.3cm}}} %
   
\newcommand{\nssec}[2]{%
  \vspace{.4mm} 
  \subsection{#1}\label{#2}\hspace{0mm}%
  \par\vspace{1.6mm} 
  \begin{picture}(0,0) 
  \put(-.50,.33){\color{white} \rule{85ex}{4ex}} 
  \put(-.4,.53){\color{black} \sc \rf{#2}. #1} 
  \end{picture} 
  \par\vspace{-4.5mm}} 

 \newcommand{\ifssec}[1]{#1} %
 \providecommand{\ifssec}[1]{} 
\ifssec{ 
  }

\allowdisplaybreaks  %
 
\newcommand{\toenote}[1]{}

\newcommand{\nichts}[1]{} %
\newcommand{\showbib}{%
  \markb{\sc References} 
  \bibliographystyle{\tpath/ecta/econometrica} 
  \bibliography{\tpath/bibtex/streuf,\tpath/bibtex/others,references}} 
\definecolor{darkred}{rgb}{.7,0,0} 

\definecolor{darkgreen}{rgb}{0,.7,0} 
   
\definecolor{darkblue}{rgb}{0,0,.97}

\definecolor{igray}{gray}{.87}

\newcommand{\rf}[1]{\ref{#1}} %
\newcommand{\pgrf}[1]{page~\pageref{#1}}

\newcommand{\ex}[1]{\mathsf{#1}} 

\newcommand{\yyin}[1]{#1}         %
\newcommand{\yyout}[1]{}          %

  \renewcommand{\lim}{\text{\normalfont{lim}}}

  \renewcommand{\max}{\text{\normalfont max}} %

  \newcommand{\nnexists}{\raisebox{0.15ex}{/}\hspace{-1.15ex}\exists}

  \newcommand{\dra}{\rotatebox[origin=c]{180}{$\Lsh$}} %
  \newcommand{\up}{_{\vartriangle}} 
   
  \newcommand{\lo}{_{\vartriangle}}

  \newcommand{\smass}[1]{\smash{#1}\rule{0ex}{2.1ex}} %

  \newcommand{\dE}{\bar{E}} 
  \newcommand{\dF}{\bar{F}} 
  
  \newcommand{\dHH}{\bar{\HH}} 
      
  \newcommand{\dI}{\bar{I}}  
  \newcommand{\dJ}{\bar{J}}

     \newcommand{\dpp}{\bar{p}} 
  \newcommand{\dQ}{\bar{Q}}

   \newcommand{\du}{\bar{u}} 
   
  \newcommand{\dW}{\bar{W}}  
  \newcommand{\dX}{\bar{X}}  
  \newcommand{\dY}{\bar{Y}}

  \newcommand{\HH}{{ \mathcal H}}

  \newcommand{\PP}{{ \mathcal P}}

  \newcommand{\QQ}{{ \mathcal Q}}

  \newcommand{\ZZ}{{ \mathcal Z}} 
  \newcommand{\ZZf}{ \mathcal Z_{\mathsf{ft}}} 
  \newcommand{\ZZi}{\mathcal Z_{\mathsf{inft}}}

  \newcommand{\eR}{\bar{\mathbb R}} %

  \newcommand{\f}[1]{\mathsf{#1}} %
    
   \newcommand{\fb}{{\f{b}}} 
   \newcommand{\fc}{{\f{c}}} 
   \newcommand{\fd}{{\f{d}}} 
   \newcommand{\fe}{{\f{e}}} 
   \newcommand{\ff}{{\f{f}}} 
   \newcommand{\fg}{{\f{g}}}

  \newcommand{\fP}{{\f{P}}}

  \newcommand{\PB}{\mathsf{P}}

  \newcommand{\TB}{\mathsf{T}}

  \newcommand{\bbJ}[2]{{^{#1}\mspace{-4mu}{#2}}} 
   
  \newcommand{\bbQ}[2]{{^{#1}\mspace{-1.5mu}{#2}}}

  \newcommand{\bbW}[2]{{^{#1}\mspace{-1mu}{#2}}} 
  \newcommand{\bbY}[2]{{^{#1}\mspace{-1mu}{#2}}}

  \newcommand{\bJ}[1]{\bbJ{#1}{J}}

  \newcommand{\bQ}[1]{\bbQ{#1}{Q}}

  \newcommand{\bW}[1]{\bbW{#1}{W}} 
  \newcommand{\bY}[1]{\bbY{#1}{Y}}

  \newcommand{\tJ}{{\bJ{t}}} 
   
  \newcommand{\tQ}{{\bQ{t}}}

  \newcommand{\tW}{{\bW{t}}} 
   
  \newcommand{\tY}{{\bY{t}}}

\usepackage[utf8]{inputenc} 
\usepackage{amssymb} 
\usepackage{newunicodechar} 
\newcommand{\nuc}[2]{\newunicodechar{#1}{#2}} %

  \nuc{α}{{ \alpha}} 
  \nuc{β}{{ \beta}} 
  \nuc{γ}{{ \gamma}} \nuc{Γ}{{ \varGamma}} 
  \nuc{δ}{{ \delta}} \nuc{Δ}{{ \varDelta}} 
  \nuc{ε}{{ \varepsilon}} 
  \nuc{ζ}{{ \zeta}} 
  \nuc{η}{{ \eta}} 
  \nuc{θ}{{ \theta}} \nuc{Θ}{{ \varTheta}} 
  \nuc{ι}{{ \iota}} 
  \nuc{κ}{{ \kappa}} 
  \nuc{λ}{{ \lambda}} \nuc{Λ}{{ \varLambda}} 
  \nuc{μ}{{ \mu}} 
  \nuc{ν}{{ \nu}} 
  \nuc{ξ}{{ \xi}} \nuc{Ξ}{{ \varXi}} 
  \nuc{π}{{ \pi}} \nuc{Π}{{ \varPi}} 
  \nuc{ρ}{{ \rho}} 
  \nuc{σ}{{ \sigma}} \nuc{Σ}{{ \varSigma}} 
  \nuc{τ}{{ \tau}} 
  \nuc{υ}{{ \upsilon}} \nuc{Υ}{{ \varUpsilon}} 
  \nuc{φ}{{ \phi}} \nuc{Φ}{{ \varPhi}} 
  \nuc{χ}{{ \chi}} 
  \nuc{ψ}{{ \psi}} \nuc{Ψ}{{ \varPsi}} 
  \nuc{ω}{{ \omega}} \nuc{Ω}{{ \varOmega}} 

  \nuc{≠}{{ \ne}} 
  \nuc{∾}{{ \sim}} %
  \nuc{≈}{{ \approx}} 
  \nuc{≡}{{ \equiv}} %
  \nuc{∈}{{ \in}} \nuc{∉}{{ \notin}} 
  \nuc{∋}{{ \ni}} \nuc{∌}{{ \not\ni}} 
  \nuc{⊊}{{ \subset}} \nuc{⊆}{{ \subseteq}} %
  \nuc{⊋}{{ \supset}} \nuc{⊇}{{ \supseteq}} %
  \nuc{≤}{{ \leq}} 
  \nuc{≥}{{ \geq}} 
  \nuc{≪}{{ \ll}} 
  \nuc{≫}{{ \gg}} 
  \nuc{≺}{{ \prec}} \nuc{≼}{{ \preccurlyeq}} 
  \nuc{≻}{{ \succ}} \nuc{≽}{{ \succcurlyeq}}
  \nuc{≲}{{ \lesssim}} %
  \nuc{≳}{{ \gtrsim}} %
  \nuc{‖}{{ \parallel}}  %
  \nuc{≊}{{ \approxeq}} %
  \nuc{≅}{\cong} %
  \nuc{≃}{\simeq} %
  \nuc{≐}{\doteq} 

  \nuc{∑}{{ \Sigma}} %
  \nuc{∫}{{ \textstyle \int\nolimits}} 
  \nuc{∏}{{ \Pi}} %
  \nuc{×}{{ \times}} 
  \nuc{÷}{{ \div}} 
  \nuc{○}{{ \circ}} 
  \nuc{⨅}{{\scalebox{1}[1.2]{\ensuremath{\cap}}}} %
  \nuc{⋂}{{\cap}}                                   %
  \nuc{⨆}{{\scalebox{1}[1.2]{\ensuremath{\cup}}}} %
  \nuc{⋃}{{\cup}}                                   %
  \nuc{⋱}{{ \setminus}} %
  \nuc{⧷}{{ \smallsetminus}} %
  \nuc{∧}{{ \wedge}} \nuc{⋀}{{ \textstyle \bigwedge\nolimits}} 
  \nuc{∨}{{ \vee}} \nuc{⋁}{{ \textstyle \bigvee\nolimits}} 
  \nuc{⊗}{{ \otimes}} 
  \nuc{⊙}{{ \odot}} \nuc{⊛}{{ \textstyle \bigodot\nolimits}} 
  \nuc{⊕}{{ \oplus}} 
  \nuc{•}{{ \bullet}} 
  \nuc{∗}{{ \star}} 
  \nuc{⁕}{{ \ast}} %
  \nuc{∙}{{ \cdot}} %

  \nuc{→}{{ \rightarrow}} 
  \nuc{⇉}{{ \rightrightarrows}} 
  \nuc{←}{{ \leftarrow}} 
  \nuc{⟷}{{ \leftrightarrow}} 
  \nuc{⇒}{{ \Rightarrow}} 
  \nuc{⇐}{{ \Leftarrow}} %
  \nuc{⇔}{{ \Leftrightarrow}} %
  \nuc{⟺}{{ \Leftrightarrow}} %
  \nuc{⇡}{{ \uparrow}} 
  \nuc{⇣}{{ \downarrow}} 

  \nuc{ℓ}{{ \ell}} 
  \nuc{∂}{{ \partial}} 
  \nuc{∞}{{ \infty}} 
  \nuc{‴}{^{ \prime\prime\prime}} %
  \nuc{″}{^{ \prime\prime}} %
  \nuc{′}{^{ \prime}} 
  \nuc{¬}{{ \neg}} 
  \nuc{∀}{{ \forall}} 
  \nuc{∃}{{ \exists}} 
  \nuc{∆}{{ \triangle}} %
  \nuc{∄}{{ \nnexists}} %
  \nuc{∅}{{ \varnothing}} 
  \nuc{□}{{ \square}} 

  \nuc{⎨}{ \lbrace } 
  \nuc{⎬}{ \rbrace } 
  \nuc{⁅}{ \langle } 
  \nuc{⁆}{ \rangle } 

  \nuc{Ṛ}{{ \mathbb R}} 
  \nuc{Ẓ}{{ \mathbb Z}} 
  \nuc{⋅}{{ \ }} %
  \nuc{˙}{{\,}} %
  \nuc{˛}{{,}\mspace{.5mu}} %
  \nuc{₵}{\text{\textcent}}%

\setlength{\unitlength}{1cm} 
 
\newrgbcolor{maroon}{.5 0 0} 
\psset{gridcolor=green} 
 
\newcommand{\myoutergrid}[1]{\ifdraft{ 
  \put(-7.5,0){\color{brown} \framebox(7.5,#1){}} 
  \put(-3.75,0){\color{brown} \framebox(7.5,#1){}} 
  \put(0,0){\color{brown} \framebox(7.5,#1){}}} } 
\psset{unit=1cm} %
  \providecommand{\forpreamble}{} \forpreamble %
  \usepackage{hyperref} \hypersetup{colorlinks=true,linkcolor=black}
  \begin{document}
  }

\ifmain{
  \newcommand{\IJGTversiononly}[1]{#1}%
  \newcommand{\arXivversiononly}[1]{}%
  \numberwithin{lemma}{section}
  \numberwithin{figure}{section}
  \numberwithin{table}{section}
  \newcommand{\hp}[1]{\smass{\hat{#1}}}%
  \newcommand{\hs}[1]{\bar{#1}}%
  \newcommand{\pj}[1]{π_{{\mathit{#1}}}}
  \newcommand{\dve}{\dra\,\,} 
  \newcommand{\klar}{\mspace{1mu}\displaystyle} %
  \newcommand{\payoa}[1]{{\tiny $\mi{\mathit{#1}}\mo$}}
  \newcommand{\payob}[2]{{\tiny $\mi{\mathit{#1}}\\{\mathit{#2}}\mo$}}
  \newcommand{\payoc}[3]{{\tiny $\mi{\mathit{#1}}\\{\mathit{#2}}\\{\mathit{#3}}\mo$}}
  \newcommand{\noab}{no-absent\-mind\-ed\-ness}
  }

\title{Specifying a Game-Theoretic Extensive Form \\[.5ex] 
  as an Abstract 5\hspace{.2mm}\raisebox{.3mm}{-}\hspace{-.15mm}ary Relation}
\date{\arXivversiononly{September 15, 2024. Supersedes arXiv:2107.10801v5 (small editorial changes).}\IJGTversiononly{September 29, 2024. Accepted for publication in {\em International Journal of Game Theory}.%
\nocite{2409-5f}}  {\em Keywords:} extensive-form game, pentaform, subgame, perfect recall. {\em Classifications:} JEL C73, MSC 91A70.%
\nichts{}
{\em Contact information:} pstreuf@uwo.ca, 519-661-3500, Economics Department, Western University, London, Ontario, N6A 5C2, Canada.}  
\thanks{This paper has benefitted greatly from the extensive comments of an anonymous associate editor and two anonymous reviewers.}
\nichts{}

\maketitle

\iffinal{\vspace{-5mm}}
{\begin{centering}  \normalsize
Peter~A. Streufert \\[-.3ex]
Economics Department \\[-.3ex]
Western University \\
\end{centering} }
\vspace{2mm}

\begin{abstract} This paper specifies an extensive form as a 5-ary relation (that is, as a set of quintuples) which satisfies eight abstract axioms.  Each quintuple is understood to list a player, a situation (that is, a name for an information set), a decision node, an action, and a successor node.  Accordingly, the axioms are understood to specify abstract relationships between players, situations, nodes, and actions.  Such an extensive form is called a ``pentaform''.  Finally, a ``pentaform game'' is defined to be a pentaform together with utility functions.  

To ground this new specification in the literature, the paper defines the concept of a ``traditional game'' to represent the literature's many specifications of finite-horizon and infinite-horizon games.  The paper's main result is to construct an intuitive bijection between pentaform games and traditional games.  Secondary results concern disaggregating pentaforms by subsets, constructing pentaforms by unions, and initial pentaform applications to Selten subgames and perfect-recall (an extensive application to dynamic programming is in Streufert 2023, arXiv:2302.03855). \end{abstract}
\nocite{2303-5dp}

\section{Introduction}\label{B234}\showit
\markb{\sc \rf{B234}. Introduction}

\nssec{New concepts and main result}{B572}

\newcommand{\noterelation}{\footnote{This concept of a relation as a set accords with Halmos 1974, Section 7, and Enderton 1977, pages 41--42.\nocite{Halmo74} \nocite{Ender77} In a similar vein, this paper regards functions (footnote~\rf{C273}) and correspondences (footnote~\rf{B359}) as sets of couples.\nichts{}}}

\newcommand{\notesit}{\footnote{\label{B598}In this paper, a ``situation'' can be either an information set (which is a set of nodes), or something else (such as a word like ``tomorrow'').  A situation can be interpreted as a name for an information set, or more abstractly, as a decision point.  Regardless of interpretation, axiom \rf{Pjw} implies a bijection between situations and information sets (footnote~\rf{D332}).  Myerson 1991 has a similar concept (footnote~\rf{E385}).\nocite{Myers91} \nichts{}}}

A 5-ary relation is merely a set of quintuples.  It is like a binary relation, which is a set of couples (ordered pairs), and also like a ternary relation, which is a set of triples.{\noterelation}  In this paper, a quintuple is denoted $⁅i˛j˛w˛a˛y⁆$.  The first element is understood to be a player, the second a situation (that is, a name for an information set),{\notesit} the third a decision node, the fourth an action, and the fifth a successor node.  Thereby, a set of quintuples is understood to specify relationships between players, situations, (two kinds of) nodes, and actions.

Such a quintuple set can specify a game-theoretic extensive form because an extensive form is, in essence, a collection of relationships between players, situations, nodes, and actions.  Correspondingly, this paper specifies an extensive form as a ``pentaform'', which is defined to be a quintuple set $Q$ which obeys eight axioms.  These axioms are formulated in terms of various projections of~$Q$.  For example, let $\pj{JI}(Q)$ denote the projection of $Q$ onto its first two coordinates, with their order reversed.  Then the first axiom requires that $\pj{JI}(Q)$ is a function.  In other words, the axiom requires that each situation $j$ is associated with exactly one player~$i$.  Intuitively, that player~$i$ is the one who controls the action at situation~$j$.  In a similar way, each of the other seven axioms formalizes one small independent feature of an extensive form.

\newcommand{\noteenderton}{\nichts{}}

\newcommand{\notediscrete}{\footnote{\label{D006}``Discreteness'' means that each decision node has a finite number of predecessors.  Non-discrete games include those in continuous time, as in Dockner, J\o rgenson, Long, and Sorger 2000; and those yet more general, as in Al\'{o}s-Ferrer and Ritzberger 2016, Chapters 1--5. \nocite{DocknJLS00} \nocite{AlosfR16} [Discreteness is defined in terms of {\em decision} nodes because Al\'{o}s-Ferrer and Ritzberger 2016, Section~6.2, admits a terminal node at the end of each infinite run (that is, each infinite play).  Such terminal nodes do not appear in the present paper.]
\nichts{}
}}

\newcommand{\noteprby}{\footnote{This paper does not formally consider probability.  If a game has a finite number of nodes, mixed strategies and expected utilities can be derived by standard means (for example, Myerson 1991, Chapter~4).  Meanwhile, if there are an infinite number of nodes, the very concepts of mixed strategy and expected utility can lead to subtle measurability issues.\nichts{}}}

Finally, a ``pentaform game'' is constructed by combining a pentaform with utility functions.{\noteprby}  To ground this new specification in the literature, the paper defines the concept of a ``traditional game'' to represent the literature's many specifications of finite-horizon and infinite-horizon games.  Such a traditional game is defined in the usual way as a tree adorned with information sets, actions, players, and utility functions.  The paper's main result is Theorem~\rf{B560}, which shows that there is a constructive and intuitive bijection{\noteenderton} from the collection of traditional games to the collection of pentaform games with information-set situations.  This suggests that pentaform games can equivalently formulate all discrete{\notediscrete} extensive-form games in the literature.
\nichts{}

\nssec{Motivation}{B573}

Pentaforms are easy to manipulate because they are sets, and because the pentaform axioms are largely compatible with the concepts of subset and union.

More precisely, Section~\rf{C577} shows that any subset of a pentaform satisfies six of the eight pentaform axioms (Proposition \rf{C372}).  This leads to a weak general condition under which a subset of a pentaform is itself a pentaform (Corollary~\rf{C579}).  This result is a powerful tool for disaggregating a pentaform.  First, Section~\rf{C577} uses the result to characterize the subsets of a pentaform that correspond to Selten 1975 subgames (Proposition~\rf{C576}).  Second, Section~\rf{C577} explains how the same result, applied to other subsets of a pentaform, is the foundation for the generalized theory of dynamic programming in Streufert 2023.  That sequel paper is the first paper to use value functions to characterize subgame perfection in arbitrary games.  

In the opposite direction, Section~\rf{C564} defines a ``block'' to be a quintuple set which satisfies all but one of the axioms.  Then it essentially shows that the union of a ``separated'' collection of blocks is itself a block (Proposition \rf{D415}), and that the union of an expanding sequence of pentaforms is itself a pentaform (Proposition~\rf{D454}).  These are convenient tools for building new pentaforms from known components, as illustrated by the finite-horizon examples of Section~\rf{C564} and the infinite-horizon example of Streufert 2023, Section~2.2.  These techniques are complementary to those of Capucci, Ghani, Ledent, and Nordvall Forsberg 2022 in the computer-science literature (footnote~\rf{D479}).\nichts{} 

In addition, pentaforms seem to have some less tangible benefits.  

(a) Other axiomatic foundations, such as the well-known foundation for consumer preferences,\nichts{} have readily fostered new extensions and results.  The same may occur with this paper's axiomatic foundation for extensive forms.  For example, the fine-grainedness of the eight axioms fostered the development of Propositions \rf{C372} and \rf{D415} for subsets and unions. 

(b) The pentaform notation is distinctly new.  Broadly, a pentaform is one high-dimensional relation, while a traditional extensive form is a list of low-dimensional relations.  To suggest the value of this unification, Section~\rf{C614} considers the concept of perfect-recall (Kuhn 1950, 1953), which simultaneously involves players, situations, nodes, and actions.  It turns out that the concept can be efficiently reformulated in terms of pentaforms.  The new formulation revolves around successor nodes rather than decision nodes, and is easily manipulated to produce a one-paragraph proof of the necessity of no-absentmindedness.

\nssec{Background}{B574}

There have been many innovations in how to specify an extensive form.  This paragraph discusses three broad strands in that literature.  (1) Eighty years ago, von Neumann and Morgenstern 1944 (pages 73--74; 78) specified each node as a set of outcomes, and noted that set inclusion arranges these nodes in a tree.  (2) Later, Kuhn 1953 specified each node as an abstract entity without any internal structure, and then separately specified the edges of the tree connecting the nodes.  (3) Next, Osborne and Rubinstein 1994 (page~200) specified each node as a sequence of past actions, and thereby created a tractable specification of infinite-horizon extensive forms.  More recently, Ritzberger 2002 (page~97) returned to the outcome-set formulation of von Neumann and Morgenstern 1944 and extended it to allow for an infinite horizon.  Then Al\'os-Ferrer and Ritzberger 2016 (Section~6.3), Kline and Luckraz 2016, and Streufert 2019\footnote{For the purposes here, lump Streufert 2019's choice-set form with its OR form, and lump its KS form with its simple form.} formally connected the three strands of literature, by building bijections between their three classes of extensive forms.  Infinite-horizon extensive forms are incorporated everywhere.\label{E419} \nocite{NeumaM44} \nocite{Kuhn5397} \nocite{OsborR94} \nocite{Ritzb02} \nocite{KlineL16} \nocite{five-1903}

\nichts{}

This paper's main result (Theorem~\rf{B560}) uses a bijection to formally connect a generic extension (Definition~\rf{C843}) of Kuhn 1953 to the new pentaform specification (Definition~\rf{C668}).  The pentaform specification has antecedents.  In some fashion, every extensive-form specification must determine a function which takes decision-node/feasible-action couples to successor nodes.  Often this function is implicitly specified by assigning actions to the edges of a tree diagram.  Alternatively, the function can be explicitly specified as a decision-node-indexed list of functions from actions to successor nodes (Kline and Luckraz 2016) or, in the reverse direction, as a pair of functions mapping each successor node to the immediately preceding node and the immediately preceding action (Mas-Colell, Whinston, and Green 1995, page~227).\footnote{A similar function from two variables to one can be found in the transition function of a stochastic game (Mertens 2002), and in the labelled transition system, i.e.\ state machine, of computer science (Blackburn, de Rijke, and Venema 2001, page~3).}  Streufert 2018 departs by viewing this function in set-theoretic terms, as a set of triples.  The present paper augments those triples with players and situations to form quintuples, and then defines an extensive form in terms of those quintuples. \nocite{MascoWG95} \nocite{Merte02} \nocite{ncp-1809} \nocite{BlackRV01}

\nichts{}

\nssec{Organization}{B575}

Section~\rf{B977} builds intuition through examples.  Section~\rf{B561} defines pentaform games.  Section~\rf{C563} discusses pentaform applications, and is unrelated to Section~\rf{C826}, which defines ``traditional games'' and relates them to pentaform games.  Finally, Appendix~\rf{E348} concerns graph-theoretic trees, and Appendices \rf{B566}--\rf{B567} provide lemmas and proofs for Sections \rf{B561}--\rf{C826}, respectively.

\section{Initial Intuition}\label{B977}\showit
\yyout{\markb{\sc \rf{B977}. Initial Intuition}} 

This brief section builds intuition through examples.  The examples suggest how an extensive form can be expressed by a quintuple set (that is, by a 5-ary relation).  This section presumes familiarity with tree diagrams.

\begin{figure}[h] \newcommand{\hgth}{2.7}
\begin{picture}(0,\hgth) \myoutergrid{\hgth} \renewcommand{\sella}[1]{#1}
  \put( 0.00, 1.10){\makebox(0,0){\scalebox{0.92}{ \begin{pspicture} \end{pspicture} }}} \end{picture}
\caption{\small By definition, $Q^1$ is the set consisting of the table's two rows, that is, $⎨⋅⁅\mspace{1.5mu}\f{Alex}˛⎨\f0⎬˛\f0˛\f{left}˛\f1\mspace{1.5mu}⁆,˙⁅\mspace{1.5mu}\ex{Alex}˛⎨\f0⎬˛\f0˛\f{right}˛\f2\mspace{1.5mu}⁆⋅⎬$.  The tree diagram provides the same data.  The set $Q^1$ is a ``pentaform'' (Section~\rf{B570}).} \label{B250} \end{figure}
 
Figure~\rf{B250}'s tree diagram has three nodes ($\f0$, $\f1$, and $\f2$), two actions ($\f{left}$ and $\f{right}$), one player ($\f{Alex}$), and one information set ($⎨\f0⎬$).  The root node ($\f0$) is underlined, and there are two edges (that is, ``arcs'' or ``twigs'').  These edges can be denoted $⁅\f0˛\f1⁆$ and $⁅\f0˛\f2⁆$ (brackets $⁅˙⁆$ will appear only on tuples).\label{E421}  Note that the action $\f{left}$ labels the edge $⁅\f0˛\f1⁆$, while the action $\f{right}$ labels the edge $⁅\f0˛\f2⁆$.  This data can be encoded within the triples $⁅\f0˛\f{left}˛\f1⁆$ and $⁅\f0˛\f{right}˛\f2⁆$.  Next, the node $\f0$ is in the information set $⎨\f0⎬$.  This (self-evident) fact can be encoded within the quadruples $⁅⎨\f0⎬˛\f0˛\f{left}˛\f1⁆$ and $⁅⎨\f0⎬˛\f0˛\f{right}˛\f2⁆$.  Finally, the player $\f{Alex}$ makes the decision at information set $⎨\f0⎬$.  This fact can be encoded within the quintuples $⁅\f{Alex}˛⎨\f0⎬˛\f0˛\f{left}˛\f1⁆$ and $⁅\f{Alex}˛⎨\f0⎬˛\f0˛\f{right}˛\f2⁆$.  In this sense, the set \begin{gather}
\zz
Q^1 = ⎨⋅⁅\f{Alex}˛⎨\f0⎬˛\f0˛\f{left}˛\f1⁆,˙⁅\f{Alex}˛⎨\f0⎬˛\f0˛\f{right}˛\f2⁆⋅⎬ \notag
\zz
\end{gather} expresses Figure~\rf{B250}'s tree diagram.  The set's two quintuples correspond to the two (non-header) rows in the figure's table (eventually the information set $⎨\f0⎬$ will be regarded as a special kind of situation).  Finally, the superscript on $Q^1$ distinguishes this first example from future examples.\label{E398}

This process can be readily generalized.  Essentially, each tree edge is changed into a quintuple.  To be more specific, each tree edge is a couple $⁅w˛y⁆$ consisting of a decision node $w$ and a successor node~$y$.  This couple is changed into the quintuple $⁅i˛j˛w˛a˛y⁆$ in which $a$ is the action labelling the edge $⁅w˛y⁆$, $j$ is the information set containing the decision node $w$, and $i$ is the player making the decision at information set~$j$.  Eventually the information set $j$ will be regarded as a special kind of situation.

\newcommand{\notestory}{\footnote{\label{B292}Figures \rf{B225} and \rf{B224} (for examples $Q^2$ and $Q^3$) correspond to Selten 1975's well-known ``horse'' game.  To tell a story for these figures, suppose a $\f{Kid}$ must decide, today, between the bad action of not doing her homework (called $\fb$) and the correct action of doing her homework (called $\fc$).  Next, tonight, if the homework has been finished (node $\f1$), a $\f{Dog}$ must decide between the dumb action of eating the homework ($\fd$) and the good action of going back to sleep ($\fg$).  Finally, tomorrow, without knowing whether the kid chose bad (node $\f2$) or the kid chose correct and the dog chose dumb (node $\f3$), the $\f{Teacher}$ must decide between excusing the kid ($\fe$) and failing the kid ($\ff$).}} 
 
Figure~\rf{B225}'s tree diagram is more complicated than the previous example because it has a nonsingleton information set.{\notestory}  Nonetheless it can be expressed as a quintuple set by the same process.  The tree has eight edges.  The edge $⁅\f3˛\f7⁆$ is changed to the quintuple $⁅˙\ex{Teacher},⎨\f2˛\f3⎬,\f3,\ff,\f7˙⁆$ to encode the facts that (i) the action $\ff$ labels the edge $⁅\f3˛\f7⁆$, (ii) the decision node $\f3$ is (self-evidently) in the information set $⎨\f2˛\f3⎬$, and (iii) the player $\ex{Teacher}$ controls the move at the information set $⎨\f2˛\f3⎬$.  By similarly changing the other seven edges $⁅w˛y⁆$ to quintuples $⁅i˛j˛w˛a˛y⁆$, one obtains all eight rows in Figure~\rf{B225}'s table.  Those rows define the set $Q^2$.

\yyin{\markb{\sc \rf{B977}. Initial Intuition}} 

\begin{figure}[t] \newcommand{\hgth}{4.6}
\begin{picture}(0,\hgth) \myoutergrid{\hgth} \renewcommand{\sellb}[1]{#1}
  \put(-0.50, 2.00){\makebox(0,0){\scalebox{0.91}{ \begin{pspicture} \end{pspicture} }}} \end{picture}
\caption{\small By definition, $Q^2$ is the set consisting of the table's rows (expressed as quintuples).  The tree diagram provides the same data.  The set $Q^2$ is a ``pentaform'' (Section~\rf{B570}).} \label{B225} \end{figure}

\begin{figure}[t] \newcommand{\hgth}{4.6}
\begin{picture}(0,\hgth) \myoutergrid{\hgth} \renewcommand{\sellc}[1]{#1}
  \put(-0.80, 2.00){\makebox(0,0){\scalebox{0.91}{ \begin{pspicture} \end{pspicture} }}} \end{picture}
\caption{\small By definition, $Q^3$ is the set of the table's rows (expressed as quintuples).  The tree diagram provides the same data.  In $Q^3$, situations are not information sets.  The set $Q^3$ is a ``pentaform'' (Section~\rf{B570}).} \label{B224} \end{figure}

\newcommand{\notesitmg}{\footnote{\label{E385}A situation $j$ is comparable to an information state $s$ in Myerson 1991, Chapter~4 (there players are assumed to have disjoint sets of information states, unlike in Myerson 1991, Chapter~2).  
\nichts{}
}}

\pagebreak
Figure~\rf{B224}'s tree diagram differs from the previous example to the extent that its information sets are named with the words ``$\f{today}$'', ``$\f{tonight}$'', and ``$\f{tomorrow}$'' (these words have meaning within the story of footnote \rf{B292}).  These three words are examples of situations.  As footnote \rf{B598} explains, situations can be either information sets (as in the first two examples) or something else (such as the three words here).  A situation can be interpreted as a name for an information set, or more abstractly, a decision point.{\notesitmg}  

This third example can be expressed as a quintuple set just as the previous two examples.  Specifically, the edge $⁅\f3˛\f7⁆$ becomes the quintuple $⁅˙\ex{Teacher}˛\ex{tomorrow}˛\f3˛\ff˛\f7˙⁆$ to encode the facts that (i) the action $\ff$ labels the edge $⁅\f3˛\f7⁆$, (ii) the decision node $\f3$ is associated with the situation $\f{tomorrow}$, and (iii) the player $\ex{Teacher}$ controls the move in the situation $\f{tomorrow}$.  By similarly changing the other seven edges to quintuples, one obtains all eight rows in Figure~\rf{B224}'s table.  These rows define the set $Q^3$.

For a fourth example, consider the quintuple set \begin{gather}
\zz
Q^4 = ⎨˙⁅\f{41}˛\f{42}˛\f{43}˛\f{44}˛\f{45}⁆,˙⁅\f{46}˛\f{47}˛\f{48}˛\f{49}˛\f{50}⁆˙⎬. \label{C782}
\zz
\end{gather} Obviously, an arbitrary quintuple set like $Q^4$ may or may not express a tree diagram.

Looking ahead, Definition~\rf{C669} will designate which quintuple sets are to be called ``pentaforms'', and Definition~\rf{C668} will define a ``pentaform game'' to be a pentaform augmented with utility functions.  In the end, Theorem~\rf{B560} (the paper's main result) will show that there is a bijection from the collection of ``traditional games'' to the collection of pentaform games with information-set situations.  The forward direction of that bijection closely resembles this section's informal process of expressing Figure~\rf{B250} and \rf{B225}'s tree diagrams as quintuple sets.

\pagebreak
\section{Pentaform Games}\label{B561}\showit
\markb{\sc \rf{B561}. Pentaform Games}
\setcounter{figure}{-1}

This Section~\rf{B561} defines pentaforms and pentaform games.  The examples of Section~\rf{B977} are used as illustrations.

\nssec{The components of a quintuple}{C594}

An arbitrary quintuple will be denoted $⁅i,j,w,a,y⁆$.  Call its first component $i$ the {\em player}, call its second component $j$ the {\em situation}, call its the third component $w$ the {\em decision node}, call its fourth component $a$ the {\em action}, and call its fifth component $y$ the {\em successor node}.  These five terms have no formal content.  They merely name the five positions in a quintuple.  For example, in the quintuple $⁅\f{46,47,48,49,50}⁆$, the player is $\f{46}$, the situation is $\f{47}$, the decision node is $\f{48}$, the action is $\f{49}$, and the successor node is $\f{50}$.  Further, let the {\em nodes} of a quintuple be its decision node and its successor node.  In other words, let the nodes of a quintuple be its third and fifth components.  For example, the nodes of $⁅\f{46,47,48,49,50}⁆$ are $\f{48}$ and $\f{50}$.\footnote{\label{D514}When necessary, a generic node will be denoted by~$x$.  Mnemonically, $w$ is before $x$ is before $y$ in the alphabet, and correspondingly, decision nodes $w$ are ``early'' nodes, and successor nodes $y$ are ``late'' nodes.}
 
This concept of quintuple is extremely (and perhaps disturbingly) abstract in the sense that no structure is imposed on the components of a quintuple.\label{E423}  For example, in the quintuple $⁅\f{46,47,48,49,50}⁆$, each of the five components is an alphanumeric string (strings use the $\f{sans⋅serif⋅font}$).  Similarly, in the quintuple $⁅\f{Teacher},\f{tomorrow},\f3,\ff,\f7⁆$ from Figure~\rf{B224}, each of the five components is an alphanumeric string.  Meanwhile, in the quintuple $⁅\f{Teacher}, ⎨\f2˛\f3⎬, \f3, \ff, \f7⎬$ from Figure~\rf{B225}, four of the five components are alphanumeric strings, but one of the components is a set of alphanumeric strings.  In general, each component can be practically anything, including a number, a set of numbers, or a sequence of numbers or strings. 
\nichts{}
\nichts{}

One advantage of this abstract formulation is that it is extremely flexible.  One use of this flexibility is to directly accommodate the literature's wide variety of notations for situations, nodes, and actions.%
\nichts{} 
For example, situations can be specified as sets of nodes (as in the information-set situation $⎨\f2˛\f3⎬$ from Figure~\rf{B225}) or without any special structure (as in the situation $\f{tomorrow}$ from Figure~\rf{B224}).  Meanwhile, nodes can be specified as sequences of actions (Osborne and Rubinstein 1994), as sets of actions (Streufert 2019), as sets of outcomes (Al\'os-Ferrer and Ritzberger 2016, Section~6.2; see footnote~\rf{D006} here), or without any special structure (as in the node $\f3$ from Figures \rf{B225} and \rf{B224}).  Finally, actions can be specified as sets of nodes (van Damme 1991, page~103), as sets of edges (Selten 1975), as sets of outcomes (Al\'os-Ferrer and Ritzberger 2016, Section~6.2; see footnote~\rf{D006} here), or without any special structure (as in the action $\ff$ from Figures \rf{B225} and \rf{B224}).  
\nichts{}  
\nocite{OsborR94} \nocite{five-1903} \nocite{AlosfR16} \nocite{Damme91} \nocite{Selte75}

\nssec{Quintuple sets and their slices}{C595}

A set of quintuples will typically be denoted by the letter~$Q$.\nichts{} Correspondingly, different quintuple sets will typically be distinguished from one another by means of markings around the letter~$Q$.  For instance, the examples in Section~\rf{B977} are denoted $Q^1$, $Q^2$, $Q^3$, and $Q^4$.  Similarly, Section \rf{C577} will consider subsets of a quintuple set $Q$ denoted $Q′⋅⊆⋅Q$ and $\bQ{t}⋅⊆⋅Q$.  Likewise, this section will consider other subsets of a quintuple set $Q$ which will be denoted $Q_j⋅⊆⋅Q$ .  

Now consider an arbitrary quintuple set $Q$, and let $J$ denote its set of situations~$j$.  In other words, let $J$ be the projection of $Q$ onto its second coordinate.  Then, for each situation $j⋅∈⋅J$, define\begin{gather}
\zz
Q_j = ⎨⋅⁅i_*,j,w_*˛a_*˛y_*⁆∈Q⋅⎬. \label{D304}
\zz
\end{gather} Thus $Q_j$ is the set of quintuples in $Q$ that have situation~$j$.  Call $Q_j$ the {\em slice of $Q$ for situation $j$}.  By inspection, $⁅Q_j⁆_{j∈J}$ is an injectively indexed partition of~$Q$.\nichts{}  Call this the {\em slice partition} of~$Q$.\nichts{}

\begin{figure}[b] \newcommand{\hgth}{6.3}
\begin{picture}(0,\hgth) \myoutergrid{\hgth} \renewcommand{\sellc}[1]{#1}
  \put(-0.20, 2.60){\makebox(0,0){\scalebox{1.04}{ \begin{pspicture} \end{pspicture} }}} \end{picture}
\caption{\small The slice partition of $Q^3$, as illustrated by both the tree diagram and the table.  The set $Q^3$ is from Figure~\rf{B224} and is a pentaform (Section~\rf{B570}).  The partition's three slices are written explicitly in eq.~(\rf{C783}).} \label{D348} \end{figure}

For example, consider example $Q^3$ from Figure~\rf{B224}.  In this example, the situation set $J^3$ is $⎨\f{today},\f{tonight},\f{tomorrow}⎬$.  Further, the slice partition $⁅Q^3_j⁆_{j∈\smash{J^3}}$ divides $Q^3$ into the three quintuple sets \subi \label{C783} \begin{align}
\zz
Q^3_{\f{today}} = ⎨˙&⁅\f{Kid˛today˛0˛c˛1}⁆, ⁅\f{Kid˛today˛0˛b˛2}⁆˙⎬, 
  \label{C784} \\
Q^3_{\f{tonight}} = ⎨˙&⁅\f{Dog˛tonight˛1˛g˛8}⁆, ⁅\f{Dog˛tonight˛1˛d˛3}⁆˙⎬,⋅\text{and} 
  \label{C785} \\
Q^3_{\f{tomorrow}} = ⎨˙&⁅\f{Teacher˛tomorrow˛2˛e˛4}⁆,⁅\f{Teacher˛tomorrow˛2˛f˛5}⁆, 
  \label{C786} \\
&⁅\f{Teacher˛tomorrow˛3˛e˛6}⁆,⁅\f{Teacher˛tomorrow˛3˛f˛7}⁆˙⎬.
  \notag
\zz
\end{align} \subo These three sets are illustrated in Figure~\rf{D348}.\footnote{\label{E399}The unusual figure number ``3.0'' is intended to simplify example references.  In the end, Figures \rf{B250} and \rf{B265} concern example $Q^1$; Figures \rf{B225} and \rf{B266} concern example $Q^2$; and Figures \rf{B224} and \rf{D348} concern example $Q^3$.}

\pagebreak
\nssec{Projections}{C596} 

Any quintuple set can be projected onto any sequence in $⎨I˛J˛W˛A˛Y⎬$.  For example, Figure~\rf{B250}'s table for $Q^1$ implies\footnote{When speaking aloud, it may be helpful to read $π_Y(Q^1)$ as ``the $Y$ of $Q^1$'' (abbreviation (\rf{C825}) shortens this to~$Y^1$).  Similarly, it may be helpful to read $\pj{JI}(Q^1)$ as ``the $JI$ of $Q^1$''.\nichts{}}
\begin{gather}
\zz
π_Y(Q^1) = ⎨˙y˙|˙(∃i˛j˛w˛a)˙⁅i˛j˛w˛a˛y⁆∈Q^1˙⎬ = ⎨\f{1,2}⎬⋅\text{and} 
  \label{C788} \\
\pj{JI}(Q^1) = ⎨˙⁅j˛i⁆˙|˙(∃w˛a˛y)˙⁅i˛j˛w˛a˛y⁆∈Q^1˙⎬ = ⎨˙⁅\f{⎨0⎬,Alex}⁆˙⎬. 
  \label{C789}
\zz
\end{gather} For another example, equation (\rf{C786}) for $Q^3_{\f{tomorrow}}$ implies  \begin{gather} 
\zz
π_A(Q^3_{\f{tomorrow}}) = ⎨˙a˙|˙(∃i˛j˛w˛y)˙⁅i˛j˛w˛a˛y⁆∈Q^3_{\f{tomorrow}}˙⎬ = ⎨\fe,\ff⎬⋅\text{and} 
  \label{C790} \\
\pj{WA}(Q^3_\f{tomorrow}) = ⎨˙⁅w˛a⁆˙|˙(∃i˛j˛y)˙⁅i˛j˛w˛a˛y⁆∈Q^3_{\f{tomorrow}}˙⎬ \label{C791} \\
˙= ⎨˙⁅\f{2˛e}⁆, ⁅\f{2˛f}⁆,⁅\f{3˛e}⁆, ⁅\f{3˛f}⁆˙⎬. \notag
\zz
\end{gather} Note that projections, like $\pj{JI}(Q^1)$ in (\rf{C789}), can re-order the coordinates.  Also note that projections of slices are well-defined simply because slices are quintuple sets.  An example is the slice $Q^3_{\f{tomorrow}}$ in (\rf{C786}), (\rf{C790}), and (\rf{C791}).
\nichts{}

Both slices and projections can be visualized by tables.  Slices select rows and then projections select columns.  For example, consider Figure~\rf{D348}'s table for $Q^3$.  Its last four rows constitute the slice $\smash{Q^3_{\f{tomorrow}}}$ in (\rf{C786}), and this slice's fourth column determines the projection $π_A(Q^3_{\f{tomorrow}})$ in (\rf{C790}).

The notation for a single-coordinate projection will often be abbreviated by replacing the letter $Q$ with the single coordinate.  Specifically, define the five abbreviations $I$, $J$, $W$, $A$, and $Y$ by \begin{gather}
\zz
I = π_I(Q),⋅J = π_J(Q),⋅W = π_W(Q),⋅A = π_A(Q),⋅\text{and}⋅Y = π_Y(Q). \label{C825}
\zz
\end{gather} These abbreviations inherit any markings on the letter~$Q$.  For example, (\rf{C788}) shows that $Y^1 = π_Y(Q^1)$ is $⎨\f1,\f2⎬$.  Similarly, (\rf{C790}) shows that $A^3_{\f{tomorrow}} = π_A(Q^3_{\f{tomorrow}})$ is~$⎨\fe,\ff⎬$.\nichts{}  

\nichts{}

Two special cases will become important after Definition~\rf{C669}.  Both concern a quintuple set $Q$ and one of its situations $j⋅∈⋅J$⋅(that is, $j⋅∈⋅π_J(Q)$).  

First, $W_j$ (that is $π_W(Q_j)$) is the decision-node set of situation~$j$.  In order to accord with standard terminology, call $W_j$ the {\em information set} of situation~$j$.  For instance, in example $Q^3$, Figure~\rf{D348}'s table\nichts{} shows that the information set of the situation $j = \f{tomorrow}$ is \begin{gather}
\zz
W^3_{\f{tomorrow}} = ⎨\f2,\f3⎬. \label{C794}
\zz
\end{gather}  Similarly, in example $Q^2$, Figure~\rf{B225}'s table shows that the information set of the situation $j = ⎨\f2,\f3⎬$ is \begin{gather}
\zz
W^2_{⎨\f2,\f3⎬} = ⎨\f2,\f3⎬. \label{C795}
\zz
\end{gather} Thus a situation $j$ may or may not equal its information set $W_j$.  This distinction will play a role in Sections~\rf{B565}--\rf{E435}, beginning with equation (\rf{D319}).

Second, $A_j$ (that is $π_A(Q_j)$) is the action set of situation~$j$.  For reasons associated with Proposition~\rf{C601}, $A_j$ can also be called the {\em feasible} action set at situation~$j$.

\newcommand{\tableQu}{
\begin{table}[t]
\centering
{\small 
\begin{tabular}{clr} 
\multicolumn{2}{l}{Pentaform $Q$} &⋅\SMALL [\rf{B570}]
  \\\hline \\[-3.6mm] 
$Q$ & set of quintuples $⁅i˛j˛w˛a˛y⁆$ &\SMALL [\rf{C594},˙\rf{C595}]\\
$I{=}π_I(Q)$ & $\dve$ set of players $i$ &\SMALL [\rf{C594},˙\rf{C596}]\\
$J{=}π_J(Q)$ & $\dve$ set of situations $j$ &\SMALL [\rf{C594},˙\rf{C596}]\\
$W{=}π_W(Q)$ & $\dve$ set of decision nodes $w$ &\SMALL [\rf{C594},˙\rf{C596}]\\
$A{=}π_A(Q)$ & $\dve$ set of actions $a$ &\SMALL [\rf{C594},˙\rf{C596}]\\
$Y{=}π_Y(Q)$ & $\dve$ set of successor nodes $y$ &\SMALL [\rf{C594},˙\rf{C596}]
\\[2mm]
$Q_j⊆Q$ & $\dve$ situation $j$'s slice of $Q$ &\SMALL [\rf{C595}]\\
$W_j{=}π_W(Q_j)$ & $\dve$ situation $j$'s decision-node set (information set)\hspace{-6mm} &\SMALL [\rf{C596}]\\
$A_j{=}π_A(Q_j)$ & $\dve$ situation $j$'s (feasible) action set &\SMALL [\rf{C596}]
\\[2mm]
$p{=}\pj{YW}(Q)$ & $\dve$ immediate-predecessor function &\SMALL [\rf{B570}]\\
$F{=}\pj{WA}(Q)$ & $\dve$ feasibility correspondence &\SMALL [\rf{B570}]
\\[2mm]
$W⋃Y$ & $\dve$ set of nodes $x$ &\SMALL [\rf{C657}]\\
$\pj{WY}(Q)$ & $\dve$ set of edges $⁅w˛y⁆$ &\SMALL [\rf{C657}]\\
$⎨r⎬{=}W⧷Y$ & $\dve$ root node $r$ &\SMALL [\rf{C657}]\\
$≼$ & $\dve$ weak precedence order & \SMALL [\rf{C657},˙\rf{E349}]\\
$≺$ & $\dve$ strict precedence order & \SMALL [\rf{C657},˙\rf{E349}] \\ 
$Y⧷W$ & $\dve$ set of end nodes $y$ & \SMALL [\rf{C657},˙\rf{E349}] \\
$\ZZ$ & $\dve$ collection of runs $Z$ &\SMALL [\rf{C657},˙\rf{E349}]
\\[2mm]
\multicolumn{2}{l}{Pentaform Game $(Q,u)$} & \SMALL [\rf{B562}]
  \\\hline \\[-3.6mm]
$u{=}⁅u_i⁆_{i∈I}$ & profile with utility function $u_i$ for each player $i$ &\SMALL [\rf{B562}]
\\[4mm]
\end{tabular} }
\caption{\small A pentaform is implicitly accompanied by its derivatives (\protect\rotatebox[origin=c]{180}{$\Lsh$}).  Definitions are in the sections in brackets {\SMALL [˙]}.} \label{C492}
\end{table} 
}

\nssec{Pentaforms}{B570}

\newcommand{\notepp}{\footnote{\label{E336}The statements $w = p(y)$, $⁅y˛w⁆⋅∈⋅\pj{YW}(Q)$, and $⁅w˛y⁆⋅∈⋅\pj{WY}(Q)$ are equivalent.}}

\newcommand{\noteaxioms}{\footnote{\label{C667}The label \rf{Pij} can be read as ``$i$ is a function of $j$''. The labels \rf{Pjw}, \rf{Pwy}, and \rf{Pay} can be read similarly.  Meanwhile, the label \rf{Pway} can be read as ``$w$ and $a$ determine $y$''.  The arrows within the labels are visual crutches in the sense that they can be removed without introducing ambiguity.}}

\newcommand{\notefcn}{\footnote{\label{C273}As in set theory (Halmos 1974, page 30; Enderton 1977, page~42), this paper identifies a function with its graph.  For readers unfamiliar with set theory, [1] a {\em function} $f$ is a set of couples such that $(∀x,y^A,y^B)$ $⁅x˛y^A⁆⋅∈⋅f$ and $⁅x,y^B⁆⋅∈⋅f$ imply $y^A = y^B$, [2] the {\em domain} of a function $f$ is $π_1f$, and its {\em range} is $π_2f$, [3] a {\em surjection from $X$ to $Y$} is a function with domain $X$ and range $Y$, [4] a {\em bijection from $X$ to $Y$} is an injective surjection from $X$ to $Y$, and [5] ``$f{:}X→Z$'' means $f$ is a function with domain $X$ and range included within~$Z$.}}

\newcommand{\notedoublecouple}{\footnote{\label{E401}The set in axiom \rf{Pway} consists of couples of the form $⁅⁅w˛a⁆˛y⁆$ whose first element is a couple of the form $⁅w˛a⁆$.  The axiom requires that the set of ``outer'' couples is a function in the sense of footnote~\rf{C273}.  Casually this means that $\pj{WAY}(Q)$ is a function from its first two coordinates.}}

\newcommand{\fcnof}{\hspace{.1ex}$\shortleftarrow$\hspace{-.9ex}%
  {\color{white}\rule{1ex}{1ex}}\hspace{-.9ex}}
\newcommand{\Pway}{\mbox{[\hspace{3.5ex}$\shortrightarrow$\hspace{-1.8ex}%
  {\color{white}\rule{0.8ex}{1ex}}\hspace{-4.3ex}Pwa\hspace{1ex}y]}}

\begin{ndef}[{\bf Pentaform}]\label{C669} A {\em pentaform} is a (possibly infinite) set $Q$ of quintuples $⁅i˛j˛w˛a˛y⁆$ such that\qquad\noteaxioms\notefcn\notedoublecouple\notepp%
\begin{picture}(0,0)%
  \put(-1.23,0){\color{white}\nichts{}\rule{7ex}{2ex}} \end{picture} %
\begin{gather}
\zz
\ttt{\mbox{[Pi{\fcnof}j]}}{Pij}%
\begin{picture}(0,0)%
  \put(-4.55,0){$^{\text{\normalfont\rf{C667}}}$} \end{picture} %
\pj{JI}(Q)⋅\text{is a function,}^{\text{\normalfont\rf{C273}}}\\
\ttt{\mbox{[Pj{\fcnof}w]}}{Pjw} \pj{WJ}(Q)⋅\text{is a function}, \\
\ttt{[Pwa]}{Pwa} (∀j∈J)⋅\pj{WA}(Q_j)⋅\text{is a Cartesian product}, \\
\ttt{\Pway}{Pway} ⎨˙⁅⁅w˛a⁆˛y⁆˙|˙⁅w˛a˛y⁆∈\pj{WAY}(Q)˙⎬⋅\text{is a function},^{\text{\normalfont\rf{E401}}} \\
\ttt{\mbox{[Pw\hspace{-.1ex}{\fcnof}y]}}{Pwy} \pj{YW}(Q)⋅\text{is a function,} \\[-.6ex]
\text{denoted by $p$ and called the {\em immediate-predecessor function},}^{\text{\normalfont\rf{E336}}} \nt
\ttt{\mbox{[Pa{\fcnof}y]}}{Pay} \pj{YA}(Q)⋅\text{is a function}, \\
\ttt{[Py]}{Py} (∀y∈Y)(∃m≥1)⋅p^m(y)⋅∉⋅Y,⋅\text{and} \\
\ttt{[Pr]}{Pr} W⧷Y⋅\text{is a singleton}
\zz
\end{gather} (where (\rf{D304}) defines each $Q_j$ and where (\rf{C825}) defines $J$, $W$, and $Y$). \end{ndef}

{\tableQu}

The remainder of this Section~\rf{B570} will discuss these eight axioms individually.  En route, Section~\rf{B977}'s examples will be reconsidered.  It will be found that examples $Q^1$, $Q^2$, and $Q^3$ satisfy the axioms, and are therefore pentaforms.  (An example of an infinite pentaform can be found in Streufert 2023, Section~2.2.)

\smallskip{\em Axiom~\rf{Pij}}.  This states that exactly one player $i$ is assigned to each situation~$j$.  This is interpreted to mean that exactly one player controls the action at each situation. To be clear, \rf{Pij} states that $\pj{JI}(Q)$ is a function, which (by footnote \rf{C273}) means that each $j⋅∈⋅J$ is associated with exactly one $i⋅∈⋅I$.  For example, (\rf{C789}) shows $\pj{JI}(Q^1) = ⎨⁅\f{⎨0⎬,Alex}⁆⎬$, and this is a function which maps $⎨\f0⎬$ to $\f{Alex}$ (here the only situation $j$ is the information set $⎨\f0⎬$).  Similarly, examples $Q^2$, $Q^3$, and $Q^4$ satisfy axiom \rf{Pij}.

\smallskip{\em Axiom~\rf{Pjw}}.  This states that exactly one situation $j$ is assigned to each decision node~$w$.  As Proposition~\rf{D328}(\rf{D329}$⟺$\rf{D330}) makes clear, this is equivalent to stating that distinct situations $j_1$ and $j_2$ have disjoint information sets $W_{j_1}$ and $W_{j_2}$.  For instance, the situation set in example $Q^3$ is $J^3 = ⎨\f{today},\f{tonight},\f{tomorrow}⎬$.  Further, (\rf{C783}) implies that the information sets are $W^3_{\f{today}} = ⎨\f0⎬$, $W^3_{\f{tonight}} = ⎨\f1⎬$, and $W^3_{\f{tomorrow}} = ⎨\f2,\f3⎬$ (the third equality already appeared in (\rf{C794})).  Since these are disjoint, Proposition~\rf{D328}(\rf{D329}$⇐$\rf{D330}) implies $Q^3$ satisfies \rf{Pjw}.  Similarly, examples $Q^1$, $Q^2$, and $Q^4$ satisfy \rf{Pjw}.

\newcommand{\noteindexp}{\footnote{\label{D332}To be clear, Proposition~\rf{D328}(\rf{D331}) means that [1] $⎨⁅j˛W_j⁆|j∈J⎬$ is an injection and [2] $⎨W_j|j∈J⎬$ partitions~$W$.  (Note that [1] implies that $⎨⁅j˛W_j⁆|j∈J⎬$ is a bijection from the situation set $J$ to the information-set collection $⎨W_j|j∈J⎬$.)\nichts{}}}

\begin{prop}\label{D328} \mbox{Let $Q$ be a quintuple set.  Then the following are equivalent.} \begin{tlist}
\yl{D329} $Q$ satisfies \rf{Pjw}.
\yl{D330} $(∀j_1∈J,j_2∈J)$ $j_1⋅≠⋅j_2$ implies $W_{j_1}⋂W_{j_2} = ∅$.
\yl{D331} $⁅W_j⁆_{j∈J}$ is an injectively indexed partition of~$W$.{\noteindexp} (Proof~\rf{D328p} in Appendix~\rf{B566}.) \end{tlist} \end{prop}  

\smallskip{\em Axiom \rf{Pwa}}.  This states that for each situation $j$, the set $\pj{WA}(Q_j)$ is a Cartesian product.  For instance, in example $Q^3$, axiom \rf{Pwa} is satisfied at $j = \f{tomorrow}$ because equation (\rf{C791}) implies $\pj{WA}(Q^3_{\f{tomorrow}}) = ⎨\f2,\f3⎬×⎨\fe,\ff⎬$.  Cartesian products can also be found at the other two situations in $Q^3$ and at all situations in $Q^1$, $Q^2$, and~$Q^4$.  Hence all four examples satisfy \rf{Pwa}.

\newcommand{\notecorres}{\footnote{\label{B359}In this paper, a correspondence is simply a set of couples.  Occasionally, the expression ``$F{:}X⇉Z$'' is used to mean ``$F$ is a correspondence such that $π_1F = X$ and $π_2F⋅⊆⋅Z$''.  (This paper does not apply the terms ``domain'' and ``range'' to correspondences.)}}

To interpret \rf{Pwa}, define the correspondence{\notecorres}\begin{gather}
\zz
F = \pj{WA}(Q). \label{C981}
\zz
\end{gather} Call $F$ the {\em feasibility} correspondence, and call $F(w)⋅⊆⋅A$ the set of {\em feasible} actions at decision node $w⋅∈⋅W$.  For example, Figure~\rf{D348}'s table implies \begin{gather}
\zz
F^3 = \pj{WA}(Q^3) = 
⎨˙⁅\f0˛\fb⁆,⁅\f0˛\fc⁆,⁅\f1˛\f{d}⁆,⁅\f1˛\fg⁆,⁅\f2˛\fe⁆,⁅\f2˛\ff⁆,⁅\f3˛\fe⁆,⁅\f3˛\ff⁆˙⎬, \notag 
\zz
\end{gather} which implies $F^3(\f0) = ⎨\fb,\fc⎬$, $F^3(\f1) = ⎨\f{d},\fg⎬$, and $F^3(\f2) = F^3(\f3) = ⎨\fe,\ff⎬$.  Below, Proposition~\rf{C601}(\rf{C602}$⟺$\rf{C604}) characterizes \rf{Pwa} by the property that the feasible set $F(w)$ is constant across the nodes $w$ in an information set.  For instance, in the example $Q^3$, equation (\rf{C794}) implies that nodes $\f2$ and $\f3$ share the information set $W^3_{\f{tomorrow}} = ⎨\f2,\f3⎬$, and the second-previous sentence shows that their feasible sets $F^3(\f2)$ and $F^3(\f3)$ are both equal to $⎨\fe,\ff⎬$.  [Further, in accord with condition (\rf{C603}) of Proposition~\rf{C601}, this common feasible set is $A^3_{\f{tomorrow}} = ⎨\fe,\ff⎬$.  For verification see (\rf{C790}).] 

\begin{prop}\label{C601} \mbox{Suppose $Q$ satisfies \rf{Pjw}.  Then the following are equivalent.} \begin{tlist}
\yl{C602} $Q$ satisfies \rf{Pwa}.
\yl{C605} $(∀j∈J)$ $\pj{WA}(Q_j) = W_j×A_j$.
\yl{C603} $(∀j∈J,w∈W_j)$ $F(w) = A_j$.
\yl{C604} $(∀j∈J,w_1∈W_j,w_2∈W_j)$ $F(w_1) = F(w_2)$. (Proof~\rf{C601p} in Appendix~\rf{B566}.) 
\end{tlist} \end{prop}

\smallskip{\em Axiom~\rf{Pway}}.\label{E422}  This axiom states that each decision-node/feasible-action couple $⁅w˛a⁆⋅∈⋅\pj{WA}(Q)$ determines a successor node~$y$.  (Intuitively, selecting a feasible action at a decision node has the effect of selecting one of the decision node's successors.)  To be clear, note that $\pj{WA}(Q)$ is the domain of the function in \rf{Pway}, and that $\pj{WA}(Q)$ can be accurately called the set of decision-node/feasible-action couples because definition (\rf{C981}) implies $⁅w˛a⁆⋅∈⋅\pj{WA}(Q)$ iff $a⋅∈⋅F(w)$.  In a different direction, it is sufficient for \rf{Pway} that each decision-node/feasible-action couple appears in exactly one quintuple in~$Q$.  By inspection, this holds in examples $Q^1$, $Q^2$, $Q^3$, and~$Q^4$.

\smallskip{\em Axioms \rf{Pwy} and \rf{Pay}}.  These axioms are similar.  Axiom \rf{Pwy} states that exactly one decision node is associated with each successor node.  Similarly, \rf{Pay} states that exactly one action is associated with each successor node.  It is sufficient for both \rf{Pwy} and \rf{Pay} that each successor node $y⋅∈⋅Y$ appears in exactly one quintuple in~$Q$.\nichts{}   By inspection, this holds in examples $Q^1$, $Q^2$, $Q^3$, and $Q^4$.

\smallskip{\em Axiom \rf{Py}}.  This can be understood in the terms of a difference equation (Luenberger 1979, page 14).  In particular, consider the difference equation $y_{k-1} = p(y_k)$, where the index $k⋅∈⋅⎨0,-1,-2,...⎬$ runs backwards, and where $p = \pj{YW}(Q)$ is the immediate-predecessor function defined in axiom \rf{Pwy} of Definition~\rf{C669}.  In this context, axiom \rf{Py} states that the backward walk starting from any $y_0⋅∈⋅Y$ eventually leaves~$Y$.  In casual terms, the set $Y$ is a ``set-source''.  For instance, consider example $Q^4$.  There definition (\rf{C782}) implies that $Y^4 = ⎨\f{45},\f{50}⎬$ and that $p^4 = \pj{YW}(Q^4) = ⎨˙⁅\f{45}˛\f{43}⁆,⁅\f{50}˛\f{48}⁆˙⎬$.  Hence $Y^4$ is a ``set-source'' since $(p^4)(\f{45}) = \f{43}⋅∉⋅Y^4$ and since $(p^4)(\f{50}) = \f{48}⋅∉⋅Y^4$.  The examples $Q^1$, $Q^2$, and $Q^3$ also satisfy \rf{Py}, though exhaustively showing so takes longer than for $Q^4$ because there can be more successor nodes in $Y$ and because the backward walks exiting $Y$ can be longer.\nocite{Luenb79}

\newcommand{\noter}{\nichts{}}

\smallskip{\em Axiom \rf{Pr}}.  This states that there is exactly one decision node which is not also a successor node.{\noter} For instance, consider example $Q^3$.  There Figure~\rf{D348}'s table implies that $W^3 = ⎨\f{0˛1˛2˛3}⎬$ and that $Y^3 = ⎨\f{1˛2˛3˛4˛5˛6˛7˛8}⎬$, and these observations imply that $W^3⧷Y^3 = ⎨\f0⎬$.  Thus $Q^3$ satisfies \rf{Pr}.  Similarly examples $Q^1$ and $Q^2$ satisfy \rf{Pr}.  Meanwhile, example $Q^4$ violates \rf{Pr} because definition~(\rf{C782}) implies $W^4 = ⎨\f{43},\f{48}⎬$, $Y^4 = ⎨\f{45},\f{50}⎬$, and $W^4⧷Y^4 = ⎨\f{43},\f{48}⎬$.  

\newcommand{\noteothers}{\footnote{\label{D318}Here are some other examples: [1] the empty set is not a pentaform because it violates \rf{Pr}, [2] a singleton set $⎨⁅i˛j˛w˛a˛y⁆⎬$ with $w⋅≠⋅y$ is a pentaform, and [3] a singleton set $⎨⁅i˛j˛w˛a˛y⁆⎬$ with $w = y$ is not a pentaform because it violates \rf{Pr} (and also \rf{Py}).}} 

\smallskip{\em Summary}. In light of the preceding paragraphs, examples $Q^1$, $Q^2$, and $Q^3$ satisfy all eight axioms, and are consequently pentaforms.  Meanwhile, example $Q^4$ satisfies all axioms except \rf{Pr}, and is consequently not a pentaform.{\noteothers}  Note that $Q^4$ proves it is possible to violate \rf{Pr} while satisfying the other seven axioms.  In fact, Table~\rf{D388} in Appendix~\rf{B566} shows that it is essentially possible to violate any one of the eight axioms while satisfying the other seven.  In this sense the eight axioms are logically independent.

To verbally summarize the definition of a pentaform, recall from Section~\rf{C594} that the terms ``player'', ``situation'', ``decision node'', ``action'', and ``successor node'' are defined to mean nothing but the five positions in a quintuple.  Then a pentaform is a set of quintuples which satisfies the following five properties.  (1) Exactly one player is assigned to each situation.  (2) Exactly one situation is assigned to each decision node.  (3) The set of actions assigned to a decision node is constant across the decision nodes assigned to each situation.  (4) The assignment of a decision-node/action couple to a successor node is a bijection.  (5) The assignment of decision nodes to successor nodes eventually takes every successor node to the unique decision node that is not a successor node.  For details, note (1) paraphrases \rf{Pij}, (2) paraphrases \rf{Pjw}, (3) paraphrases \rf{Pwa} with the help of \rf{Pjw} and Proposition~\rf{C601}(\rf{C602}$⇔$\rf{C604}), (4) paraphrases \rf{Pway}, \rf{Pwy}, and \rf{Pay}, and finally, (5) is based on \rf{Pwy} and paraphrases \rf{Py} and \rf{Pr}.

\nssec{The out-tree implied by a pentaform}{C657}

An out-tree is a couple $(X,E)$ consisting of a node set $X$ and an edge set $E$ which together satisfy the usual properties of a discrete game tree.  The relevant definitions from graph theory are reviewed in the first paragraph of Section~\rf{E349} in Appendix~\rf{E348}.  The term ``out-tree'' is being used instead of ``directed rooted tree'' or ``divergent arborescence'' only because it is the shortest of these three synonymous terms.

\begin{prop}\label{C632} Suppose $Q$ is a quintuple set.  Then (a) $Q$ satisfies \rf{Pwy}, \rf{Py}, and \rf{Pr} iff $(W⋃Y,\pj{WY}(Q))$ is a nontrivial out-tree.  Further (b) if $Q$ satisfies \rf{Pwy}, \rf{Py}, and \rf{Pr}, then the unique element of $W⧷Y$ is the root of the out-tree $(W⋃Y,\pj{WY}(Q))$. (Proof~\rf{C632p}.) \end{prop}

\newcommand{\noteX}{\footnote{In another paper, one might define a special symbol (like ``$X$'') for $Q$'s node set $W⋃Y$.  In this paper it seems better to avoid the additional notation.}}

Consider a pentaform~$Q$, or more narrowly, a quintuple set $Q$ which satisfies \rf{Pwy}, \rf{Py}, and \rf{Pr}.  Proposition~\rf{C632} implies that $(W⋃Y,\pj{WY}(Q))$ is a nontrivial out-tree, and that the unique element of $W⧷Y$ is the root node of this out-tree.  Correspondingly, define $Q$'s {\em root node} $r$ by{\noteX}  \begin{gather}
\zz
⎨r⎬ = W⧷Y. \label{C982} 
\zz
\end{gather} The remainder of this paragraph derives some familiar entities from $Q$'s out-tree $(W⋃Y,\pj{WY}(Q))$.  First, derive $Q$'s {\em weak precedence} relation $≼$ and {\em strict precedence} relation $≺$ from the out-tree $(W⋃Y,\pj{WY}(Q))$ via (\rf{E351}) in Section~\rf{E349}.  Second, let $Q$'s set of {\em end nodes} be \begin{gather}
\zz
Y⧷W \label{E356}
\zz
\end{gather} (the same can be derived from $(W⋃Y,\pj{WY}(Q))$ via (\rf{E354}) in Section~\rf{E349}).  Third, derive $Q$'s {\em run} (or play) collection $\ZZ$ from the out-tree $(W⋃Y,\pj{WY}(Q))$ via (\rf{E355}) in Section~\rf{E349}.

\nssec{Pentaform games}{B562}

\newcommand{\noteinfiniteutility}{\footnote{\label{C666}Infinite utility numbers are included because, in economics, many popular utility functions generate $-∞$ utility when some consumption level is zero.  Such a utility function is often part of a consumer dynamic optimization problem, and such a problem can be specified as a one-player game.}}

Suppose $Q$ is a pentaform with its player set $I$ (from abbreviation (\rf{C825})) and its run collection $\ZZ$ (from the previous paragraph).  A {\em utility function} for player $i$ is a function of the form $u_i{:}\ZZ→\eR$, where $\eR$ is the extended real line $Ṛ⋃⎨{-}∞,∞⎬$.{\noteinfiniteutility}  A {\em utility-function profile} is a $u = ⁅u_i⁆_{i∈I}$ which lists a utility function $u_i$ for each player $i⋅∈⋅I$.  The following definition defines a pentaform game to be a couple listing a pentaform $Q$ and a utility-function profile~$u$.  Figures \rf{B265} and \rf{B266} provide two relatively simple examples (Streufert 2023, Section~2.2 provides an infinite-horizon example).\nichts{} 

\begin{ndef}[{\bf Pentaform Game}]\label{C668} A {\em pentaform game} is a couple $(Q,u)$ such that $Q$ is a pentaform (Definition~\rf{C669}) and $u$ is of the form $⁅u_i{:}\ZZ→\eR⁆_{i∈I}$ (where $Q$ determines $I$ and $\ZZ$, as summarized in Table~\rf{C492}). \end{ndef}

A less general formulation would be to augment each quintuple in a pentaform with a profile that specifies each player's reward for reaching the quintuple's successor node~$y$ (that is, for transiting the quintuple's edge $⁅w˛y⁆$).  Then a player's utility from a play $Z$ could be calculated by summing their rewards from reaching the successor nodes in the play.  In this formulation, the entire game is a set of tuples.  This formulation is less general than Definition~\rf{C668} when there is an infinite horizon (an infinite sum of rewards might be ill-defined, and further, there are infinite-horizon utility functions which cannot be expressed as sums).  
\nichts{}

\begin{figure}[h] \newcommand{\hgth}{3.0}
\begin{picture}(0,\hgth) \myoutergrid{\hgth} \renewcommand{\sellf}[1]{#1}
  \put(-0.70, 1.50){\makebox(0,0){\scalebox{1}{ \begin{pspicture} \end{pspicture} }}} \end{picture}
\caption{\small The pentaform game $(Q^1,u^1)$.  The set $Q^1$ is from Figure~\rf{B250}, and consists of the upper table's rows (expressed as quintuples).  The utility function $u^1_{\f{Alex}}{:}\ZZ^1→\eR$ is defined by the lower table.  The tree diagram provides the same data.} \label{B265} \end{figure}

\begin{figure}[h] \newcommand{\hgth}{6.5}
\begin{picture}(0,\hgth) \myoutergrid{\hgth} \renewcommand{\sellg}[1]{#1}
  \put(-1.40, 3.30){\makebox(0,0){\scalebox{1}{ \begin{pspicture} \end{pspicture} }}} \end{picture}
\caption{\small The pentaform game $(Q^2,u^2)$.  The set $Q^2$ is from Figure~\rf{B225}, and consists of the upper table's rows.  The utility-function profile $u^2 = ⁅u^2_{i}{:}\ZZ^2→\eR⁆_{i∈I^2}$ is defined by the lower table.  The tree diagram provides the same data.} \label{B266} \end{figure}

\section{Some Pentaform Tools and Applications}\label{C563}\showit
\markb{\sc \rf{C563}. Some Pentaform Tools and Applications}

Sections \rf{C577} and \rf{C564} develop tools for subsets and unions, and show their application to Selten subgames, dynamic programming, and pentaform construction.
\nichts{}  
Separately, Section~\rf{C614} applies pentaforms to perfect-recall and no-absentmindedness.
\nichts{}  
(Section~\rf{C826} will not depend on this Section~\rf{C563}.)

\nssec{Subsets of Pentaforms}{C577}

Part (\rf{D176}) of Proposition~\rf{C372} shows that any subset of a pentaform satisfies six of the eight pentaform axioms.  This and the definition of pentaform immediately imply part~(\rf{C578}).  Further, the proof of part~(\rf{D176}) is intuitive.  First consider the five axioms \rf{Pij}, \rf{Pjw}, \rf{Pway}, \rf{Pay}, and \rf{Pwy}.  Each states that some multidimensional projection of $Q$ is a function, and any subset of a function is also a function (by the definition of function in footnote~\rf{C273}).  So if one of these five axioms holds for $Q$, it almost obviously holds for any $Q′⋅⊆⋅Q$.  Meanwhile, axiom \rf{Py} means that $Y$ is a ``set-source'' for the difference equation defined by $p$, as discussed in Section~\rf{B570}.  If this holds for $Y$, it must also hold for $Y′⋅⊆⋅Y$.  Thus it is intuitive that \rf{Py} for $Q$ implies \rf{Py} for any $Q′⋅⊆⋅Q$ (the proposition's proof addresses some details about $p′$).

\begin{prop}\label{C372} Suppose $Q$ is a pentaform and $Q′⋅⊆⋅Q$.  Then \ttr{a}{D176} $Q′$ satisfies \rf{Pij}, \rf{Pjw}, \rf{Pway}, \rf{Pay}, \rf{Pwy}, and \rf{Py}. Thus \ttr{b}{C578} $Q′$ is a pentaform iff it satisfies \rf{Pwa} and \rf{Pr}.  (Proof~\rf{C372p}.) \end{prop}

\nichts{}

As discussed in Section~\rf{C595}, each quintuple set $Q$ is partitioned by the collection $⎨Q_j|j∈J⎬$ of its slices $Q_j$.  The following corollary states that the union of a subcollection of a pentaform's slice partition satisfies the first seven pentaform axioms.  This follows from Proposition~\rf{C372}(\rf{D176}) if it can be shown that such a union of slices satisfies \rf{Pwa}.  This holds because $Q$ satisfies \rf{Pwa} by assumption and because \rf{Pwa} is defined in terms of individual slices.   

\begin{ncrly}\label{C579} Suppose $Q$ is a pentaform and $J′⋅⊆⋅J$.  Then \ttr{a}{D413} $⨆⎨Q_j|j∈J′⎬$ satisfies \rf{Pij}, \rf{Pjw}, \rf{Pwa}, \rf{Pway}, \rf{Pay}, \rf{Pwy}, and \rf{Py}. Hence \ttr{b}{D414} $⨆⎨Q_j|j∈J′⎬$ is a pentaform iff it satisfies \rf{Pr}.  (Proof above.) \end{ncrly}

This remainder of this Section~\rf{C577} uses Corollary~\rf{C579} to construct the pentaforms for the Selten subgames of an arbitrary pentaform game.  It also explains how the same Corollary~\rf{C579} leads to Streufert 2023's new results on dynamic programming.

Consider a pentaform~$Q$.  Then for any $w⋅∈⋅W$, define\begin{gather}
\zz
\bQ{w} = ⎨˙⁅i_*˛j_*˛w_*˛a_*˛y_*⁆∈Q˙|˙w≼w_*˙⎬. \label{D316}
\zz
\end{gather} To put this in other words, say that a quintuple is {\em weakly after} $w$ iff its decision node weakly succeeds~$w$.  Then $\bQ{w}$ is the set of quintuples that are weakly after~$w$.  A {\em (Selten) subroot} is a member of\begin{gather}
\zz
T = ⎨⋅t∈W⋅|⋅\tJ⋅\text{and}⋅π_J(Q⧷\tQ)⋅\text{are disjoint}⋅⎬, \label{D305}
\zz
\end{gather} where $\tJ$ abbreviates $π_J(\tQ)$ by the sentence following (\rf{C825}).  In other words, a decision node $t⋅∈⋅W$ is a subroot iff each situation listed in a quintuple weakly after $t$ is not listed in a quintuple anywhere else. \nichts{}
Lemma~\rf{C585} shows this is equivalent to $\tQ$ being the union of a subcollection of $Q$'s slice partition.

Because of this, Corollary~\rf{C579}(\rf{D414}) can be applied to each $\tQ$.  The result is the following proposition, which shows that the $\tQ$ associated with each Selten subroot $t⋅∈⋅T$ is a pentaform.  As a consequence, the pentaform $\tQ$ can serve as the extensive form of the Selten subgame starting at $t$ (Streufert 2023, Section~4.2, completes the subgame by defining its utility functions). 

\begin{prop}\label{C576} Suppose $Q$ is a pentaform and $t⋅∈⋅T$.  Then $\tQ$ is a pentaform.  (Proof~\rf{C576p}.) \end{prop}

Corollary~\rf{C579} also plays a central role in the dynamic-programming theory of Streufert 2023, which is the first paper to use value functions to characterize subgame perfection in arbitrary games (only pure strategies are considered there).  For an analogy, consider a repeated game.  There, the whole-game extensive form (that is, the supergame's extensive form) combines many replicas of a single-stage extensive form (which is typically very simple).  In this paper's terminology, the ``subroots'' of the whole game are the nodes starting the replicas of the stage form.  Now imagine that different subroots have different stage forms.  In Streufert 2023's terminology, these generalized stage forms are called ``piece forms''.  Each of the various piece forms has some combination of [a] finite runs, each of which terminates in a subsequent subroot or whole-form endnode, and [b] infinite runs, each of which fails to reach a subsequent subroot or whole-form endnode.  

Streufert 2023 specifies an arbitrary pentaform game and shows [a] that the piece-form collection partitions the pentaform and [b] that this piece-form partition is coarser than the pentaform's slice partition.  Thus it can use Corollary~\rf{C579}(\rf{D414}) to show that each piece form is a pentaform.  On this foundation the paper is able to build a notion of ``piecewise Nashness'' which generalizes dynamic-programming's Bellman equation to arbitrary games.

\nssec{Unions of blocks}{C564}

\newcommand{\noteGhani}{\footnote{\label{D479}This Section~\rf{C564} is related to the\nichts{} ongoing work of Ghani, Kupke, Lambert, and Nordvall Forsberg 2018; Bolt, Hedges, and Zahn 2023; and Capucci, Ghani, Ledent, and Nordvall Forsberg 2022.  Both that literature and this Section~\rf{C564} seek to systematically construct games out of game fragments.  A precise comparison is elusive because the mathematical foundations are very different.  More is said there about utility.  The relative advantages here include constructing general infinite-horizon games, using the relatively simple operation of union, and using relatively finely-grained axioms.
\nichts{}
\nocite{CapucGLN21v2}\nocite{BoltHZ1923}\nocite{GhaniKLN18}
\nichts{}
}}

This Section~\rf{C564} shows how to construct pentaforms as unions of ``blocks''.{\noteGhani}  For the purposes of this section, let a {\em (penta)block} be a quintuple set $Q$ satisfying the first seven pentaform axioms, namely \begin{gather}
\zz
\text{\rf{Pij}, \rf{Pjw}, \rf{Pwa}, \rf{Pway}, \rf{Pwy}, \rf{Pay}, and \rf{Py}}. \label{D464}
\zz
\end{gather} Thus a pentaform is equivalent to a block which satisfies the final pentaform axiom~\rf{Pr}.  To put this in other words, consider an arbitrary quintuple set $Q$ and call $W⧷Y$ the set of $Q$'s {\em start nodes}.  Then \rf{Pr} is equivalent to $Q$ having exactly one start node.  Hence a pentaform is equivalent to a block with exactly one start node.

In this section's terminology, Corollary~\rf{C579}(\rf{D413}) shows that if $J′⋅⊆⋅J$ is a subset of a pentaform $Q$'s situation set, then $⨆⎨Q_j|j∈J′⎬$ is a block.  This easily implies that each slice $Q_j$ of a pentaform $Q$ is a block.  For example, reconsider Figure~\rf{D348}, which illustrates that $⎨Q^3_{\f{today}},Q^3_{\f{tonight}},Q^3_{\f{tomorrow}}⎬$ is the slice partition of $Q^3$.  Since $Q^3$ is a pentaform, each of these three slices is a block. 

\newcommand{\noteiss}{\footnote{Incidentally, suppose $\QQ$'s member sets have information-set situations (\rf{D319}).  Then, in (\rf{D418}) and (\rf{D419}), the condition $J^A⋂J^B = ∅$ is redundant.  Specifically, the disjointness of $W^A$ and $W^B$ implies the disjointness of $\PP(W^A)$ and $\PP(W^B)$, which implies the disjointness of the information-set collections $⎨W^A_j|j∈J^A⎬⋅⊆⋅\PP(W^A)$ and $⎨W^B_j|j∈J^B⎬⋅⊆⋅\PP(W^B)$, which by (\rf{D319}) implies the disjointness of $⎨j|j∈J^A⎬$ and $⎨j|j∈J^B⎬$, which by inspection implies the disjointness of $J^A$ and $J^B$.}}

Now consider an arbitrary collection $\QQ$ of quintuple sets~$Q$.  Then $\QQ$ is said to be {\em weakly separated} iff its member sets do not share situations,{\noteiss} decision nodes, or successor nodes.  In other words, $\QQ$ is weakly separated iff, for all distinct $Q^A$ and $Q^B$ in $\QQ$, \begin{gather}
\zz
J^A⋂J^B = ∅,⋅W^A⋂W^B = ∅,⋅\text{and}⋅Y^A⋂Y^B = ∅. \label{D418} 
\zz
\end{gather} Further, $\QQ$ is said to be {\em strongly separated} iff its member sets do not share situations or nodes.  In other words, $\QQ$ is strongly separated iff, for all distinct $Q^A$ and $Q^B$ in $\QQ$, \begin{gather}
\zz
J^A⋂J^B = ∅⋅\text{and}⋅(W^A⋃Y^A)˙⋂˙(W^B⋃Y^B) = ∅. \label{D419}
\zz
\end{gather} By inspection, (\rf{D419}) implies (\rf{D418}).  Intuitively, not sharing nodes in general implies not sharing decision nodes in particular and also not sharing successor nodes in particular.

\nichts{}

\nichts{}

\newcommand{\notearb}{\footnote{Incidentally, Proposition~\rf{D415} is proved via Lemma~\rf{D427}, which shows that the union of an arbitrary weakly separated collection of blocks satisfies all axioms except \rf{Py} and \rf{Pr}.  This relatively unstructured lemma appears to have independent value as an alternative to Proposition~\rf{D415}.  \nichts{}}}

\begin{prop}\label{D415} \ttr{a}{D416} Suppose $⎨Q^A,Q^B⎬$ is a weakly separated collection of blocks such that the start-node set $W^A⧷Y^A$ is disjoint from the end-node set $Y^B⧷W^B$.  Then $Q^A⋃Q^B$ is a block whose start-node set is the union of \begin{gather}
\zz
W^A⧷Y^A⋅\text{and}⋅(W^B⧷Y^B)⧷(Y^A⧷W^A) \notag
\zz
\end{gather} and whose end-node set is the union of \begin{gather}
\zz
(Y^A⧷W^A)⧷(W^B⧷Y^B)⋅\text{and}⋅Y^B⧷W^B. \notag
\zz
\end{gather} \ttr{b}{D417} Suppose $\QQ$ is a strongly separated collection of blocks.  Then $⨆\QQ$ is a block with start-node set $⨆_{Q∈\QQ}(π_W(Q)⧷π_Y(Q))$ and end-node set $⨆_{Q∈\QQ}(π_Y(Q)⧷π_W(Q))$.  (Proofs \rf{D416p} and \rf{D417p}.){\notearb} \end{prop}

\newcommand{\glty}{guilty}
\newcommand{\inct}{innocent}
{\newcommand{\Qglty}{Q^{\rm \glty}}
\newcommand{\Qinct}{Q^{\rm \inct}}
\newcommand{\notemorekid}{\footnotetext{$\Qglty$ and $\Qinct$ can be interpreted as continuations of the story in footnote~\rf{B292}.  In $\Qglty$, the kid did not do her homework, the teacher has not yet reported a verdict, and the kid must decide whether to say something to influence her parents' reaction to the coming verdict ($\f{s}$ denotes the action of saying something, and $\f{\tilde{s}}$ denotes the opposite).  Similarly, in $\Qinct$, the dog ate the kid's homework, the teacher has not yet reported a verdict, and the kid must decide whether to say something to influence her parents' reaction.}}

To explore this proposition, consider three quintuple sets:\ the example $Q^3$ from Figure \rf{D348}; \begin{align}
\zz
\Qglty = &⋅⎨
⁅\f{Kid}˛\f{\glty}˛\f4˛\f{s}˛\f{11}⁆,
⁅\f{Kid}˛\f{\glty}˛\f4˛\f{\tilde{s}}˛\f{12}⁆, \label{D423} \\
&⋅⋅˙⁅\f{Kid}˛\f{\glty}˛\f5˛\f{s}˛\f{13}⁆,
⁅\f{Kid}˛\f{\glty}˛\f5˛\f{\tilde{s}}˛\f{14}⁆⎬;⋅\text{and} \notag\\[.4ex]
\Qinct = &⋅⎨
⁅\f{Kid}˛\f{\inct}˛\f6˛\f{s}˛\f{15}⁆,
⁅\f{Kid}˛\f{\inct}˛\f6˛\f{\tilde{s}}˛\f{16}⁆, \label{D424} \\
&⋅⋅˙⁅\f{Kid}˛\f{\inct}˛\f7˛\f{s}˛\f{17}⁆,
⁅\f{Kid}˛\f{\inct}˛\f7˛\f{\tilde{s}}˛\f{18}⁆⎬.\footnotemark \notag
\zz
\end{align} \notemorekid These three quintuple sets are illustrated in Figure~\rf{D422}.  All three are blocks.  In particular, (i) $Q^3$ is a block because it is a pentaform and because every pentaform is a block by the block definition (\rf{D464}), (ii) $\Qglty$ is a block because it is like the slice $Q^3_{\f{tomorrow}}$ from the pentaform $Q^3$ and because every slice of every pentaform is a block by Corollary~\rf{C579}(\rf{D413}), and (iii) $\Qinct$ is a block because it also is like the slice $Q^3_{\f{tomorrow}}$.  Further, the three blocks are weakly separated (\rf{D418}).  To see this, note that the three blocks do not share situations, do not share decision nodes, and do not share successor nodes.

\begin{figure}[h] \newcommand{\hgth}{8.0}
\begin{picture}(0,\hgth) \myoutergrid{\hgth} \renewcommand{\sellc}[1]{#1}
  \put( 0.00, 3.70){\makebox(0,0){\scalebox{0.97}{ \begin{pspicture} \end{pspicture} }}} \end{picture}
\caption{\small Three blocks{:} $Q^3$ from Figure~\rf{D348}, $Q^{\rm guilty}$ from equation (\rf{D423}), and $Q^{\rm innocent}$ from equation (\rf{D424}).} \label{D422} \end{figure}

Proposition~\rf{D415}(\rf{D416}) assumes that $Q^A$ and $Q^B$ are weakly separated blocks and that no start node of $Q^A$ is also an end node of $Q^B$.  This admits the possibility that an end node of $Q^A$  is also a start node of $Q^B$.  Roughly, $Q^A$ can precede $Q^B$, but not vice versa.
\nichts{}  
For example, consider $(Q^A,Q^B) = (Q^3,\Qglty)$ and ignore $\Qinct$.  The only start node of $Q^A = Q^3$ is $\f0$, and this is not in $⎨\f{11},\f{12},\f{13},\f{14}⎬$, which is the end-node set of $Q^B = \Qglty$.  Thus Proposition~\rf{D415}(\rf{D416}) implies that $Q^3⋃\Qglty$ is a block whose start nodes are \subi \begin{gather}
\zz
\text{the start nodes of}⋅Q^A = Q^3⋅\text{together with} \label{D420} \\[-.8ex]
\text{the start nodes of}⋅Q^B = \Qglty⋅\text{that are not also end nodes of}⋅Q^A = Q^3. \label{D421}
\zz
\end{gather} \subo In (\rf{D421}), the start nodes of $\Qglty$ are $\f4$ and $\f5$, which are also end nodes of $Q^3$.  Thus (\rf{D421}) contributes nothing.  Meanwhile, in (\rf{D420}), the only start node of $Q^3$ is~$\f0$.  Therefore the only start node of $Q^3⋃\Qglty$ is $\f0$, which obviously implies that $Q^3⋃\Qglty$ has exactly one start node, which by this Section~\rf{C564}'s first paragraph implies that $Q^3⋃\Qglty$ is a pentaform.  

In a similar fashion, Proposition~\rf{D415}(\rf{D416}) can be applied at $(Q^A,Q^B) = (Q^3,\Qinct)$ to show that $Q^3⋃\Qinct$ is a pentaform.  Further, using this technique twice shows that $Q^3⋃\Qglty⋃\Qinct$ is a pentaform.  This can be accomplished by constructing the union as $(Q^3⋃\Qglty)⋃\Qinct$ or as $(Q^3⋃\Qinct)⋃\Qglty$.\nichts{}

Proposition~\rf{D415}(\rf{D417}) provides another way to prove that $Q^3⋃\Qglty⋃\Qinct$ is a pentaform.  Since $\Qglty$ and $\Qinct$ do not share nodes, $⎨\Qglty,\Qinct⎬$ is strongly separated (\rf{D419}).  Thus Proposition~\rf{D415}(\rf{D417}) implies that $\Qglty⋃\Qinct$ is a block whose start-node set is \begin{gather}
\zz
⨆_{Q∈⎨\Qglty,\Qinct⎬}˙(π_W(Q)⧷π_Y(Q)) = ⎨\f4,\f5⎬⋃⎨\f6,\f7⎬ \notag
\zz
\end{gather} and whose end-node set is \begin{gather}
\zz
⨆_{Q∈⎨\Qglty,\Qinct⎬}˙(π_Y(Q)⧷π_W(Q)) = ⎨\f{11,12,13,14}⎬⋃⎨\f{15,16,17,18}⎬. \notag
\zz
\end{gather} Then, as in the previous two paragraphs, Proposition~\rf{D415}(\rf{D416}) can be applied at \linebreak $(Q^A,Q^B) = (Q^3,\Qglty⋃\Qinct)$ to show that $Q^3⋃(\Qglty⋃\Qinct)$ is a block with exactly one start node.  Thus the union is a pentaform.

Roughly, the previous paragraph showed how to augment the pentaform $Q^3$ with a ``layer'' $\QQ = ⎨\Qglty,\Qinct⎬$ of two additional blocks whose start nodes were among the end nodes of $Q^3$.  This same technique can be used to augment any finite-horizon\nichts{} pentaform with any layer $\QQ$ of additional blocks whose start nodes are among the end nodes of the original pentaform.  The layer's blocks do not need to be similar to one another.  Also, the layer can have arbitrarily many (and possibly uncountably many) blocks, which freely allows arbitrarily many end nodes of the original pentaform to be connected with successor nodes in the layer's additional blocks.

Further, layer after layer can be added to generate an infinite expanding sequence of pentaforms.  Proposition~\rf{D454} shows that the union of such an expanding sequence is a pentaform.  This provides a straightforward way to construct an infinite-horizon pentaform.  For instance, Streufert 2023's motivating example is a partially infinitely repeated game of cry-wolf.  Its pentaform is built via Propositions \rf{D415} and \rf{D454} (details in Streufert 2023, Lemma~A.6).

\begin{prop}\label{D454} Suppose $⁅Q^n⁆_{n≥0}$ is an infinite sequence of pentaforms such that $(∀n≥1)$ $Q^{n-1}⋅⊆⋅Q^n$ and $r^n = r^0$.  Then $⨆_{n≥0}Q^n$ is a pentaform with root $r^0$.  (Proof~\rf{D454p}.) \end{prop}

\nichts{}

\nichts{}

\nssec{Perfect-recall in terms of pentaforms}{C614}

Although perfect-recall will remain a relatively subtle concept, it seems helpful to re-express the concept in terms of pentaforms.  This section does so, and is unconnected with Sections \rf{C577} and \rf{C564}.

\nichts{}

The concept of perfect-recall was introduced in two different but equivalent ways by Kuhn 1950, page~575, and Kuhn 1953, page~213.\label{E424}  Later it was characterized in other ways by Okada 1987, page~87; Ritzberger 1999, Theorems 1 and~2; and Al\'{o}s-Ferrer and Ritzberger 2017, Theorems 1 and~2.  Three formulations especially close to the approach here are Kreps and Wilson 1982, Equation (2.3); Myerson 1991, page 43; and Selten 1975, page~27 (Selten differs by having a specialized notion of actions).  Essentially, these formulations consider any decision nodes $w_0$, $w_1$, and $w_2$ such that \subi \label{E640} \begin{gather}
\zz
\text{$w_1$ and $w_2$ are in the same information set,} \label{E638} \\[-.6ex]
\text{$w_0$ precedes $w_1$, and} \notag \\[-.6ex]
\text{$w_0$ and $w_1$ are controlled by the same player.} \notag
\zz 
\end{gather} Then perfect-recall is the necessity of a decision node $w_{00}$ such that \begin{gather}
\zz
\text{$w_0$ and $w_{00}$ are in the same information set,} \label{E639} \\[-.6ex]
\text{$w_{00}$ precedes $w_2$, and}  \notag \\[-.6ex]
\text{the path from $w_0$ to $w_1$ starts with the same action as the path from $w_{00}$ to $w_2$}. \notag 
\zz
\end{gather} \subo Unfortunately, it is notationally awkward to express the first action on the way to $w_1$ and likewise the first action on the way to $w_2$.  This is one reason that many people regard perfect-recall as a relatively subtle concept.\nocite{Kuhn50} \nocite{Okada87} \nocite{Ritzb99} \nocite{KrepsW82} \nocite{AlosfR17-IJGT}

\nichts{}

\newcommand{\noteija}{\footnote{To be explicit, let $a_y = \pj{YA}(Q)(y)$, where $\pj{YA}(Q)$ is a function by \rf{Pay}.  Then let $j_y = \pj{WJ}(Q)(p(y))$ where $p = \pj{YW}(Q)$ is a function by \rf{Pwy} and $\pj{WJ}(Q)$ is a function by \rf{Pjw}.  Finally, let $i_y = \pj{JI}(Q)(j_y)$ where $\pj{JI}(Q)$ is a function by \rf{Pij}.}}

Pentaforms allow an alternative formulation which revolves around successor nodes rather than decision nodes.  To begin, consider a pentaform $Q$, and note that the axioms \rf{Pij}, \rf{Pjw}, \rf{Pwy}, and \rf{Pay} imply that each successor node $y⋅∈⋅Y$ determines its entire quintuple $⁅i˛j˛w˛a˛y⁆⋅∈⋅Q$.  Thus we may speak of $y$'s player~$i_{\klar y}$, $y$'s situation~$j_{\klar y}$, and $y$'s action~$a_{\klar y}$.{\noteija} For intuitive purposes, it is important to be reminded that $y$ is a successor node.  Toward that end, say that a successor node $y$ is obtained ``by'' player $i_y$, that $y$ is obtained ``from'' situation $j_y$, and that $y$ is obtained ``via'' action $a_y$.

Then say that a pentaform $Q$ satisfies {\em perfect-recall} iff, for all successor nodes $y_0$, $y_1$, and $y_2$ in $Y$, the conditions \subi\label{E528} \begin{gather}
\zz
j_{\klar y_1} = j_{\klar y_2},⋅y_0⋅≺⋅y_1,⋅\text{and}⋅i_{\klar y_0} = i_{\klar y_1}, \label{E529}
\zz
\end{gather} imply the existence\footnote{By inspection, (\rf{E530}) holds with $y_{00}$ equaling $y_0$ iff $y_0⋅≺⋅y_2$.  Thus perfect-recall becomes interesting precisely when $y_0$ does not precede $y_2$.} of a successor node $y_{00}⋅∈⋅Y$ such that \begin{gather}
\zz
y_{00}⋅≺⋅y_2⋅\text{and}⋅⁅j_{\klar y_{00}},a_{\klar y_{00}}⁆ = ⁅j_{\klar y_0},a_{\klar y_0}⁆. \label{E530}
\zz
\end{gather}\subo This translates (\rf{E640}) into quintuples.  In English, suppose that $y_1$ and $y_2$ are obtained from the same situation, and that $y_1$ has a predecessor $y_0$ which is obtained by the same player.\footnote{\label{E488}Equation (\rf{E529})'s first equality and \rf{Pij} imply that $y_1$ and $y_2$ are obtained by the same player, which by (\rf{E529})'s second equality implies that $y_0$, $y_1$, and $y_2$ are all obtained by the same player.}  Then perfect-recall requires that $y_2$ has a predecessor $y_{00}$ which is obtained from the same situation,\footnote{This and axiom \rf{Pij} imply that $y_{00}$ and $y_0$ are obtained by the same player, which by footnote~\rf{E488} implies that all four nodes are obtained by the same player.} and via the same action, as~$y_0$.  Note that the actions $a_{\klar y_0}$ and $a_{\klar y_{00}}$ are easy to express because the new formulation revolves around successor nodes rather than decision nodes. This is an advantage over the traditional formulation discussed earlier. 

\begin{figure}[t] \newcommand{\hgth}{3}
\begin{picture}(0,\hgth)  \myoutergrid{\hgth}
  \put(.2,1.3){\makebox(0,0){\scalebox{1}{ \begin{pspicture} \end{pspicture} }}} \end{picture}
\caption{The definition (\rf{E528}) of perfect-recall assumes the solid lines and then requires the dashed lines.} \label{C550} \end{figure}

For an example, modify the pentaform $Q^3$ from Figure~\rf{D348} by replacing both the player $\f{Kid}$ and the player $\f{Teacher}$ with a new player named $\f{KidTeacher}$.  Here the new formulation of perfect-recall fails at $(y_0,y_1,y_2) = (\f2,\f5,\f7)$.  In particular, $\f5$ and $\f7$ are obtained from the same situation ($\f{tomorrow}$), and $\f5$ has a predecessor $\f2$ which is obtained by the same player ($\f{KidTeacher}$).  Yet, $\f7$ does not have a predecessor which is obtained from the same situation, and via the same action, as $\f2$.  Specifically, $\f7$ is preceded by $\f3$ and~$\f1$ (the root node $\f0$ is excluded because it is not a successor node).  [a] The predecessor $\f3$ is obtained from $\f{tonight}$ while $\f2$ is obtained from $\f{today}$.  [b] The predecessor $\f1$ is obtained from $\f{today}$ as $\f2$ is obtained from $\f{today}$, but, $\f1$ is obtained via the action~$\fc$ while $\f2$ is obtained via the action~$\fb$.  Intuitively, when $\f{KidTeacher}$ is at situation $\f{tomorrow}$ (from which they obtain $\f5$ and~$\f7$), they remember the situation $\f{today}$ (from which they obtain predecessors of $\f5$ and $\f7$) but not the action which they took there (since they obtain the predecessors of $\f5$ and $\f7$ via different actions).

Lastly, the remainder of this section suggests that the new version of perfect-recall can be easily manipulated.  It does so by considering the relationship between perfect-recall and {\noab}.  The concept of {\noab} is [a] left unnamed and imposed as part of the definition of a game by Kuhn 1953, page~195, [b] called ``linearity'' by Isbell 1957, page~86, [c] called ``nonrepetition'' by Alpern 1988, page~470, and [d] called ``no-absentmindedness'' by Piccione and Rubinstein 1997, pages~9--10, and Ritzberger 1999, page~72.
The concept is relatively simple.  It says that one decision node cannot strictly precede another decision node in the same information set.  Correspondingly, say that a pentaform $Q$ satisfies {\em {\noab}} iff \begin{gather}
\zz
(∄y_1∈Y,y_2∈Y)⋅y_1⋅≺⋅y_2⋅\text{and}⋅j_{\klar y_1} = j_{\klar y_2}. \label{E510}%
\zz
\end{gather} This says that no successor node strictly precedes another successor node from the same situation.\nocite{Isbel57} \nocite{Alper88} \nocite{PicciR97} 

The proof of the following proposition is new.  The result is stated without recorded proof in Piccione and Rubinstein 1997, pages 3--5, and Hillas and Kvasov 2020, Section~2 (a self-evident result in Isbell 1953, page 86, uses a different definition of perfect-recall).  Meanwhile, there is a recorded proof in Ritzberger 1999, pages 73--74, which differs from the proposition's proof by using actions, nontrivial feasible sets, and strategies.\label{E425} \nocite{HillaK20}

\nichts{} 
\nichts{}

\begin{prop}\label{D178} Perfect-recall (\rf{E528}) implies {\noab} (\rf{E510}). \end{prop} 

\newcommand{\notedetail}{\nichts{}}  

\begin{pf} To prove the contrapositive, suppose that $Q$ is a pentaform which violates no-absentmindedness.   Then definition (\rf{E510}) implies there are successor nodes $y_0$ and $y_1$ such that $y_0⋅≺⋅y_1$ and $j_{\klar y_0} = j_{\klar y_1}$.  The equality and \rf{Pij} imply $i_{\klar y_0} = i_{\klar y_1}$.  Next, let $y_2$ be the earliest weak predecessor of $y_0$ and $y_1$ which is from the same situation as $y_0$ and $y_1$ (it is irrelevant that $y_2$ might equal $y_0$).{\notedetail} Note that $j_{\klar y_2} = j_{\klar y_1}$.  Thus $y_0$, $y_1$, and $y_2$ satisfy the hypotheses of perfect-recall (\rf{E529}).  Yet the definition of $y_2$ implies that there is no $y_{00}$ such that $y_{00}⋅≺⋅y_2$ and $j_{\klar y_{00}} = j_{\klar y_0}$.  Hence the conclusion of perfect-recall (\rf{E530}) cannot be satisfied. \end{pf}

\nichts{}

\newcommand{\smidge}{\\[.4ex]}
\newcommand{\novert}{\multicolumn{2}{c}{}}
\newcommand{\tablestd}{
\begin{table}[t]
\centering

{\small 
\begin{tabular}{cclrc|cc}
&&&& \rule{5mm}{0mm} & \rule{1mm}{0mm} & \\[110mm]
&&&&\novert& \\[-119mm] 

&&&
  &\novert& \begin{minipage}{18ex}\begin{center}
  All in \\ {\underline{pentaform terms}} 
  \end{center}\end{minipage} \\[-3mm]
&\multicolumn{2}{l}{Out-tree $(X,E)$}&\SMALL[\rf{C665}\rf{D321},˙\rf{E349}]
  &\novert& \\[-3.8mm]
&\multicolumn{3}{l}{\underline{\hspace{10.1cm}}} 
  &\novert& \\[1mm] 
&$X$ & set of nodes $x$ &\SMALL [\rf{C665}\rf{D321},˙\rf{E349}]
  &\novert& $\hp{W}⋃\hp{Y}$ \smidge
&$E$ & set of edges $⁅w˛y⁆$ &\SMALL [\rf{C665}\rf{D321},˙\rf{E349}]
  &\novert& $\pj{WY}(\hp{Q})$ \smidge
&$r$ & $\dve$ root node &\SMALL[\rf{C665}\rf{D321},˙\rf{E349}]
  &\novert& $⎨\hp{r}⎬{=}\hp{W}⧷\hp{Y}$ \smidge
&$π_1E$ & $\dve$ set of decision nodes $w$ &\SMALL[\rf{C665}\rf{D321}]
  &\novert& $\hp{W}{=}π_W(\hp{Q})$ \smidge
&$π_2E$ & $\dve$ set of successor nodes $y$ &\SMALL[\rf{C665}\rf{D321}]
  &\novert& $\hp{Y}{=}π_Y(\hp{Q})$ \smidge
&$E^{-1}$ & $\dve$ immediate-predecessor function &\SMALL[\rf{C665}\rf{D321}]
  &\novert& $\hp{p}{=}\pj{YW}(\hp{Q})$ \smidge
&$≼$ & $\dve$ weak precedence order & \SMALL [\rf{C665}\rf{D321},˙\rf{E349}]
  &\novert& $\hp{≼}$ \smidge
&$≺$ & $\dve$ strict precedence order & \SMALL [\rf{C665}\rf{D321},˙\rf{E349}]
  &\novert& $\hp{≺}$ \smidge
&$\ZZ$ & $\dve$ collection of runs Z &\SMALL [\rf{C665}\rf{D321},˙\rf{E349}]
  &\novert& $\hp{\ZZ}$ \\[3.5mm]

&\multicolumn{2}{l}{Traditional Game $(X,E,\HH,α,ι,u)$}&\SMALL[\rf{C665}]
  &\novert& \\[-3.8mm]
\rule{0mm}{0mm}&\multicolumn{3}{l}{\underline{\hspace{10.1cm}}}
  &\novert& \\[1mm] 
&$\HH$ & collection of information sets $H$ &\SMALL[\rf{C665}\rf{D322}]
  &\novert& $\hp{J}{=}π_J(\hp{Q})$ \\[.7ex] 
&$α$ & action-assigning function  &\SMALL[\rf{C665}\rf{D323}]
  &\novert& \SMALL $\miz ⎨⁅⁅w˛y⁆˛a⁆| \\ ⁅w˛y˛a⁆∈\pj{WYA}(\hp{Q})⎬ \moz$ \\[2.5ex]
&$A$ & $\dve$ set of actions $a$ &\SMALL[\rf{C665}\rf{D323}]
  &\novert& $\hp{A}{=}π_A(\hp{Q})$ \smidge
&$F$ & $\dve$ feasibility correspondence &\SMALL[\rf{C665}\rf{D323}]
  &\novert& $\hp{F}{=}\pj{WA}(\hp{Q})$ \smidge
&$ι$ & player-assigning function &\SMALL[\rf{C665}\rf{D324}] 
  &\novert& $\pj{WI}(\hp{Q})$ \smidge
&$I$ & $\dve$ set of players $i$ &\SMALL[\rf{C665}\rf{D324}]
  &\novert& $\hp{I}{=}π_I(\hp{Q})$ \smidge
&$u{=}⁅u_i⁆_{i∈I}$ & profile with utility function $u_i$ for each $i$\hspace{-2.5cm} &\SMALL[\rf{C665}\rf{D325}] 
  &\novert& $\hp{u}{=}⁅\hp{u}_i⁆_{i∈\hp{I}}$ \\[1.5mm]

\multicolumn{4}{c}{}\\[-0mm] %
\end{tabular} }

\caption{\small {\em Left-hand Side:} Out-trees and traditional games are implicitly accompanied by their derivatives (\protect\rotatebox[origin=c]{180}{$\Lsh$}).  Definitions are in the sections in brackets~{\SMALL[˙]}.  {\em Right-hand Side:} The same entities in terms of the pentaform game $\PB(X,E,\HH,α,ι,u) = (\hp{Q},\hp{u})$ (Proposition~\rf{C699}).
} \label{C670}
\end{table} }

\section{Equivalence with Traditional Games}\label{C826}\showit
\markb{\sc \rf{C826}. Equivalence with Traditional Games}

This Section~\rf{C826} compares pentaform games with traditional games.  It is unconnected with Section~\rf{C563}. 

\nssec{Definition of traditional games}{C665}

Definition~\rf{C843} will define a ``traditional game'' to be a tree that has been adorned with information sets, actions, players, and utility functions in a more-or-less usual way.  The phrase ``more-or-less'' acknowledges that there is no generally accepted way
 
\noindent to define an extensive-form game.  Rather, substantially different definitions appear in social science, mathematics, computer science, logic, and engineering; and further, there is often substantial variety within any one of these fields.  Somewhere in the middle is Definition~\rf{C843}, which might best be called the definition of a ``representative'' traditional game.  In developing this definition, the author's guiding principles were to head for the middle, to avoid specialized assumptions, to use explicit notation, and to use standard mathematics whenever possible.  (In the context of Section~\rf{B574}'s literature review, Definition~\rf{C843} is a generic extension of Kuhn 1953, and fairly close to the ``KS'' formulations in Kline and Luckraz 2016 and Streufert 2019.)

The following five paragraphs define the symbols and terms that appear in Definition~\rf{C843}.  The paragraphs have been given letters to facilitate referencing, and the material is summarized in the left-hand side of Table~\rf{C670}.  Because of the diversity in the literature, almost every reader will be unfamiliar with something, and a number of specific hazards are discussed in footnotes \rf{C671}--\rf{C675}. (Incidentally, it can be helpful to notice, from a mathematical perspective, that paragraphs \rf{D323}--\rf{D325} are almost independent of one another.  The only link is that the players from paragraph \rf{D324} index the utility functions in paragraph~\rf{D325}.)  

\newcommand{\notestepa}{\footnote{\label{C671}This first step may be unfamiliar for any of several reasons.  First, game trees can be specified as undirected rooted trees rather than out-trees.  Second, game trees can be specified via order theory rather than graph theory.  Third, some popular specifications use specialized notations which implicitly impose some of the properties of game trees.  For example, some game-tree properties are implicitly imposed if nodes are specified as sequences of past actions, or sets of past actions, or sets of outcomes.  Finally, it is common to specify game trees by diagrams rather than by notation.}} 

\newcommand{\noteW}{\footnote{The reader might expect a special symbol (like ``$W$'') for the decision-node set $π_1E$, and another special symbol (like ``$p$'') for the immediate-predecessor function $E^{-1}$.  In this paper, it seems better to avoid the additional notation.}}

\smallskip {\em Paragraph \ttz{a}{D321}}.  The first component of a traditional game (Definition~\rf{C843}) is a nontrivial out-tree $(X,E)$.{\notestepa}  By definition, such an out-tree consists of a node set $X$ and an edge set $E$ which together satisfy the usual properties of a discrete game tree (full details in the first paragraph of Section~\rf{E349} in Appendix~\rf{E348}).  The term ``out-tree'' is being used instead of ``directed rooted tree'' or ``divergent arborescence'' only because it is the shortest of these three synonymous terms.  By usual constructions, an out-tree determines its root node $r$ (first paragraph of Section~\rf{E349}), its decision-node set $π_1E$,{\noteW} its successor-node set $π_2E$, its immediate-predecessor function $E^{-1}$ (Lemma~\rf{E320}(\rf{E344})), its weak and strict precedence orders $≼$ and $≺$ (equation (\rf{E351})), its end-node set $π_2E⧷π_1E$ (equation (\rf{E354})), and its collection $\ZZ$ of runs or equivalently ``plays'' (equation (\rf{E355})).

\smallskip {\em Paragraph \ttz{b}{D322}}. The next component of a traditional game is a partition $\HH$ of the decision-node set $π_1E$.  The members (i.e.\ cells) of this partition are called {\em information sets}.  (Intuitively, a player will be informed that they are in an information set, but not that they are at a particular node in that set.) 

\newcommand{\notestepc}{\footnote{\label{C673}Action-assigning functions may be unfamiliar for any of several reasons.  First, actions might be assigned to successor nodes $y⋅∈⋅π_2E$ rather than to edges $⁅w˛y⁆⋅∈⋅E$. \nichts{} The two are equivalent because there is a bijection between $π_2E$ and $E$ (this bijection follows from Lemma~\rf{E320}(\rf{E344})).  Second, some popular specifications use specialized notations which implicitly assign actions to edges or successor nodes.  For example, if nodes are specified as sequences of past actions, then the action assigned to a successor node is simply the node's most recent past action.  Third, it is common to assign actions to edges in a tree diagram rather than through formal notation.}}

\newcommand{\notemeasure}{\footnote{\label{D353}The term ``measurability'' is new in this context.  The paper does not use measurability theory in any substantial way.  Rather, ``measurability'' is just a mathematically ordinary and precise name for conditions (\rf{D350}) and (\rf{D351}).  (The same constructions are called ``continuity'' in Streufert 2021.\nocite{gm-2105})\nichts{} }}

\smallskip{\em Paragraph \ttz{c}{D323}}. Next, each edge of the tree is assigned an action.  Formally, this assignment is accomplished by a function $α$ from the edge set~$E$.  Call $α$ the {\em action-assigning function} (or ``labelling function'').{\notestepc}  Then let $A$ be the range of $α$, that is, let \begin{gather}
\zz
A = ⎨˙α(w˛y)˙|˙⁅w˛y⁆∈E˙⎬, \label{E378}
\zz
\end{gather} and call an element of $A$ an {\em action}.  (Intuitively, each edge's action is what the controlling player would ``do'' to choose the edge.)  It will be assumed that $α$ is {\em locally injective} in the sense that, for any two edges of the form $⁅w˛y_1⁆⋅∈⋅E$ and $⁅w˛y_2⁆⋅∈⋅E$, \begin{gather}
\zz
y_1⋅≠⋅y_2⋅\text{implies}⋅α(w˛y_1)⋅≠⋅α(w˛y_2). \label{D347}
\zz
\end{gather}\nichts{} 
Thus local injectivity means that two different edges from one decision node $w$ cannot be assigned the same action.  Further, from the action-assigning function $α$, derive the correspondence $F{:}π_1E⇉A$ by\begin{gather}
\zz
(∀w∈π_1E)⋅F(w) = ⎨˙a˙|˙(∃y)˙α(w˛y){=}a˙⎬. \label{D169}
\zz
\end{gather} Thus each $F(w)$ is the set of actions that label the edges leaving~$w$.  Call $F(w)$ the set of actions that are {\em feasible} at~$w$.  Then let $⁅F(w)⁆_{w∈π_1E}˙{:}˙π_1E˙→˙\PP(A)$ be the associated set-valued function.  It will be assumed that $⁅F(w)⁆_{w∈π_1E}$ is {\em measurable}{\notemeasure} in the sense that it is measurable as a function from $π_1E$ (endowed with the $σ$-algebra whose elements are arbitrary unions of cells from the information-set partition $\HH$) into the collection of the subsets of $A$ (endowed with the discrete $σ$-algebra).  This is equivalent to \begin{gather}
\zz
(∀H∈\HH,w_1∈H,w_2∈H)⋅F(w_1) = F(w_2), \label{D350}
\zz
\end{gather} which requires that two nodes in one information set have the same feasible-action set.

{\tablestd}

\newcommand{\notestepd}{\footnote{\label{C674}There are other ways to specify where players move.  First, players could be assigned to information sets in $\HH$ rather than to decision nodes in $π_1E$.\nichts{}  This would implicitly impose the measurability of (\rf{D351}).  Second, more than one player could be assigned to the same decision node.  This can specify simultaneous moves, given suitable modifications elsewhere [as in Al\'os-Ferrer and Ritzberger 2016, page 138, condition (DEF2)].  Traditionally, simultaneous moves are specified by multiple information sets [as in Osborne and Rubinstein 1994, page 202].}}

\smallskip
\pagebreak
{\em Paragraph \ttz{d}{D324}}. Next, each decision node of the tree is assigned to a player.  Formally, this assignment is accomplished by a function $ι$ from the decision-node set $π_1E$.  \linebreak Call $ι$ the {\em player-assigning function}.{\notestepd}  Then let $I$ be the range of $ι$, that is, let \begin{gather}
\zz
I = ⎨˙ι(w)˙|˙w∈π_1E˙⎬, \label{E377}
\zz
\end{gather} and call an element of $I$ a {\em player}.  (Intuitively, the role of $ι$ is to specify which player controls the action at each decision node.)  It will be assumed that $ι$ is {\em measurable} (footnote~\rf{D353}) in the sense that it is measurable as a function from $π_1E$ (endowed with

\pagebreak
 
\noindent the $σ$-algebra whose elements are arbitrary unions of cells from the information-set partition $\HH$) into the player set $I$ (endowed with the discrete $σ$-algebra). This is equivalent to \begin{gather}
\zz
(∀H∈\HH,w_1∈H,w_2∈H)⋅ι(w_1) = ι(w_2), \label{D351}
\zz
\end{gather} which requires that two nodes in one information set are controlled by the same player.

\smallskip{\em Paragraph \ttz{e}{D325}}. Finally, each player is given a utility function defined over the runs of the tree.  Formally, this is accomplished by a profile $u = ⁅u_i⁆_{i∈I}$ which lists a function $u_i{:}\ZZ→\eR$ for each player~$i$.  Here $\ZZ$ is the run collection (\rf{E355}) of the out-tree $(X,E)$ as defined in paragraph~\rf{D321}, and $\eR$ is the set of extended real numbers $Ṛ⋃⎨{-}∞,∞⎬$ as discussed in Section~\rf{B562}.  Call $u_i$ the {\em utility function} (or ``payoff function'') of player~$i$.\footnote{\label{C675}This specification of utility may be slightly unfamiliar.  First, it is common, in finite-horizon games, to map end nodes to utility numbers.  In this special case, there is a bijection between the end-node set $π_2E⧷π_1E$ and the run collection $\ZZ$ (here $\ZZ$ equals the collection $\ZZf$ of finite runs defined in Section~\rf{E349}'s last two paragraphs).  Second, footnote \rf{C666} explains the slight benefit of allowing utility numbers in $\eR = Ṛ⋃⎨{-}∞,∞⎬$ rather than the more familiar~$Ṛ$.}

\begin{ndef}[{\bf Traditional Game}]\label{C843} A {\em traditional game} is a tuple $(X,E,\HH,$ $α,ι,u)$ such that\begin{gather}
\zz
(X,E)⋅\text{is a nontrivial out-tree (Section~\rf{E349})}, \ttt{[T1]}{T1} \nt
\HH⋅\text{is a partition of}⋅π_1E, \ttt{[T2]}{T2} \nt
α⋅\text{is a locally injective (\rf{D347}) function from}⋅E, \ttt{[T3]}{T3} \nt
⁅F(w)⁆_{w∈π_1E}⋅\text{is measurable (\rf{D350})}, \ttt{[T4]}{T4} \nt
ι⋅\text{is a measurable (\rf{D351}) function from}⋅π_1E,⋅\text{and} \ttt{[T5]}{T5} \nt
u⋅\text{is of the form}⋅⁅u_i{:}\ZZ→\eR⁆_{i∈I} \ttt{[T6]}{T6} \notag
\zz
\end{gather} (where $α$ determines $F$ (\rf{D169}), $ι$ determines $I$ (\rf{E377}), and $(X,E)$ determines $\ZZ$ (\rf{E355})). \end{ndef}

For an example, consider Figure~\rf{B265}'s tree diagram.  It has already served other purposes.  Nonetheless, it also illustrates the traditional game $(X^5,E^5,\HH^5,α^5,ι^5,u^5)$ defined by $X^5 = ⎨\f0,\f1,\f2⎬$, $E^5 = ⎨⁅\f0˛\f1⁆,⁅\f0˛\f2⁆⎬$, $\HH^5 = ⎨⎨\f0⎬⎬$,\begin{gather}
\zz
α^5 = ⎨⋅⁅⁅\f0˛\f1⁆˛\f{left}⁆,˙⁅⁅\f0˛\f2⁆˛\f{right}⁆⋅⎬, \nt
ι^5 = ⎨⋅⁅\f0˛\f{Alex}⁆⋅⎬,⋅\text{and} \label{C714} \\
u^5_{\f{Alex}} = ⎨⋅⁅⎨\f0˛\f1⎬˛2⁆,˙⁅⎨\f0,\f2⎬˛4⁆⋅⎬\notag
\zz
\end{gather} (footnote \rf{C273} explains that, in this paper, a function is a set of couples).  This definition implies that $π_1E^5 = ⎨\f0⎬$, that $\ZZ^5 = ⎨⎨\f0˛\f1⎬,⎨\f0˛\f2⎬⎬$, that $F^5(\f0) = ⎨\f{left},\f{right}⎬$, that $A^5 = ⎨\f{left},\f{right}⎬$, and that $I^5 = ⎨\f{Alex}⎬$ (these derivative entities inherit the example's superscript $^5$˙).

\nssec{``Pentaforming'' a traditional game}{B565}

\newcommand{\notecod}{\nichts{}}

The remainder of this Section~\rf{C826} develops a bijection between traditional games (Definition~\rf{C843}) and certain kinds of pentaform games (Definition~\rf{C668}).  To begin, this Section~\rf{B565} constructs an operator $\PB$ which maps each traditional game to a pentaform game.  Mnemonically, $\PB$ ``pentaforms'' a traditional game.  To be specific, let $\PB$ be the operator (equivalently function){\notecod} that takes each traditional game $(X,E,\HH,α,ι,u)$ to the $\PB(X,E,\HH,α,ι,u) = (\hp{Q},\hp{u})$ defined by \subi \label{C678} \begin{gather}
\zz
\hp{Q} = ⎨˙⁅ι(w)˛H_w˛w˛α(w˛y)˛y⁆˙|˙⁅w˛y⁆∈E˙⎬⋅\text{and} \label{C679} \\
\hp{u} = u, \label{C680}
\zz
\end{gather} \subo where in (\rf{C679}), for each decision node $w⋅∈⋅π_1E$, $H_w$ is the cell of the information-set partition $\HH$ that contains $w$ ($\HH$ partitions $π_1E$ by \rf{T2} in Definition~\rf{C843}).  

Say that a quintuple set $Q$ has {\em information-set situations} iff \begin{gather}
\zz
(∀j∈J)⋅W_j = j. \label{D319}
\zz
\end{gather} For example, equation (\rf{C795}) shows that the pentaform $Q^2$ (Figure~\rf{B225} or \rf{B266}) has information-set situations, while equation (\rf{C794}) shows that the pentaform $\smash{Q^3}$ (Figure~\rf{B224} or \rf{D348}) does not.  Thus the set of pentaform games with information-set situations is a proper subset of the set of pentaform games.  Theorem~\rf{C689} shows that the operator $\PB$ maps traditional games into this proper subset of pentaform games.  This reflects the fact that the information sets of traditional games are less general than the situations of pentaform games.%
\footnote{The lesser generality of traditional games is an artifact of Definition~\rf{C843} rather than a disadvantage of the diverse literature that Definition~\rf{C843} represents.  For example, Myerson 1991 specifies a game with information states like the situations here (footnote \rf{E385}), and such a construction has been left out of Definition~\rf{C843} because it is unusual in the literature (other constructions have been left out as well, as explained in the first paragraph of Section~\rf{C665}).} 

\begin{nthm}\label{C689} The operator $\PB$ takes each traditional game (Definition~\rf{C843}) to a pentaform game (Definition~\rf{C668}) with information-set situations (\rf{D319}). (Proof:\ Lemma~\rf{B405}.)  \end{nthm}

The definition (\rf{C678}) of $\PB$ accords with Section~\rf{B977}'s informal process of expanding edges into quintuples.  To explore this, consider the traditional game $(X^5,E^5,\HH^5,α^5,ι^5,u^5)$ defined in equation (\rf{C714}).  It is illustrated by the tree diagram on Figure~\rf{B265}'s left-hand side.  Then \begin{gather}
\zz
\PB(X^5,E^5,\HH^5,α^5,ι^5,u^5) = (Q^1,u^1), \label{E430}
\zz
\end{gather} where the couple $(Q^1,u^1)$ shown in the figure's right-hand side.  Transparently, the lower table $u^1$ is a mere rearrangement of the utilities in the tree diagram, in accord with (\rf{C680}).  Less transparently, the upper table $Q^1$ is derived from the tree diagram, in accord with (\rf{C679}).  To see this, first note that the quintuples (rows) in the upper table can be indexed by the edges $⁅w˛y⁆$ that they contain.  This indexing accords with the dummy variable $⁅w˛y⁆$ in (\rf{C679}).  Then consider, for example, the quintuple containing the edge $⁅w˛y⁆ = ⁅\f0˛\f2⁆$.  As in (\rf{C679}), and as in Section~\rf{B977}'s informal process, this edge is expanded into the quintuple $⁅\f{Alex},⎨\f0⎬,\f0,\f{right},\f2⁆$ by including its action $α^5(\f0˛\f2) = \f{right}$, its decision node's information set $H_{\f0} = ⎨\f0⎬⋅∈⋅\HH˙$, and its decision node's player $ι^5(\f0) = \f{Alex}$.  The other quintuple in the upper table is derived in the same way. 
\nichts{}

The practical import of Theorem~\rf{C689} is that pentaform games are general enough to accommodate any discrete extensive-form game.  Although Section~\rf{B977}'s examples are finite, pentaforms can have trees with arbitrary degree (which admits decision nodes with uncountably many immediate successors), and up to countably infinite height (which admits an infinite horizon, as in the example pentaform of Streufert 2023, Section~2.2).\label{E420}

\nssec{``Traditionalizing'' a pentaform game}{C690}

\newcommand{\notetwotoone}{\footnotetext{\label{E436}Footnote \rf{E401} explains a similar construction.  Intuitively, this is a function from two variables to one variable.}}

This Section~\rf{C690} constructs the operator $\TB$ from pentaform games (Definition~\rf{C668}) to traditional games (Definition~\rf{C843}).  Mnemonically, $\TB$ ``traditionalizes'' a pentaform game.  To be specific, let $\TB$ be the operator that takes a pentaform game $(Q,u)$ to the $\TB(Q,u) = (\hs{X},\hs{E},\hs{\HH},\hs{α},\hs{ι},\hs{u})$ defined by %
\subi \label{C677} \begin{align}
\zz
\hs{X} =&⋅W⋃Y, \label{C693}\\
\hs{E} =&⋅\pj{WY}(Q), \label{C694}\\
\hs{\HH} =&⋅⎨˙W_j˙|˙j∈J˙⎬, \label{C695}\\
\qquad\hs{α} =&⋅⎨˙⁅⁅w˛y⁆˛a⁆˙|˙⁅w˛y˛a⁆∈\pj{WYA}(Q)˙⎬,\footnotemark \nichts{^{\text{\rf{E436}}}} \label{C696}\\
\hs{ι} =&⋅\pj{WI}(Q),⋅\text{and} \label{C697}\\
\hs{u} =&⋅u \label{C698}
\zz
\end{align} \subo (where $J$, $W$, $Y$, and $W_j$ abbreviate the projections $π_J(Q)$, $π_W(Q)$, $π_Y(Q)$, and $π_W(Q_j)$, in accord with (\rf{C825}) and the sentence thereafter).  \notetwotoone

\begin{nthm}\label{C691} The operator $\TB$ takes each pentaform game (Definition~\rf{C668}) to a traditional game (Definition~\rf{C843}). (Proof:\ Lemma~\rf{B331}.) \end{nthm}

Broadly speaking, a pentaform $Q$ is a high-dimensional relation, while a traditional extensive form is a list of low-dimensional relations.  Correspondingly, a pentaform describes the relationships between players, situations, nodes, and actions all at once, while a traditional extensive form describes the same relationships one by one.  In this light, it is reasonable that equations (\rf{C693})--(\rf{C697}) would use various projections of a pentaform $Q$ to build the first five components of a traditional game.  To be completely precise, [a] definitions (\rf{C693}), (\rf{C694}), (\rf{C696}), and (\rf{C697}) use projections of $Q$ itself, while [b] each information set $W_j$ in (\rf{C695}) is the projection of a {\em slice} of $Q$ (but, if $Q$ has information-set situations (\rf{D319}), then the right-hand side of (\rf{C695}) reduces to $⎨W_j|j∈J⎬ = ⎨j|j∈J⎬ = J$, which is yet another projection of $Q$ itself).

The practical import of Theorem~\rf{C691} is that each pentaform game can be interpreted as a traditional game.  In other words, the pentaform formulation does not introduce anything new.

\nssec{Bijection}{C692} 

The bijection of the following theorem is the paper's main result.  The bijection is illustrated in Figure~\rf{D126} by the two opposing arrows between the thick vertical bars.  As claimed in Section~\rf{B572}, this bijection is constructive and intuitive.  The construction is provided by equations (\rf{C678}) and (\rf{C677}).  The intuition is provided by Section~\rf{B565}'s second-last paragraph, by Section~\rf{C690}'s second-last paragraph, and by the remainder of this Section~\rf{C692}.\footnote{\label{E429}Incidentally, Table~\rf{C670}, Figure~\rf{D126}, and all the equations of Sections~\rf{C692} and \rf{E435} place traditional games on the left and pentaform games on the right.  This consistency seems helpful.}

\begin{figure}[h] \newcommand{\hgth}{3.3}
\begin{picture}(0,\hgth)  \myoutergrid{\hgth}
  \put(.6,1.4){\makebox(0,0){\scalebox{1}{ \begin{pspicture} \end{pspicture} }}} \end{picture}
\caption{The operators $\PB$ and $\TB$} \label{D126} \end{figure}

\begin{nthm}[{{\bf Main Theorem}}]\label{B560} $\PB$ is a bijection from the collection of traditional games to the collection of pentaform games with information-set situations (\rf{D319}).  Its inverse is the restriction of $\TB$ to the collection of pentaform games with information-set situations. (Proof~\rf{B560p}.) \end{nthm}

For example, consider the traditional game $(X^5,E^5,\HH^5,α^5,ι^5,u^5)$ from (\rf{C714}) and the pentaform game $(Q^1,u^1)$ from Figure~\rf{B265}'s tables.  Equation~(\rf{E430}) shows that $\PB(X^5,E^5,\HH^5,α^5,ι^5,u^5) = (Q^1,u^1)$.  By applying $\TB$ to both sides, we have \begin{gather}
\zz
\TB\PB(X^5,E^5,\HH^5,α^5,ι^5,u^5) = \TB(Q^1,u^1), \notag 
\zz
\end{gather} which by Theorem~\rf{B560} implies $(X^5,E^5,\HH^5,α^5,ι^5,u^5) = \TB(Q^1,u^1)$.  Thus the operators $\PB$ and $\TB$ toggle back and forth between $(X^5,E^5,\HH^5,α^5,ι^5,u^5)$ and $(Q^1,u^1)$.  In this sense, the two games are equivalent in spite of their superficial differences.

\newcommand{\notethree}{\footnote{\label{D041}As Figure~\rf{D126} suggests, the composition $\PB\TB$ changes any pentaform game into a pentaform game with information-set situations.  For example, consider $(Q^3,u^2)$, where $Q^3$ is from Figure~\rf{D348} and $u^2$ is from Figure~\rf{B266} (this mixture of examples is a well-defined pentaform game because $I^3 = I^2$ and $\ZZ^3 = \ZZ^2$).  Definition~(\rf{C677}) implies $\TB(Q^3,u^2) = \TB(Q^2,u^2)$, which implies $\PB\TB(Q^3,u^2) = \PB\TB(Q^2,u^2)$, which by Theorem~\rf{B560} and $Q^2$'s information-set situations implies $\PB\TB(Q^3,u^2) = (Q^2,u^2)$.  Hence the composition $\PB\TB$ changes $(Q^3,u^2)$, which does not have information-set situations, into $(Q^2,u^2)$, which does.}}

For another example, consider the pentaform game $(Q^2,u^2)$ in Figure~\rf{B266}.  Since this game has information-set situations,{\notethree} Theorem~\rf{B560} implies $\PB\TB(Q^2,u^2) = (Q^2,u^2)$.  In this sense, the traditional game $\TB(Q^2,u^2)$ and the pentaform game $(Q^2,u^2)$ are equivalent.  Further, this $\TB(Q^2,u^2)$ could be explicitly derived by applying the projections of definition (\rf{C677}) to $(Q^2,u^2)$.

\nssec{Derivative entities}{E435}

As discussed earlier, Table~\rf{C670}'s left-hand side summarizes the notation and terminology of an arbitrary traditional game.  More specifically, the table's left-hand side lists the 6 components of a traditional game, together with 10 of its derivatives.  Now consider the table's right-hand side.  It re-expresses the same 16 traditional entities using the terms of the traditional game's pentaform equivalent.  To be precise, the 16 rows of the table reproduce the 16 conclusions of the following proposition. 

\begin{prop}\label{C699} Suppose $(X,E,\HH,α,ι,u)$ is a traditional game, and let \linebreak $\PB(X,E,\HH,α,ι,u) = (\hp{Q},\hp{u})$.  Then the following hold. 
\tts{1}{C7x}~$X = \hp{W}⋃\hp{Y}$.
\tts{2}{C7e}~$E = \pj{WY}(\hp{Q})$.
\tts{3}{C7r}~$r = \hp{r}$.
\tts{4}{C7w}~$π_1E = \hp{W}$.
\tts{5}{C7y}~$π_2E = \hp{Y}$.
\tts{6}{C7pp}~$E^{-1} = \hp{p}$.
\tts{7}{C7wp}~$≼ = \hp{≼}$.
\tts{8}{C7sp}~$≺ = \hp{≺}$.
\tts{9}{C7z}~$\ZZ = \hp{\ZZ}$.
\tts{10}{C7h}~$\HH = \hp{J}$.
\tts{11}{C7L}~$α = ⎨⁅⁅w˛y⁆˛a⁆|⁅w˛y˛a⁆∈\pj{WYA}(\hp{Q})⎬$ (footnote~\rf{E401}).
\tts{12}{C7a}~$A = \hp{A}$.
\tts{13}{C7f}~$F = \hp{F}$.
\tts{14}{C7M}~$ι = \pj{WI}(\hp{Q})$.
\tts{15}{C7i}~$I = \hp{I}$.
\tts{16}{C7u}~$u = \hp{u}$.
(Proof \rf{C777p}.) \end{prop}

Conversely, Proposition~\rf{C777} interprets the derivatives of a pentaform game in terms of the pentaform game's traditional equivalent.  This proposition starts with an arbitrary pentaform game.  This setting is more general than that of Proposition~\rf{C699}, because the range of $\PB$ (in Proposition~\rf{C699}) consists of the special pentaform games with information-set situations (this fact is from Theorem~\rf{B560} and Figure~\rf{D126}).  To reconcile the propositions' conclusions, note that conclusion \rf{E453} of Proposition~\rf{C777} reduces to conclusion \rf{C7h} of Proposition~\rf{C699} in the special case of information-set situations (\rf{D319}). 
 
\nichts{}

\begin{prop}\label{C777} Suppose $(\hp{Q},\hp{u})$ is a pentaform game, and let $(X,E,\HH,α,ι,u) = \TB(\hp{Q},\hp{u})$.  Then all the conclusions of Proposition~\rf{C699} hold, except that conclusion~\rf{C7h} is replaced by \tts{10*}{E453} $\HH = ⎨˙\hp{W}_j˙|˙j∈\hp{J}˙⎬$. (Proof~\rf{C777p}.) \end{prop}

\appendix %

\vspace{2mm}
\section{Out-Trees}\label{E348}
\markb{\sc Appendix~\rf{E348}. Out-Trees}

\nssec{Familiar Definitions}{E349}

\newcommand{\notepaths}{\footnote{\label{E358}There is a related but distinct concept in the literature.  Let a ``path in a graph'' be a walk in a graph which visits each of its nodes {\em exactly once}.  Thus, within an arbitrary graph, a ``path in the graph'' is a special kind of a walk in the graph.  Nonetheless, if the graph is an out-tree, a ``path in the graph'' is equivalent to a walk in the graph.  This can be confusing:  It is important that walks rather than ``paths'' appear in the definition of an out-tree, and at the same time, within an out-tree, the distinction between walks and ``paths'' is irrelevant.}}

\newcommand{\notealtdef}{\footnote{Similar definitions in the game-theory literature include Perea 2001, page~10; Maschler, Solon, and Zamir 2013, page~41; and Kline and Luckraz 2016, page~88. \nocite{Perea01} \nocite{MaschSZ13} Similar definitions in the graph-theory literature include Tutte 1984, page~126; and Bang-Jensen and Gutin 2009, page~21. \nocite{Tutte84} \nocite{BangjG09} (Beware of variations in terminology.)  Several of these definitions apply only to finite trees, and several definitions explicitly include antisymmetry or irreflexivity.  In contrast, the definition here applies equally well to infinite trees, and the definition here implies antisymmetry and thus irreflexivity.\nichts{}}}

\newcommand{\noteuniquer}{\nichts{}}

A {\em directed graph} is a couple $(X,E)$ such that $E⋅⊆⋅X^2$.  The elements of $X$ are called {\em nodes}, and the elements of $E$ are called {\em edges}.  A {\em walk in a directed graph $(X,E)$ from $w$ to $y$} is a sequence $⁅x_0,x_1,x_2,...,x_m⁆$ in $X$ such that \begin{gather}
\zz
x_0 = w,⋅x_m = y,⋅\text{and}⋅(∀ℓ{<}m)⋅⁅x_ℓ˛x_{ℓ+1}⁆⋅∈⋅E. \label{E426}
\zz
\end{gather} Note that a walk{\notepaths} can visit a node more than once.  

An {\em out-tree}{\notealtdef} is a directed graph $(X,E)$ such that \begin{gather}
\zz
(∃r∈X)(∀x∈X)⋅\text{there exists a unique walk in $(X,E)$ from}⋅r⋅\text{to}⋅x. \label{E302}
\zz
\end{gather} The term ``out-tree'' is synonymous with the terms ``(outwardly) directed rooted tree'' and ``diverging arborescence''.  It can be shown that each out-tree has exactly one node $r$ that satisfies (\rf{E302}).{\noteuniquer}  Call this the {\em root node} of the out-tree.  For example, $(⎨\f5⎬,∅)$ is a one-node out-tree because the one-node sequence $⁅\f5⁆$ is the unique walk in $(⎨\f5⎬,∅)$ from the root node $r = \f5$ to itself.  An out-tree is said to be {\em nontrivial} iff it has more than one node.\nichts{}

\newcommand{\noteantisym}{\nichts{}}

Consider a nontrivial out-tree $(X,E)$ with its root node~$r$.  The remainder of this section will define its precedence relation and its collection of runs (that is, plays).

Let $≼$ and $≺$ be the binary relations on $X$ defined by \subi \label{E351} \begin{gather}
\zz
x⋅≼⋅y⋅\text{iff there is a walk in $(X,E)$ from $x$ to $y$, and} \label{E352} \\
x⋅≺⋅y⋅\text{iff ($x⋅≼⋅y$ and $x⋅≠⋅y$)}. \label{E353}
\zz
\end{gather} \subo Note $≼$ is reflexive because the right-hand side of (\rf{E352}) admits the trivial walk $(⎨x⎬,∅)$ from a node $x$ to itself.  Call $≼$ and $≺$ the {\em weak} and {\em strict precedence} orders, respectively.  It can be shown that $≼$ is a partial order on $X${\noteantisym} and that $≺$ is the asymmetric part of~$≼$.\nichts{}

Runs will be defined in three steps.  First, let an {\em end node} (or leaf) be a member of \begin{gather}
\zz
π_2E⧷π_1E, \label{E354}
\zz
\end{gather} and let a {\em finite run} (or finite play) be a walk in $(X,E)$ from the root node $r$ to an end node.  Second, let an {\em infinite run} (or infinite play) be a sequence $⁅x_0,x_1,...⁆$ in $X$ such that $x_0 = r$ and $(∀ℓ)$ $⁅x_ℓ˛x_{ℓ+1}⁆⋅∈⋅E$.  Third, let a {\em run} (or play) be a finite run or an infinite run.

Although each run is a {\em sequence} of nodes, each run can be characterized by the {\em set} of nodes that it visits.\footnote{One proof of this result is built on the fact that each walk in an out-tree must also be a ``path'' in the out-tree (footnote~\rf{E358}).\nichts{}}  Accordingly, a run will be regarded as a set $Z⋅⊆⋅X$.  Let $\ZZf$ be the collection of finite runs, let $\ZZi$ be the collection of infinite runs, and let \begin{gather}
\zz
\ZZ = \ZZf⋃\ZZi \label{E355}
\zz
\end{gather} be the collection of runs.  There are three possibilities:\ (i) $\ZZ = \ZZf$ and $\ZZi = ∅$, as in the out-trees of Section~\rf{B977}'s figures, (ii) $\ZZ = \ZZi$ and $\ZZf = ∅$, in which case there are no end nodes, and (iii) both $\ZZf$ and $\ZZi$ are nonempty, as in the out-tree of an infinite centipede game.\nichts{}

\nssec{Two Lemmas}{E350}

\newcommand{\noteKnuth}{\footnote{\label{E400}Lemmas \rf{E320} and \rf{E337} can be combined to characterize nontrivial out-trees:\nichts{}  {\em Let $E$ be a set of couples.  Then $(X,E)$ is a nontrivial out-tree iff (a) $X = π_1E⋃π_2E$, (b) $E^{-1}$ is a function, (c) $(∀y∈π_2E)(∃m≥1)$ $E^{-m}(y)⋅∉⋅π_2E$, and (d) $π_1E⧷π_2E$ is a singleton.}  This way of expressing an out-tree is unusual.  A similar concept is in Streufert 2018, equation (1).  A related but less similar concept is in Knuth 1997, page 373.\nocite{Knuth97}}}

Lemmas~\rf{E320} and \rf{E337} depart slightly from the literature.{\noteKnuth}  They are used to prove Proposition~\rf{C632}.\nichts{}  Later, Proposition~\rf{C632} is used to prove Theorems \rf{C689}--\rf{C699}.  
\nichts{}
\nichts{}

\begin{lemma}\label{E320} Suppose $(X,E)$ is a nontrivial out-tree and its root is~$r$.  Then \subi \begin{gather}
\zz
X = π_1E⋃π_2E, \label{E347} \\ 
E^{-1}⋅\text{is a function}, \label{E344} \\
(∀y∈π_2E)(∃m≥1)⋅E^{-m}(y)⋅∉⋅π_2E,⋅\text{and} \label{E345} \\
⎨r⎬ = π_1E⧷π_2E. \label{E346} 
\zz
\end{gather}\subo
\IJGTversiononly{(Proof: Streufert 2024, Lemma~\rf{E320}.)}
\end{lemma}

\arXivversiononly{

\begin{pf}  See Claims~\rf{E338}--\rf{E343}. \begin{cllist}

\yl{E340} $r⋅∈⋅π_1E$. Nontriviality implies there is a node $x$ distinct from~$r$.  Thus (\rf{E302}) implies there is a walk of length greater than zero from $r$ to $x$; which by the walk definition (\rf{E426}) implies there is an $x_1$ such that $⁅r˛x_1⁆⋅∈⋅E$; which implies $r⋅∈⋅π_1E$.

\yl{E339} $r⋅∉⋅π_2E$. Suppose $r⋅∈⋅π_2E$.  Then there is $w⋅∈⋅X$ such that $⁅w˛r⁆⋅∈⋅E$ (it is irrelevant whether $w$ and $r$ are distinct).  Meanwhile, (\rf{E302}) implies there is a walk from $r$ to~$w$.  Thus $⁅w˛r⁆⋅∈⋅E$ implies there is a still longer walk from $r$ to $w$, back to $r$ via $⁅w˛r⁆$, and on to $w$ for a second time.  The previous two sentences  contradict the uniqueness in (\rf{E302}). 

\yl{E338} {\em (\rf{E347}) holds.}  Since $(X,E)$ is a directed graph, we have $E⋅⊆⋅X^2$, which implies $X⋅⊇⋅π_1E⋃π_2E$.  To show $X⋅⊆⋅π_1E⋃π_2E$, take an arbitrary $x⋅∈⋅X$.  On the one hand, if $x = r$, then Claim~\rf{E340} implies $x⋅∈⋅π_1E$.  On the other hand, if $x⋅≠⋅r$, then (\rf{E302}) implies there is a walk of length greater than zero from $r$ to $x$; which by the walk definition (\rf{E426}) implies there is a sequence $⁅x_0,x_1,...,x_m⁆$ with $x_0 = r$, $x_m = x$, $m⋅≥⋅1$, and $(∀ℓ{<}m)$ $⁅x_ℓ,x_{ℓ+1}⁆⋅∈⋅E$; which implies $⁅x_{m-1},x⁆⋅∈⋅E$; which implies $x⋅∈⋅π_2E$.

\yl{E341} {\em (\rf{E344}) holds.} Since $(X,E)$ is a directed graph, we have $E⋅⊆⋅X^2$, which implies $E$ is a relation.  Thus it suffices to show that there are not $y$, $w^1$, and $w^2$ such that $⁅y˛w^1⁆⋅∈⋅E^{-1}$ and $⁅y˛w^2⁆⋅∈⋅E^{-1}$ and $w^1⋅≠⋅w^2$ (footnote~\rf{C273}).  Suppose there were.  Then we would have \subi \begin{gather}
\zz
⁅w^1˛y⁆⋅∈⋅E, \label{E317} \\
⁅w^2˛y⁆⋅∈⋅E,⋅\text{and} \label{E318} \\
w^1⋅≠⋅w^2. \label{E319}  
\zz
\end{gather} \subo By (\rf{E302}) at $x = w^1$, and by (\rf{E317}), there is a walk from $r$ to $w^1$ and then on to $y$ in one step.  Similarly, by (\rf{E302}) at $x = w^2$, and by (\rf{E318}), there is a walk from $r$ to $w^2$ and then on to $y$ in one step.  Yet, by (\rf{E302}) at $x = y$, there is a unique walk from $r$ to $y$, which contradicts the previous two sentences because $w^1$ and $w^2$ are distinct by (\rf{E319}).

\yl{E342} {\em (\rf{E345}) holds.}  Take $y⋅∈⋅π_2E$.  Then Claim~\rf{E339} implies $y⋅≠⋅r$; which by (\rf{E302}) implies there is a walk of length greater than zero from $r$ to $y$; which implies there is a sequence $⁅x_0,x_1,...,x_m⁆$ such that  \begin{gather}
\zz
x_0 = r,⋅x_m = y,⋅m⋅≥⋅1,⋅\text{and}⋅(∀ℓ{<}m)⋅⁅x_ℓ,x_{ℓ+1}⁆⋅∈⋅E; \notag 
\zz
\end{gather} which by Claim~\rf{E341} (\rf{E344}) implies $x_0 = r$, $x_m = y$, $m⋅≥⋅1$, and $(∀ℓ{<}m)$ $x_ℓ = E^{-1}(x_{ℓ+1})$; which implies $r = E^{-m}(y)$; which by Claim~\rf{E339} implies $E^{-m}(y)⋅∉⋅π_2E$.  

\yl{E343} {\em (\rf{E346}) holds.}  For $⎨r⎬⋅⊆⋅π_1E⧷π_2E$, Claim~\rf{E340} implies $r⋅∈⋅π_1E$, and Claim~\rf{E339} implies $r⋅∉⋅π_2E$.  For $⎨r⎬⋅⊇⋅π_1E⧷π_2E$, it suffices to show  \begin{gather}
\zz
π_1E⧷⎨r⎬⋅⊆⋅π_2E. \notag 
\zz
\end{gather} Take $w⋅∈⋅π_1E⧷⎨r⎬$.  Then $w⋅∈⋅π_1E$ and Claim~\rf{E338} (\rf{E347}) imply $w⋅∈⋅X$.  Thus $w⋅≠⋅r$ and (\rf{E302}) imply that there is a walk of length greater than zero from $r$ to $w$; which implies there is $⁅x_0,x_1,...,x_m⁆$ such that $x_0 = r$, $x_m = w$, $m⋅≥⋅1$, and $(∀ℓ{<}m)$ $⁅x_ℓ,_{ℓ+1}⁆⋅∈⋅E$; which implies $⁅x_{m-1}˛w⁆⋅∈⋅E$; which implies $w⋅∈⋅π_2E$. \end{cllist} \unskipcl \end{pf} 

}%

\begin{lemma}\label{E337} Let $E$ be a set of couples.  Then $(X,E)$ is a nontrivial out-tree if \subi 
 \begin{gather}
\zz
X = π_1E⋃π_2E, \label{E324} \\
E^{-1}⋅\text{is a function}, \label{E321} \\
(∀y∈π_2E)(∃m≥1)⋅E^{-m}(y)⋅∉⋅π_2E,⋅\text{and} \label{E322} \\
π_1E⧷π_2E⋅\text{is a singleton}. \label{E323} 
\zz
\end{gather}\subo%
\IJGTversiononly{(Proof: Streufert 2024, Lemma~\rf{E337}.)}
\end{lemma}

\arXivversiononly{

\begin{pf} Suppose $(X,E)$ satisfies (\rf{E324})--(\rf{E323}).  By (\rf{E323}) define $r$ by \begin{gather}
\zz
⎨r⎬ = π_1E⧷π_2E. \label{E325}
\zz
\end{gather} For future reference, we argue that \begin{gather}
\zz
(∀x∈X,m≥0,n≥0)⋅E^{-m}(x) = E^{-n}(x) = r⋅\text{implies}⋅m = n. \label{E326}
\zz
\end{gather} Suppose $E^{-m}(x) = E^{-n}(x) = r$ and yet $m⋅≠⋅n$.  Assume without loss of generality that $m > n$.  Then the equalities and manipulation imply $r = E^{-m}(x) = E^{-(m-n)}(E^{-n}(x)) = E^{-(m-n)}(r)$, which by $m > n$ implies $⁅E^{-1}(r),r⁆⋅∈⋅E$, which implies $r⋅∈⋅π_2E$, which violates (\rf{E325}). 

Since $E$ is assumed to be a set of couples, (\rf{E324}) implies $E⋅⊆⋅X^2$, which implies $(X,E)$ is a directed graph.  Thus it suffices to show that $r$ satisfies (\rf{E302}).  Toward that end, take an arbitrary node $x⋅∈⋅X$.  

On the one hand, suppose $x = r$.  By inspection, $⁅r⁆$ is a one-node walk in $(X,E)$ from $x = r$ to itself. For uniqueness, note that any other walk in $(X,E)$ from $r$ to itself would be a sequence $⁅y_0,y_1,...,y_n⁆$ with $y_0=r$, $y_n=r$, $n⋅≥⋅1$, and $(∀k{<}n)$ $⁅y_k,y_{k+1}⁆⋅∈⋅E$.  This implies $⁅y_{n-1},r⁆⋅∈⋅E$, which implies $r⋅∈⋅π_2E$, which violates (\rf{E325}).

On the other hand, suppose $x⋅≠⋅r$.  Then (\rf{E325}) implies $x⋅∉⋅π_1E⧷π_2E$; which by (\rf{E324}) implies $x⋅∈⋅π_2E$; which by (\rf{E322}) implies that there is an $m⋅≥⋅1$ such that $E^{-m}(x)⋅∉⋅π_2E$; which by (\rf{E324}) implies $E^{-m}(x)⋅∈⋅π_1E⧷π_2E$; which by (\rf{E325}) implies \subi \begin{gather}
\zz
E^{-m}(x) = r. \label{E335}
\zz
\end{gather} This implies that \begin{gather}
\zz
⁅x_ℓ⁆^m_{ℓ=0} = ⁅E^{-(m-ℓ)}(x)⁆^m_{ℓ=0} \label{E332}  
\zz
\end{gather} \subo is a sequence such that $x_0 = r$ and $x_m = x$.  By (\rf{E332}) and manipulation, \begin{gather}
\zz
(∀ℓ{<}m)⋅x_ℓ = E^{-(m-ℓ)}(x) = E^{-1}(E^{-(m-(ℓ+1))}(x)) = E^{-1}(x_{ℓ+1});\notag 
\zz
\end{gather} which implies $(∀ℓ{<}m)$ $⁅x_ℓ˛x_{ℓ+1}⁆⋅∈⋅E$; which by the previous sentence implies that $⁅x_ℓ⁆^m_{ℓ=0}$ is a walk in $(X,E)$ from $r$ to~$x$.

For uniqueness, suppose that $⁅y_k⁆^n_{k=0}$ is an arbitrary walk in $(X,E)$ from $r$ to~$x$.  Then $y_0 = r$, $y_n = x$, and $(∀k{<}n)$ $⁅y_k˛y_{k+1}⁆⋅∈⋅E$; which implies $y_0 = r$, $y_n = x$, and $(∀k{<}n)$ $y_k = E^{-1}(y_{k+1})$; which implies \subi  \begin{gather}
\zz
y_0 = r⋅\text{and} \label{E333} \\
⁅y_k⁆^n_{k=0} = ⁅E^{-(n-k)}(x)⁆^n_{k=0}. \label{E334}
\zz
\end{gather} \subo By (\rf{E332}) and (\rf{E334}), the walks $⁅x_ℓ⁆^m_{ℓ=0}$ and $⁅y_k⁆^n_{k=0}$ are equal provided that $m = n$ (note $ℓ$ and $k$ are just dummy variables).  Thus it suffices to show that $m = n$.  Toward that end, note that (\rf{E333}) and (\rf{E334}) at $k = 0$ imply that $E^{-n}(x) = r$.  This and (\rf{E335}) imply that the hypothesis of (\rf{E326}) is met, which by (\rf{E326}) implies $m = n$.  \end{pf}

}%

\newcommand{\qbox}[1]{
  & & & & & & & & \\[-2.5ex] 
  $ \begin{array}{l} ⎨˙⁅\f{Ann˛now˛0˛b˛1}⁆, \\ #1 \end{array} $}
\newcommand{\tch}{\mspace{12mu}}
\newcommand{\noo}{\raisebox{-.4ex}{\Large $\circ$}}
\newcommand{\twoline}{\\[1.7ex] \hline}
\newcommand{\threeline}{\\[3.2ex] \hline}
\newcommand{\ir}{\nichts{}}
\newcommand{\ib}{\nichts{}}
\newcommand{\iu}{\nichts{}}
\newcommand{\iB}{\nichts{}} 
\newcommand{\iC}{\nichts{}} 
\newcommand{\iD}{\nichts{}}
\newcommand{\iR}[1]{\nichts{}}

\newcommand{\tablecx}{
\begin{table}[t]
{\Small 
\yyin{\vspace{7mm}}
\begin{tabular}{l|ccc|ccc|cc}
$\mspace{80mu} Q$ & \rf{Pij} & \rf{Pjw} & \rf{Pwa} & \rf{Pway} & \rf{Pwy} & \rf{Pay}
  & \rf{Py} & \rf{Pr} \\[1ex] \hline
\qbox{ 
  \tch⁅\f{{\ir Bob}˛now˛0˛b˛1}⁆˙⎬ }
&\noo&+\iC&+\iR{1x1}     &+\iC&+\iC&+\iC &+&+\iR{0} \twoline
\qbox{ 
  \tch⁅\f{Ann˛{\ir later}˛0˛b˛1}⁆˙⎬ }
&+\iB&\noo&+\iR{1x1,1x1} &+\iC&+\iC&+\iC &+&+\iR{0} \twoline
\qbox{
  \tch{\ib ⁅\f{Ann˛now˛0˛c˛2}⁆}, \\
  \tch{\ib ⁅\f{Ann˛now˛1˛b˛3}⁆}˙⎬ }
&+\iC&+\iC&\noo          &+\iD&+\iD&+\iD &+&+\iR{0} \threeline
\qbox{
  \tch⁅\f{Ann˛now˛0˛b˛{\ir 2}}⁆˙⎬ }
&+\iC&+\iC&+\iR{1x1}     &\noo&+\iB&+\iB &+&+\iR{0} \twoline
\qbox{
  \tch⁅\f{Ann˛now˛{\ir 1}˛b˛1}⁆˙⎬ }
&+\iC&+\iB&+\iR{2x1} &+\iB&\noo&+\iC &+$\footnotemark$&+\iR{0} \twoline
\qbox{
  \tch⁅\f{Ann˛now˛0˛{\ir c}˛1}⁆˙⎬ }
&+\iC&+\iC&+\iR{1x2}     &+\iD&+\iC&\noo &+&+\iR{0} \twoline
\qbox{
  \tch{\iu ⁅\f{Ann˛now˛2˛b˛3}⁆}, \\
  \tch⁅\f{Ann˛now˛3˛b˛2}⁆˙⎬ }
&+\iC&+\iC&+\iR{3x1}     &+\iD&+\iD&+\iB &\noo&+\iR{0} \threeline
\qbox{
  \tch{\iu ⁅\f{Ann˛now˛2˛b˛3}⁆}˙⎬ }
&+\iC&+\iC&+\iR{2x1}     &+\iD&+\iD&+\iB &+&\noo\iR{0,2} \twoline  
\multicolumn{4}{c}{}\\
\end{tabular} } 
\caption{Eight examples, each of which violates exactly one axiom.  This can be efficiently verified by inspecting each column.  (See footnote \rf{D389} about one exceptional cell.)\nichts{}}
\label{D388}
\end{table}}

\newcommand{\notetablecx}{\footnotetext{\label{D389}When axiom \rf{Pwy} fails, the $p = \pj{YW}(Q)$ defined within \rf{Pwy} fails to be a function, which then leaves axiom \rf{Py} ill-defined.  This affects one cell in Table~\rf{D388}{:} in that context, $p = \pj{YW}(Q)$ is taken to be a correspondence, and \rf{Py} is taken to mean that $(∀y∈Y)(∃m≥1)⋅p^m(y)⧷Y⋅≠⋅∅$.\nichts{}}}

\renewcommand{\sectitle}{For Pentaform Definition (Section~\rf{B561})}
\section{\sectitle}\label{B566}\showit
\markb{\sc Appendix \rf{B566}. \sectitle}

\begin{lemma}[from set theory]\label{D327}  Suppose $G$ is a set of couples $⁅x˛y⁆$.  Then define $X = π_1G$, define $Y = π_2G$, and for each $y⋅∈⋅Y$ define $G^{-1}(y) = ⎨x|⁅x˛y⁆∈G⎬$.  Then the following are equivalent.  \begin{tlist}
\yl{C844} $G$ is a function.  
\yl{C848} $(∀y_1∈Y,y_2∈Y)$ $y_1⋅≠⋅y_2$ implies $G^{-1}(y_1)⋂G^{-1}(y_2) = ∅$. 
\yl{C847} $⁅G^{-1}(y)⁆_{y∈Y}$ is an injectively indexed partition of~$X$. 
\end{tlist} \end{lemma}

\nichts{} 

\nichts{}

\begin{lemma}\label{D333} Suppose $Q$ is a quintuple set.  Then the following hold. \begin{tlist}
\yl{D366} $(∀j∈J)$ $W_j = ⎨w|⁅w˛j⁆∈\pj{WJ}(Q)⎬$.  
\yl{D367} $(∀j∈J)$ $Y_j = ⎨y|⁅y˛j⁆∈\pj{YJ}(Q)⎬$. 
\yl{D445} $(∀j∈J)$ $\pj{WA}(Q_j) = ⎨⁅w˛a⁆|⁅w˛a˛j⁆∈\pj{WAJ}(Q)⎬$.  
\end{tlist} \end{lemma}

\begin{pf} {\em (\rf{D366})}.  By $Q_j$'s definition~(\rf{D304}), $π_W(Q_j)$ equals $π_W(˙⎨⁅i_*,j,w_*,a_*,y_*⁆∈Q⎬˙)$, which by the definition of $π_W$ equals \begin{gather}
\zz
⎨˙w˙|˙(∃i_*,a_*,y_*)˙⁅i_*,j,w,a_*,y_*⁆∈Q˙⎬, \notag 
\zz
\end{gather} which by the definition of $\pj{WJ}(Q)$ equals $⎨˙w˙|˙⁅w˛j⁆∈\pj{WJ}(Q)˙⎬$. 

{\em (\rf{D367})}.  This is proved as part (\rf{D366}), except for reversing the roles of $w$ and~$y$. 

{\em (\rf{D445})}.  Definition~(\rf{D304}) implies $\pj{WA}(Q_j)$ equals $\pj{WA}(˙⎨⁅i_*,j,w_*,a_*,y_*⁆∈Q⎬˙)$, which by the definition of $\pj{WA}$ equals \begin{gather}
\zz
⎨˙⁅w˛a⁆˙|˙(∃i_*,y_*)˙⁅i_*,j,w,a,y_*⁆∈Q˙⎬, \notag 
\zz
\end{gather} which by the definition of $\pj{WAJ}(Q)$ equals $⎨˙⁅w˛a⁆˙|˙⁅w˛a˛j⁆∈\pj{WAJ}(Q)˙⎬$. \end{pf}

\begin{npf}[{{\bf for Proposition~\rf{D328}}}]\label{D328p} ⋅\begin{cllist}

\yl{E183} {\em For each $j⋅∈⋅J$, the inverse image $(\pj{WJ}(Q))^{-1}(j)$ equals the information set~$W_j$.}  Fix a situation $j⋅∈⋅J$.  Then its inverse image $(\pj{WJ}(Q))^{-1}(j)$ by definition equals $⎨˙w˙|˙⁅w˛j⁆∈\pj{WJ}(Q)˙⎬$, which by Lemma~\rf{D333}(\rf{D366}) equals the information set $W_j$. \end{cllist}

{\em Conclusion}. Consider Lemma~\rf{D327} with $G$ there equal to $\pj{WJ}(Q)$ here.  Thus $X$, $Y$, and $G^{-1}(y)$ there equal $W$, $J$, and $(\pj{WJ}(Q))^{-1}(j)$ here.  So the lemma implies that the following are equivalent. \begin{tlist}[*] 
\yl{E184} $\pj{WJ}(Q)$ is a function.
\yl{E185} $(∀j_1∈J,j_2∈J)$ $j_1⋅≠⋅j_2$ implies $(\pj{WJ}(Q))^{-1}(j_1)⋅⋂⋅(\pj{WJ}(Q))^{-1}(j_2) = ∅$.  
\yl{E186} $⁅˙(\pj{WJ}(Q))^{-1}(j)˙⁆_{j∈J}$ is an injectively indexed partition of~$W$.  
\end{tlist} Condition (\rf{E184}*) is equivalent to the proposition's (\rf{D329}) by the definition of \rf{Pjw} in Definition~\rf{C669}.  Meanwhile, (\rf{E185}*) is equivalent to the proposition's (\rf{D330}) by Claim~\rf{E183}.  Similarly, (\rf{E186}*) is equivalent to the proposition's (\rf{D331}) by Claim~\rf{E183}. \end{npf}

\begin{lemma}\label{D368} Suppose $Q$ is a quintuple set which satisfies \rf{Pjw} and \rf{Pwy}.  Then $⁅Y_j⁆_{j∈J}$ is an injectively indexed partition of~$Y$. \end{lemma}

\begin{pf} Consider Lemma~\rf{D327} with $G$ there equal to $\pj{YJ}(Q)$ here.  Then $X$, $Y$, and $G^{-1}(y)$ there equal $Y$, $J$, and $(\pj{YJ}(Q))^{-1}(j)$ here.  Therefore, since the lemma assumptions \rf{Pjw} and \rf{Pwy} imply that $\pj{YJ}(Q)$ is a function, Lemma~\rf{D327}(\rf{C844}$⇒$\rf{C847}) implies that $⁅\pj{YJ}(Q)^{-1}(j)⁆_{j∈J}$ is an injectively indexed partition of~$Y$.  Each $\pj{YJ}(Q)^{-1}(j)$ by definition equals $⎨˙y˙|˙⁅y˛j⁆∈\pj{YJ}(Q)˙⎬$, which by Lemma~\rf{D333}(\rf{D367}) equals $Y_j$. \end{pf}

\begin{npf}[{{\bf for Proposition~\rf{C601}}}]\label{C601p}⋅ 

\newcommand{\notegslice}{\footnote{\label{E388}Proof~\rf{C601p} uses a general slice operator of the form $T|_{C:c}$.  Specifically, suppose that $T$ is a set of tuples with the same set of coordinates, and that $C$ is one of those coordinates.  Then let $T|_{C:c}$ be the set consisting of those tuples in $T$ that have the value $c$ in the coordinate~$C$.  For example, any situation slice $Q_j$ could be more explicitly expressed as $Q|_{J:j}$.}}

\newcommand{\notegproj}{\footnote{\label{E386}To be clear, consider the expression $π_W(\pj{WA}(Q_j))$.  The inner projection is a projection of a set of quintuples, as discussed in Section~\rf{C596}.  Meanwhile, the outer projection is a projection of a set of couples, which is taken to be self-explanatory.  Similar self-explanatory projections appear throughout Proof~\rf{C601p}.}} 

{\em (\rf{C602}$⇒$\rf{C605})}. Assume (\rf{C602}), and take a situation $j⋅∈⋅J$.  Then $\pj{WA}(Q_j)$ is a Cartesian product, which implies $\pj{WA}(Q_j) = π_W(\pj{WA}(Q_j))˙×˙π_A(\pj{WA}(Q_j))$.{\notegproj} Note $π_W(\pj{WA}(Q_j))$ is $π_W(Q_j)$, which by abbreviation (\rf{C825}) is $W_j$.  Similarly $π_A(\pj{WA}(Q_j))$ is $π_A(Q_j)$, which by abbreviation (\rf{C825}) is $A_j$. 

{\em (\rf{C605}$⇒$\rf{C603})}.  Assume (\rf{C605}).  Take a situation $j⋅∈⋅J$ and a decision node $w$ in $j$'s information set $W_j$.  Express the node's feasible action set $F(w)$ as $π_A(F|_{W:w})$.{\notegslice}  By $F$'s definition (\rf{C981}), this is $π_A(\pj{WA}(Q)|_{W:w})$, which by inspection is $π_A(\pj{WA}(Q|_{W:w}))$, which by the assumption $w⋅∈⋅W_j$ and the axiom \rf{Pjw} is $π_A(\pj{WA}(Q_j|_{W:w}))$, which by inspection is $π_A(\pj{WA}(Q_j)|_{W:w})$, which by part (\rf{C605}) is $π_A((W_j×A_j)|_{W:w})$, which by the assumption $w⋅∈⋅W_j$ is $π_A(⎨w⎬×A_j)$, which by inspection is $A_j$. 

{\em (\rf{C603}$⇒$\rf{C604})}.  This holds by inspection.

{\em (\rf{C604}$⇒$\rf{C602})}.  Assume (\rf{C604}).  Take a situation $j⋅∈⋅J$.  It must be shown that $\pj{WA}(Q_j)$ is a Cartesian product.  Note $π_W(\pj{WA}(Q_j))$ is $π_W(Q_j)$, which by abbreviation (\rf{C825}) is $W_j$.  Thus it suffices to show that \begin{gather}
\zz
(∀w_1∈W_j,w_2∈W_j)⋅⎨a|⁅w_1˛a⁆∈\pj{WA}(Q_j)⎬ = ⎨a|⁅w_2˛a⁆∈\pj{WA}(Q_j)⎬.\notag
\zz
\end{gather} Toward that end, take decision nodes $w_1⋅∈⋅W_j$ and $w_2⋅∈⋅W_j$.  Then the left-hand side is $π_A(˙⎨⁅w_1˛a⁆∈\pj{WA}(Q_j)⎬˙)$, which can be expressed using the generalized slice operator of footnote~\rf{E388} as $π_A(\pj{WA}(Q_j)|_{W:w_1})$, which by assumption $w_1⋅∈⋅W_j$ and axiom \rf{Pjw} is equal to $π_A(\pj{WA}(Q)|_{W:w_1})$, which by $F$'s definition (\rf{C981}) is equal to $π_A(F|_{W:w_1})$, which is an alternative expression for $F(w_1)$.  By the same reasoning, the right-hand side is $F(w_2)$.  Finally, $F(w_1) = F(w_2)$ by assumption~(\rf{C604}). \end{npf}

\begin{lemma}\label{E359} Suppose $Q$ is a quintuple set which satisfies \rf{Pwy}, \rf{Py}, and \rf{Pr}.  Then $(W⋃Y,\pj{WY}(Q))$ is a nontrivial out-tree. \end{lemma} 

\begin{pf} ⋅\begin{cllist}
 
\yl{E361} $W = π_1(\pj{WY}(Q))$.  By abbreviation (\rf{C825}), $W$ equals $π_W(Q)$, which by inspection equals $π_1(\pj{WY}(Q))$.  

\yl{E362} $Y = π_2(\pj{WY}(Q))$.  By abbreviation (\rf{C825}), $Y$ equals $π_Y(Q)$, which by inspection equals $π_2(\pj{WY}(Q))$. 

\yl{E363} {\em (a) $\pj{YW}(Q) = (\pj{WY}(Q))^{-1}$. (b) $p = (\pj{WY}(Q))^{-1}$.}  Part (a) holds by inspection.  Part (b) holds by (a) and the definition of $p$ in axiom \rf{Pwy}. \end{cllist}

{\em Conclusion.} By inspection $\pj{WY}(Q)$ is a set of couples.  Thus by Lemma~\rf{E337}, it suffices to show that \subi\begin{gather}
\zz
W⋃Y = π_1(\pj{WY}(Q))⋃π_2(\pj{WY}(Q)), \label{E410} \\
(\pj{WY}(Q))^{-1}⋅\text{is a function}, \label{E411} \\
(∀y∈π_2(\pj{WY}(Q)))(∃m≥1)⋅(\pj{WY}(Q))^{-m}(y)⋅∉⋅π_2(\pj{WY}(Q)),⋅\text{and} \label{E412} \\
π_1(\pj{WY}(Q))⧷π_2(\pj{WY}(Q))⋅\text{is a singleton}. \label{E413}
\zz
\end{gather}\subo Claims \rf{E361} and \rf{E362} imply (\rf{E410}).  Axiom \rf{Pwy} and Claim~\rf{E363}(a) imply (\rf{E411}).  Axiom \rf{Py} and Claims \rf{E362} and \rf{E363}(b) imply (\rf{E412}). Axiom \rf{Pr} and Claims \rf{E361} and \rf{E362} imply (\rf{E413}).
 \end{pf}

\tablecx

\begin{lemma}\label{E360} Suppose $Q$ is a quintuple set such that $(W⋃Y,\pj{WY}(Q))$ is a nontrivial out-tree.  Let $r$ denote its root.  Then $Q$ satisfies \rf{Pwy}, \rf{Py}, \rf{Pr}, and $⎨r⎬ = W⧷Y$. \end{lemma}

\begin{pf} ⋅\begin{cllist}

\yl{E369} $π_1(\pj{WY}(Q)) = W$.  By inspection, $π_1(\pj{WY}(Q))$ equals $π_W(Q)$, which by abbreviation (\rf{C825}) equals~$W$.

\yl{E370} $π_2(\pj{WY}(Q)) = Y$.  By inspection, $π_2(\pj{WY}(Q))$ equals $π_Y(Q)$, which by abbreviation (\rf{C825}) equals~$Y$. 

\yl{E371} $(\pj{WY}(Q))^{-1} = \pj{YW}(Q)$.  This holds by inspection. \end{cllist}

\notetablecx

{\em Conclusion}.  Because $(W⋃Y,\pj{WY}(Q))$ is a nontrivial out-tree by assumption, \linebreak Lemma~\rf{E320}(\rf{E344})--(\rf{E346}) implies \subi\begin{gather}
\zz
(\pj{WY}(Q))^{-1}⋅\text{is a function}, \label{E414} \\
(∀y∈π_2(\pj{WY}(Q)))(∃m≥1)⋅(\pj{WY}(Q))^{-m}(y)⋅∉⋅π_2(\pj{WY}(Q)),⋅\text{and} \label{E415} \\
⎨r⎬ = π_1(\pj{WY}(Q))⧷π_2(\pj{WY}(Q)). \label{E416} 
\zz
\end{gather}\subo Equation (\rf{E414}) and Claim~\rf{E371} imply axiom \rf{Pwy}.  Axiom~\rf{Pwy} defines the \linebreak immediate-predecessor function $p$ to be $\pj{YW}(Q)$, which by Claim~\rf{E371} implies $p = (\pj{WY}(Q))^{-1}$, so that equation (\rf{E415}) and Claim~\rf{E370} imply axiom \rf{Py}.  Equation~(\rf{E416}) and Claims~\rf{E369} and \rf{E370} imply $⎨r⎬ = W⧷Y$.  This implies axiom \rf{Pr}. 
\end{pf}

\begin{npf}[{{\bf for Proposition~\rf{C632}}}]\label{C632p} Part (a) follows by inspection from Lemmas \rf{E359} and \rf{E360}.  For part (b), suppose $Q$ satisfies \rf{Pjw}, \rf{Py}, and \rf{Pr}.  Then Lemma~\rf{E359} implies $(W⋃Y,\pj{WY}(Q))$ is a nontrivial out-tree.  Let $r$ be its root.  Then Lemma~\rf{E360} implies $⎨r⎬ = W⧷Y$. \end{npf}

\begin{lemma}\label{D478} Suppose $Q$ is a quintuple set which satisfies \rf{Pwy} and \rf{Py}.  Then the following hold. \begin{tlist}
\yl{E531} $(∀y∈Y)(∃ℓ≥1)$ $p^ℓ(y)⋅∈⋅W⧷Y$.  
\yl{E532} If $Q$ also satisfies \rf{Pr}, then $(∀y∈Y)(∃ℓ≥1)$ $p^ℓ(y) = r$.  \end{tlist} \end{lemma}

\begin{pf} {\em (\rf{E531})}. Take a successor node $y⋅∈⋅Y$.  Then \rf{Py} implies there is $ℓ⋅≥⋅1$ such that $p^ℓ(y)⋅∉⋅Y$.  Further, since \rf{Pwy} defines $p$ to be $\pj{YW}(Q)$, the range of $p$ is $π_W(Q)$, which by abbreviation (\rf{C825}) is~$W$.  Thus $p^ℓ(y)⋅∈⋅W⧷Y$.  

{\em (\rf{E532})}.  If \rf{Pr} also holds, then (\rf{C982}) defines $r$ to be the sole member of $W⧷Y$.  Hence (\rf{E532}) follows from (\rf{E531}). \end{pf}

\section{For Pentaform Tools and Applications (Section~\rf{C563})}\label{E189}
\markb{\sc Appendix \rf{E189}.  For Pentaform Tools and Applications (Section~\rf{C563})}

\begin{npf}[{{\bf for Proposition~\rf{C372}}}]\label{C372p}  {\em (\rf{D176})}. First consider axiom \rf{Pij}.  
Since a function is being regarded as a set of couples (see footnote \rf{C273} on \pgrf{C273}), we have that \lic{D128} any subset of a function is itself a function.  Meanwhile, the assumption $Q′⋅⊆⋅Q$ implies \il{D127} $\pj{JI}(Q′)⋅⊆⋅\pj{JI}(Q)$.  Then in steps, \rf{Pij} for $Q$ implies $\pj{JI}(Q)$ is a function, which by \rf{D128} and \rf{D127} implies $\pj{JI}(Q')$ is a function, which implies \rf{Pij} for~$Q′$.  

Next consider each of the axioms \rf{Pjw}, \rf{Pway}, \rf{Pay}, and \rf{Pwy}.  Here the axiom for $Q$ implies the corresponding axiom for $Q′$ by an argument similar to that of the previous paragraph for \rf{Pij}.

Finally consider axiom \rf{Py}.  First note that \rf{Pwy} for $Q′$ has already been derived, and that this implies $p′ = \pj{YW}(Q′)$ is a well-defined function with domain $Y′$ and range $W′$.  Now take $y⋅∈⋅Y′$.   Then $p′(y)⋅∈⋅W′$ is well-defined.  [Step~1] If $p′(y)⋅∉⋅Y′$, then $p′(y)⋅∈⋅W′⧷Y′$ and the argument is complete.  Else $p′(y)⋅∈⋅Y′$ so $(p′)^2(y)⋅∈⋅W′$ is well-defined.  [Step~2] If $(p′)^2(y)⋅∉⋅Y′$, then $(p′)^2(y)⋅∈⋅W′⧷Y′$ and the argument is complete.  Else $(p′)^2(y)⋅∈⋅Y′$ so $(p′)^3(y)⋅∈⋅W′$ is well-defined.  By repeating this process indefinitely, either the argument finishes at some step or $(∀m≥1)$ $(p′)^m(y)⋅∈⋅Y′$.  

To rule out the latter contingency, suppose it held.  Note $Q′⋅⊆⋅Q$ implies $p′ = \pj{YW}(Q′)$ is a restriction of $p = \pj{YW}(Q)$.  Hence the supposition implies that $(∀m≥1)$ $p^m(y)⋅∈⋅Y′$.  Also note $Q′⋅⊆⋅Q$ implies $Y′⋅⊆⋅Y$.  Thus the second-previous sentence implies [1] $(∀m≥1)$ $p^m(y)⋅∈⋅Y$, and further, the definition of $y$ in the previous paragraph implies [2] $y⋅∈⋅Y$.  These two observations  contradict \rf{Py} for~$Q$.

{\em (\rf{C578})}. This follows from part (\rf{D176}) and the pentaform definition (Definition~\rf{C669}). \end{npf}

\newcommand{\notetQk}{\footnote{To be clear, $(\tQ)_j$ is the situation-$j$ slice (definition (\rf{D304})) of the quintuple set $\tQ$.}}

\begin{lemma}\label{C366} Suppose $Q$ is a pentaform and $t⋅∈⋅T$ (where $T$ is the subroot set (\rf{D305})).  Then the following hold. \begin{tlist}
\yl{C457} $\tW = ⎨w∈W|t≼w⎬$.
\yl{C458} $\tY = ⎨y∈Y|t≺y⎬$.
\yl{C519} $(∀j∈\tJ)$ $(\tQ)_j = Q_j$.\notetQk\nichts{}
\yl{C987} $\tQ = ⨆⎨Q_j|j∈\tJ⎬$.
\end{tlist}\end{lemma} 

\begin{pf} {\em (\rf{C457})}.  In steps, $\tW$ by abbreviation (\rf{C825}) is $π_W(\tQ)$, which by definition (\rf{D316}) is $π_W(⎨˙⁅i˛j˛w˛a˛y⁆∈Q˙|$\hspace{0mm}$˙t≼w˙⎬)$, which by inspection is $⎨w∈π_W(Q)|t≼w⎬$, which by abbreviation (\rf{C825}) is $⎨w∈W|t≼w⎬$.  

{\em (\rf{C458})}. In steps, $\tY$ abbreviates $π_Y(\tQ)$, which by definition (\rf{D316}) is $π_Y(⎨˙⁅i˛j˛w˛a˛y⁆∈Q˙|$\hspace{0mm}$˙t≼w˙⎬)$, which by inspection is  \begin{gather}
\zz
⎨˙y∈Y˙|˙(∃w)˙t≼w, ⁅w˛y⁆∈\pj{WY}(Q)˙⎬. \notag 
\zz
\end{gather} Thus it suffices to show that this set's entrance requirement is equivalent to $t⋅≺⋅y$.  The entrance requirement implies $t⋅≼⋅w$ and $w⋅≺⋅y$, which implies $t⋅≺⋅y$.  Conversely, $t⋅≺⋅y$ implies that $t⋅≼⋅p(y)$ and $p(y)⋅≺⋅y$, which implies the entrance requirement at $w = p(y)$.

{\em (\rf{C519})}.  Take a situation $j⋅∈⋅\tJ$.  Simply, $\tQ⋅⊆⋅Q$ implies $(\tQ)_j⋅⊆⋅Q_j$.  For the reverse inclusion, suppose there was a quintuple in $Q_j$ that was not in $(\tQ)_j$.  Then the quintuple being in $Q_j$ implies there is $⁅i_*˛w_*˛a_*˛y_*⁆$ such that $⁅i_*,j,w_*˛a_*˛y_*⁆$ is in $Q_j$ but not $(\tQ)_j$.  Hence two applications of the slice definition (\rf{D304}) implies $⁅i_*,j,w_*˛a_*˛y_*⁆$ is in $Q$ but not $\tQ$, which implies $⁅i_*,j,w_*˛a_*˛y⁆⋅∈⋅Q⧷\tQ$, which implies $j⋅∈⋅π_J(Q⧷\tQ)$.  This and the assumption $j⋅∈⋅\tJ$ imply that $\tJ$ and $π_J(Q⧷\tQ)$ intersect.  This contradicts the lemma's assumption that $t⋅∈⋅T$, by $T$'s definition (\rf{D305}).

{\em (\rf{C987})}.  For the forward inclusion, take a quintuple $⁅i˛j˛w˛a˛y⁆⋅∈⋅\tQ$.  Then projection implies $j⋅∈⋅π_J(\tQ)$, which by abbreviation (\rf{C825}) implies $j⋅∈⋅\tJ$.  Further, $⁅i˛j˛w˛a˛y⁆⋅∈⋅Q$ and $Q_j$'s definition (\rf{D304}) imply $⁅i˛j˛w˛a˛y⁆⋅∈⋅Q_j$.  For the reverse inclusion, take a situation $j⋅∈⋅\tJ$.  Then part (\rf{C519}) implies $Q_j$ is equal to $(\tQ)_j$, which by slice definition~(\rf{D304}) is a subset of $\tQ$. \end{pf}

\begin{lemma}\label{C585} Suppose $Q$ is a pentaform and $w⋅∈⋅W$.  Then $w⋅∈⋅T$ iff there is $J′⋅⊆⋅J$ such that $\bQ{w} = ⨆⎨Q_j|j∈J′⎬$. {\normalfont (This lemma is cited in the text.)}\end{lemma} 

\begin{pf} For the forward direction, suppose $w⋅∈⋅T$ (that is, suppose $w$ is a subroot).  Then Lemma~\rf{C366}(\rf{C987}) implies $\bQ{w} = ⎨Q_j|j∈J′⎬$ for $J′ = \bJ{w}$.  

For the reverse direction, suppose $w⋅∉⋅T$.  Then the assumption $w⋅∈⋅W$ and $T$'s definition (\rf{D305}) imply there is a $j_*⋅∈⋅J$ which is listed both in a quintuple of $\bQ{w}$ and in a quintuple of $Q⧷\bQ{w}$.  Thus definition (\rf{D304}) implies that the slice $Q_{j_*}$ intersects both $\bQ{w}$ and $Q⧷\bQ{w}$.  Thus, since $⎨Q_j|j∈J⎬$ is a partition, it does not have a subcollection $⎨Q_j|j∈J′⎬$ whose union is $\bQ{w}$. \end{pf}

\begin{npf}[{{\bf for Proposition~\rf{C576}}}]\label{C576p} ⋅\begin{cllist}

\yl{C590} {\em $\tQ$ satisfies \rf{Pr}.} It suffices to show $\tW⧷\tY = ⎨t⎬$.  

For the forward inclusion, it suffices to show that $\tW⧷⎨t⎬⋅⊆⋅\tY$.  Toward that end, suppose $w⋅∈⋅\tW⧷⎨t⎬$, that is, suppose $w$ is a decision node in $\tW$ other than $t$ itself.  Then Lemma~\rf{C366}(\rf{C457}) implies $t⋅≼⋅w$ and $t⋅≠⋅w$, which by definition (\rf{E353}) imply \ilc{E431} $t⋅≺⋅w$.  Further, \rf{E431} with the order definitions (\rf{E352}) and (\rf{E353}) imply there is a nontrivial walk from $t$ to $w$, which by the walk definition (\rf{E426}) implies \il{E432} $w⋅∈⋅Y$.  Finally, Lemma~\rf{C366}(\rf{C458}) with \rf{E431} and \rf{E432} implies $w⋅∈⋅\tY$.  

For the reverse inclusion, first note that $T$'s definition (\rf{D305}) implies $t⋅∈⋅W$, which by Lemma~\rf{C366}(\rf{C457}) implies $t⋅∈⋅\tW$.  Thus it remains to show $t⋅∉⋅\tY$.  If $t⋅∈⋅\tY$ did hold, then Lemma~\rf{C366}(\rf{C458}) would imply $t⋅≺⋅t$, which is impossible by $≺$'s definition (\rf{E353}). \end{cllist}

{\em Conclusion}.  Lemma~\rf{C366}(\rf{C987}) (or the forward direction of Lemma~\rf{C585}) implies that there is $J′⋅⊆⋅J$ such that $\tQ = ⨆⎨Q_j|j∈J′⎬$.  Thus Corollary~\rf{C579}(\rf{D414}) and Claim~\rf{C590} imply that $\tQ$ is a pentaform. \end{npf}

\begin{lemma}\label{D427} Suppose $\QQ$ is a weakly separated (\rf{D418}) collection of quintuple sets which satisfy \rf{Pij}, \rf{Pjw}, \rf{Pwa}, \rf{Pway}, \rf{Pwy}, and \rf{Pay}.  Then $⨆\QQ$ satisfies the same six axioms. \end{lemma}

\begin{pf} This holds by the following claims. \begin{cllist}

\yl{C593} {\em $⨆\QQ$ satisfies \rf{Pij}.}  By inspection, \lic{D393} $\pj{JI}(⨆\QQ) = ⨆_{Q∈\QQ}˙\pj{JI}(Q)$.  Also, each $Q⋅∈⋅\QQ$ satisfies \rf{Pij} by assumption, which implies that \il{D392} $(∀Q∈\QQ)$ $π_{JI}(Q)$ is a function.  Also, weak separation (\rf{D418}) implies that \li{D391} the members $Q$ of $\QQ$ have distinct situations~$j$.  These three facts imply that $\pj{JI}(⨆\QQ)$ is the union of a set of functions with disjoint domains, which implies that $\pj{JI}(⨆\QQ)$ is itself a function, which implies that $⨆\QQ$ satisfies \rf{Pij}. 

\yl{D346} {\em $⨆\QQ$ satisfies \rf{Pjw}, \rf{Pway}, \rf{Pwy}, and \rf{Pay}.}  The arguments for these axioms mimic Claim~\rf{C593}'s argument for \rf{Pij}.  More precisely, the arguments for \rf{Pjw} and \rf{Pway} rely on the members of $\QQ$ having distinct decision nodes $w$, and the arguments for \rf{Pay} and \rf{Pwy} rely on the members of $\QQ$ having distinct successor nodes~$y$.

\yl{C591} {\em $⨆\QQ$ satisfies \rf{Pwa}.}  It suffices to show that\begin{gather}
\zz
(∀j∈π_J(⨆\QQ))⋅\pj{WA}((⨆\QQ)_j) = π_W((⨆\QQ)_j)×π_A((⨆\QQ)_j). \notag
\zz
\end{gather} Toward that end, take a situation $j^*⋅∈⋅π_J(⨆\QQ)$.  Weak separation (\rf{D418}) implies the members of $\QQ$ have distinct situations, which implies there is a unique member $Q^*⋅∈⋅\QQ$ such that $j^*⋅∈⋅π_J(Q^*)$.  Hence the union's slice $(⨆\QQ)_{j^*}$ equals the member's slice $Q^*_{j^*}$.  Thus it suffices to show that\begin{gather}
\zz
\pj{WA}(Q^*_{j^*}) = π_W(Q^*_{j^*})×π_A(Q^*_{j^*}). \notag
\zz
\end{gather} This holds because $Q^*$ satisfies \rf{Pwa} by assumption. \end{cllist} \unskipcl \end{pf}

\setlength{\emergencystretch}{2em}

\begin{npf}[{{\bf for Proposition~\rf{D415}(\rf{D416})}}]\label{D416p} \mbox{The result follows from Claims \rf{D428}--\rf{D431}.} \begin{cllist} 

\newcommand{\notemidsplit}{\footnote{For intuition, the identity $W = (W⧷Y)⋃(W⋂Y)$ splits any decision-node set $W$ into the start-node set $W⧷Y$ and the ``middle''-node set $W⋂Y$.  Similarly, the identity $Y = (Y⧷W)⋃(Y⋂W)$ splits any successor-node set $Y$ into the end-node set $Y⧷W$ and the ``middle''-node set $Y⋂W$.  The proof of Claim~\rf{D430} uses the first identity to split $W^A$ and the second identity to split $Y^B$; the proof of Claim~\rf{D429} uses the second identity to split $Y^A$; and the proof of Claim~\rf{D431} uses the first identity to split $W^B$.}}

\yl{D430} {\em $W^A$ and $Y^B$ are disjoint.}  Since $W^A = (W^A⧷Y^A)⋃(W^A⋂Y^A)$,{\notemidsplit} it suffices to show both (a) $(W^A⧷Y^A)⋂Y^B = ∅$ and (b) $(W^A⋂Y^A)⋂Y^B = ∅$.  Note (b) holds because weak separation implies $Y^A⋂Y^B = ∅$.  Now consider (a).  Since $Y^B = (Y^B⧷W^B)⋃(Y^B⋂W^B)$, it suffices to show both \begin{gather}
\zz
(W^A⧷Y^A)⋂(Y^B⧷W^B) = ∅⋅\text{and}⋅(W^A⧷Y^A)⋂(Y^B⋂W^B) = ∅.  \notag
\zz
\end{gather} The former is assumed by part (\rf{D416}).  The latter holds because weak separation implies $W^A⋂W^B = ∅$.

\yl{D433} {\em $Q^A⋃Q^B$ satisfies \rf{Pij}, \rf{Pjw}, \rf{Pwa}, \rf{Pway}, \rf{Pwy}, and \rf{Pay}.}  By the block definition (\rf{D464}), the blocks $Q^A$ and $Q^B$ satisfy these six axioms.  Thus weak separability and Lemma~\rf{D427} imply that their union does.

\yl{D432} {\em $Q^A⋃Q^B$ satisfies \rf{Py}.}  For notational ease, define $\dpp = \pj{YW}(Q^A⋃Q^B)$.  To be clear, $\dpp$ is a function since $Q^A⋃Q^B$ satisfies \rf{Pwy} by Claim~\rf{D433}.  By inspection, it is a superset (equivalently an extension) of both $p^A = \pj{YW}(Q^A)$ and $p^B = \pj{YW}(Q^B)$ (these equalities are two instances of the definition of the immediate-predecessor function in axiom \rf{Pwy}).  Further, Lemma~\rf{D478}(\rf{E531}) applied to the block $Q^A$ implies $(∀y∈Y^A)(∃ℓ^A≥1)$ $(p^A)^{ℓ^A}(y)⋅∈⋅W^A⧷Y^A$, which by Claim~\rf{D430} implies \begin{gather}
\zz
(∀y∈Y^A)(∃ℓ^A≥1)⋅(p^A)^{ℓ^A}(y)⋅∈⋅W^A⧷(Y^A⋃Y^B). \label{D434}
\zz
\end{gather} 

To show that $Q^A⋃Q^B$ satisfies \rf{Py}, take an arbitrary successor node $y⋅∈⋅Y^A⋃Y^B$.  It must be shown that\begin{gather}
\zz
(∃ℓ≥1)⋅\dpp^ℓ(y)⋅∉⋅Y^A⋃Y^B. \label{D435} 
\zz
\end{gather} Obviously $y⋅∈⋅Y^A$ or $y⋅∈⋅Y^B$ (intuitively, $Q^A$ ``precedes'' $Q^B$, so backward walks from nodes in $Y^A$ will tend to be shorter than backward walks from nodes in $Y^B$).  First suppose $y⋅∈⋅Y^A$.  Then (\rf{D434}) implies there is $ℓ^A⋅≥⋅1$ such that $(p^A)^{ℓ^A}(y)⋅∉⋅Y^A⋃Y^B$, which by $p^A⋅⊆⋅\dpp$ implies (\rf{D435}).  Second suppose $y⋅∈⋅Y^B$.  Then \rf{Py} for $Q^B$ implies there is $ℓ^B⋅≥⋅1$ such that $(p^B)^{ℓ^B}(y)⋅∉⋅Y^B$.  If $(p^B)^{ℓ^B}(y)⋅∉⋅Y^A$, the previous sentence implies $(p^B)^{ℓ^B}(y)⋅∉⋅Y^A⋃Y^B$, which by $p^B⋅⊆⋅\dpp$ implies (\rf{D435}).  Otherwise $(p^B)^{ℓ^B}(y)⋅∈⋅Y^A$, which by (\rf{D434}) implies there is $ℓ^A⋅≥⋅1$ such that $(p^A)^{ℓ^A}((p^B)^{ℓ^B}(y))⋅∉⋅Y^A⋃Y^B$, which by $p^A⋅⊆⋅\dpp$ and $p^B⋅⊆⋅\dpp$ implies $\dpp^{ℓ^A+ℓ^B}(y)⋅∉⋅Y^A⋃Y^B$, which implies (\rf{D435}).

\yl{D428} {\em $Q^A⋃Q^B$ is a block.} This follows from Claims \rf{D433} and \rf{D432} and the block definition (\rf{D464}).

\yl{D429} {\em $Q^A⋃Q^B$'s start-node set is the union of}\begin{gather}
\zz
W^A⧷Y^A⋅\text{and}⋅(W^B⧷Y^B)⧷(Y^A⧷W^A). \notag
\zz
\end{gather} By inspection, $π_W(Q^A⋃Q^B) = W^A⋃W^B$ and $π_Y(Q^A⋃Q^B) = Y^A⋃Y^B$.  So $Q^A⋃Q^B$'s start-node set is $(W^A⋃W^B)⧷(Y^A⋃Y^B)$, which by inspection is the union of \begin{gather}
\zz
W^A⧷(Y^A⋃Y^B)⋅\text{and}⋅W^B⧷(Y^A⋃Y^B). \notag
\zz
\end{gather} The first set is equal to $W^A⧷Y^A$ by Claim~\rf{D430}.  The second set is equal to $(W^B⧷Y^B)⧷Y^A$, which is equal to $(W^B⧷Y^B)⧷[(Y^A⧷W^A)⋃(Y^A⋂W^A)]$, which by $W^B⋂W^A = ∅$ (from weak separation) is equal to $(W^B⧷Y^B)⧷(Y^A⧷W^A)$.

\yl{D431} {\em $Q^A⋃Q^B$'s end-node set is the union of} \begin{gather}
\zz
(Y^A⧷W^A)⧷(W^B⧷Y^B)⋅\text{\em and}⋅Y^B⧷W^B. \notag
\zz
\end{gather} By inspection, $π_Y(Q^A⋃Q^B) = Y^A⋃Y^B$ and $π_W(Q^A⋃Q^B) = W^A⋃W^B$.  Thus $Q^A⋃Q^B$'s end-node set is $(Y^A⋃Y^B)⧷(W^A⋃W^B)$, which by inspection is the union of \begin{gather}
\zz
Y^A⧷(W^A⋃W^B)⋅\text{and}⋅Y^B⧷(W^A⋃W^B). \notag
\zz
\end{gather} The second set is equal to $Y^B⧷W^B$ by Claim~\rf{D430}.  The first set is equal to $(Y^A⧷W^A)⧷W^B$, which is equal to $(Y^A⧷W^A)⧷[(W^B⧷Y^B)⋃(W^B⋂Y^B)]$, which by $Y^A⋂Y^B = ∅$ (from weak separation) is equal to $(Y^A⧷W^A)⧷(W^B⧷Y^B)$.\nichts{} \end{cllist} \unskipcl \end{npf}

\pagebreak

\begin{npf}[{{\bf for Proposition~\rf{D415}(\rf{D417})}}]\label{D417p} This follows from Claims \rf{D444}--\rf{D443}. \begin{cllist}

\yl{D437} {\em $⨆\QQ$ satisfies \rf{Pij}, \rf{Pjw}, \rf{Pwa}, \rf{Pway}, \rf{Pwy}, and \rf{Pay}.}  Since strong separation (\rf{D419}) implies weak separation (\rf{D418}), $\QQ$ is weakly separated.  Further, the block definition (\rf{D464}) implies that every block in $\QQ$ satisfies the six axioms.  Thus Lemma~\rf{D427} implies that $⨆\QQ$ satisfies the six axioms.

\yl{D436} {\em $⨆\QQ$ satisfies \rf{Py}.}  For notational ease, let $\dpp = \pj{YW}(⨆\QQ)$.  To be clear, $\dpp$ is a function since $⨆\QQ$ satisfies \rf{Pwy} by Claim~\rf{D437}.  Thus it suffices to show \begin{gather}
\zz
(∀y∈π_Y(⨆\QQ))(∃ℓ≥1)⋅\dpp^ℓ(y)⋅∉⋅π_Y(⨆\QQ).\notag
\zz
\end{gather} Toward that end, take an arbitrary successor node $y⋅∈⋅π_Y(⨆\QQ)$.  Since $π_Y(⨆\QQ) = ⨆_{Q∈\QQ}π_Y(Q)$ by inspection, there is a block $Q^*⋅∈⋅\QQ$ such that $y⋅∈⋅Y^*$.  In accord with axiom \rf{Pwy}, define $p^* = \pj{YW}(Q^*)$, which by $Q^*⋅∈⋅\QQ$ implies $p^*⋅⊆⋅\pj{YW}(⨆\QQ)$, which by $\dpp$'s definition above implies $p^*⋅⊆⋅\dpp$ (equivalently $p^*$ is a restriction of $\dpp$).  Further, Lemma~\rf{D478}(\rf{E531}) applied to the block $Q^*$ implies there is $ℓ⋅≥⋅1$ such that $(p^*)^ℓ(y)⋅∈⋅W^*⧷Y^*$, which by $p^*⋅⊆⋅\dpp$ implies \begin{gather}
\zz
\dpp^ℓ(y)⋅∈⋅W^*⧷Y^*. \label{D438} 
\zz
\end{gather} Since (\rf{D438}) implies $\dpp^ℓ(y)⋅∈⋅W^*$, this $\dpp^ℓ(y)$ is a node of $Q^*$, which by strong separation implies that $\dpp^ℓ(y)$ is not a node of $⨆(\QQ⧷⎨Q^*⎬)$, which implies that $\dpp^ℓ(y)$ is not a successor node of $⨆(\QQ⧷⎨Q^*⎬)$, which by $\dpp^ℓ(y)⋅∉⋅Y^*$ from (\rf{D438}) implies that $\dpp^ℓ(y)$ is not a successor node of $⨆\QQ$, which is equivalent to $\dpp^ℓ(y)⋅∉⋅π_Y(⨆\QQ)$.

\yl{D444} {\em $⨆\QQ$ is a block.}  This holds by Claims \rf{D437} and \rf{D436} and the block definition~(\rf{D464}). 

\yl{D439} {\em $⨆\QQ$'s start-node set is $⨆_{Q∈\QQ}(π_W(Q)⧷π_Y(Q))$.}  Since $⨆\QQ$'s start-node set is $π_W(⨆\QQ)⧷π_Y(⨆\QQ)$, it suffices to show\begin{gather}
\zz
π_W(⨆\QQ)⧷π_Y(⨆\QQ) = ⨆_{Q∈\QQ}(π_W(Q)⧷π_Y(Q)). \notag
\zz
\end{gather} For the forward inclusion, consider \ilc{D441} $w⋅∈⋅π_W(⨆\QQ)⧷π_Y(⨆\QQ)$.  Note \rf{D441} implies $w⋅∈⋅π_W(⨆\QQ)$, which implies there is a block $Q^*⋅∈⋅\QQ$ such that $w⋅∈⋅π_W(Q^*)$.  Further, \rf{D441} implies $w⋅∉⋅π_Y(⨆\QQ)$, which by $Q^*⋅∈⋅\QQ$ implies $w⋅∉⋅π_Y(Q^*)$.  The previous two sentences imply $w⋅∈⋅π_W(Q^*)⧷π_Y(Q^*)$, which by $Q^*⋅∈⋅\QQ$ implies $w⋅∈⋅⨆_{Q∈\QQ}(π_W(Q)⧷π_Y(Q))$.  

Conversely, for the reverse inclusion, suppose there is an individual block $Q′⋅∈⋅\QQ$ with start node \il{D440} $w⋅∈⋅π_W(Q′)⧷π_Y(Q′)$.  Then \rf{D440} implies $w⋅∈⋅π_W(Q′)$, which by $Q′⋅∈⋅\QQ$ implies \il{D442} $w⋅∈⋅π_W(⨆\QQ)$.  Also, \rf{D440} implies $w⋅∈⋅π_W(Q′)$, which (in a different direction) implies that $w$ is a node of $Q′$, which by strong separation implies that $w$ is not a node of $⨆(\QQ⧷⎨Q′⎬)$, which implies that $w$ is not a successor node of $⨆(\QQ⧷⎨Q′⎬)$, which is equivalent to $w⋅∉⋅π_Y(⨆\QQ⧷⎨Q′⎬)$, which by $w⋅∉⋅π_Y(Q′)$ from \rf{D440} implies $w⋅∉⋅π_Y(⨆\QQ)$.  This and \rf{D442} imply $w⋅∈⋅π_W(⨆\QQ)⧷π_Y(⨆\QQ)$.

\yl{D443} {\em $⨆\QQ$'s end-node set is $⨆_{Q∈\QQ}(π_Y(Q)⧷π_W(Q))$.} This can be proved like Claim~\rf{D439} was proved.  Replace ``start'' with ``end'', switch $W$ and $Y$, replace $w$ with $y$, and replace ``successor node'' with ``decision node''.
\nichts{}
\end{cllist} \unskipcl \end{npf}

\pagebreak

\begin{npf}[{{\bf for Proposition~\rf{D454}}}]\label{D454p} For notational ease, define\begin{gather}
\zz
\dQ = ⨆_{n≥0}Q^n. \label{D471}
\zz
\end{gather}\unskipcl  \begin{cllist} 

\yl{D396} {\em $\dQ$ satisfies \rf{Pij}}.  Suppose $\dQ$ violates \rf{Pij}.  Then there exists a situation $j⋅∈⋅\dJ$ and players $i_1⋅∈⋅\dI$ and $i_2⋅∈⋅\dI$ such that $i_1⋅≠⋅i_2$ and both $⁅j˛i_1⁆$ and $⁅j˛i_2⁆$ are in $\pj{JI}(\dQ)$.  Note $⁅j˛i_1⁆$ being in $\pj{JI}(\dQ)$ implies there is $⁅w_1˛a_1˛y_1⁆$ such that $⁅i_1˛j˛w_1˛a_1˛y_1⁆⋅∈⋅\dQ$, which by $\dQ$'s definition (\rf{D471}) implies there is $n_1⋅≥⋅0$ such that\begin{gather}
\zz
⁅i_1˛j˛w_1˛a_1˛y_1⁆⋅∈⋅Q^{n_1}. \notag
\zz
\end{gather} Similarly, $⁅j˛i_2⁆$ being in $\pj{JI}(\dQ)$ implies there is $⁅w_2˛a_2˛y_2⁆$ such that $⁅i_2˛j˛w_2˛a_2˛y_2⁆$ is in $\dQ$, which by $\dQ$'s definition (\rf{D471}) implies there is $n_2⋅≥⋅0$ such that\begin{gather}
\zz
⁅i_2˛j˛w_2˛a_2˛y_2⁆⋅∈⋅Q^{n_2}. \notag
\zz
\end{gather} Let $n_* = \max⎨n_1,n_2⎬$.  Then the assumption $(∀n≥1)⋅Q^{n-1}⋅⊆⋅Q^n$ implies that both $⁅i_1˛j˛w_1˛a_1˛y_1⁆$ and $⁅i_2˛j˛w_2˛a_2˛y_2⁆$ are in $Q^{n_*}$, which implies that both $⁅j˛i_1⁆$ and $⁅j˛i_2⁆$ are in $\pj{JI}(Q^{n_*})$, which by $i_1⋅≠⋅i_2$ implies $Q^{n_*}$ violates \rf{Pij}, which violates the assumption that $Q^{n_*}$ is a pentaform.

\yl{D397} {\em $\dQ$ satisfies \rf{Pjw}, \rf{Pway}, \rf{Pwy}, and \rf{Pay}.} Each of these four axioms is derived as Claim~\rf{D396} derived \rf{Pij}. 

\yl{D401} {\em $\dQ$ satisfies \rf{Pwa}.}  Suppose $\dQ$ violates \rf{Pwa}.  Then there exists a situation $j⋅∈⋅\dJ$ with a decision node $w_1⋅∈⋅\dW_j$, another decision node $w_2⋅∈⋅\dW_j$, and an action~$a$, such that \ilc{D402} $⁅w_1˛a⁆⋅∈⋅\pj{WA}(\dQ_j)$ and  \il{D403} $⁅w_2˛a⁆⋅∉⋅\pj{WA}(\dQ_j)$.  We will find a contradiction.  (Looking ahead, \rf{D402} will be used immediately and \rf{D403} will be used at the end.)

Note that \rf{D402} and Lemma~\rf{D333}(\rf{D445}) imply $⁅j˛w_1˛a⁆⋅∈⋅\pj{JWA}(\dQ)$, which implies there is $⁅i_1˛y_1⁆$ such that $⁅i_1˛j˛w_1˛a˛y_1⁆⋅∈⋅\dQ$, which by $\dQ$'s definition (\rf{D471}) implies there is $n_1⋅≥⋅0$ such that \begin{gather}
\zz
⁅i_1˛j˛w_1˛a˛y_1⁆⋅∈⋅Q^{n_1}. \notag
\zz
\end{gather} Meanwhile, $w_2⋅∈⋅\dW_j$ and Lemma~\rf{D333}(\rf{D366}) imply $⁅j˛w_2⁆⋅∈⋅\pj{JW}(\dQ)$, which implies there is $⁅i_2˛a_2˛y_2⁆⋅∈⋅\dQ$ such that $⁅i_2˛j˛w_2˛a_2˛y_2⁆⋅∈⋅\dQ$, which by $\dQ$'s definition (\rf{D471}) implies there is $n_2⋅≥⋅0$ such that\begin{gather}
\zz
⁅i_2˛j˛w_2˛a_2˛y_2⁆⋅∈⋅Q^{n_2}. \notag
\zz
\end{gather} Let $n_* = \max⎨n_1,n_2⎬$.  Then the assumption $(∀n≥1)⋅Q^{n-1}⋅⊆⋅Q^n$ implies that both $⁅i_1˛j˛w_1˛a˛y_1⁆$ and $⁅i_2˛j˛w_2˛a_2˛y_2⁆$ are in $Q^{n_*}$.  The first implies $⁅j˛w_1˛a⁆⋅∈⋅\pj{JWA}(Q^{n_*})$, which by Lemma~\rf{D333}(\rf{D445}) implies 
\il{D411} $⁅w_1˛a⁆⋅∈⋅\pj{WA}(Q^{n_*}_j)$.  The second implies $⁅j˛w_2⁆⋅∈⋅\pj{JW}(Q^{n_*})$, which by Lemma~\rf{D333}(\rf{D366}) implies $w_2⋅∈⋅W^{n_*}_j$, which by abbreviation (\rf{C825}) implies $w_2⋅∈⋅π_W(Q^{n_*}_j)$, which by inspection implies \il{D412} $w_2⋅∈⋅π_W(\pj{WA}(Q^{n_*}))$.

By assumption, $Q^{n_*}$ is a pentaform, which implies $Q^{n_*}$ satisfies \rf{Pwa}, which implies that $\pj{WA}(Q^{n_*}_j)$ is a rectangle, which by \rf{D411} and \rf{D412} implies $⁅w_2˛a⁆⋅∈⋅\pj{WA}(Q^{n_*}_j)$, which by Lemma~\rf{D333}(\rf{D445}) implies $⁅j˛w_2˛a⁆⋅∈⋅\pj{JWA}(Q^{n_*})$, which implies there is $⁅i\lo˛y\lo⁆$ such that $⁅i\lo˛j˛w_2˛a˛y\lo⁆⋅∈⋅Q^{n_*}$, which by $\dQ$'s definition (\rf{D471}) implies $⁅i\lo˛j˛w_2˛a˛y\lo⁆⋅∈⋅\dQ$, which implies $⁅j˛w_2˛a⁆⋅∈⋅\pj{JWA}(\dQ)$, which by Lemma~\rf{D333}(\rf{D445}) implies $⁅w_2˛a⁆⋅∈⋅\pj{WA}(\dQ_j)$, which contradicts \rf{D403}.

\yl{D406} {\em (a) $\dW = ⨆_{n≥0}W^n$. (b) $\dY = ⨆_{n≥0}Y^n$.}  First consider (a).  In steps, $\dW$ by abbreviation (\rf{C825}) is $π_W(\dQ)$, which by $\dQ$'s definition (\rf{D471}) is $π_W(⨆_{n≥0}Q^n)$, which by inspection equals $⨆_{n≥0}π_W(Q^n)$, which by abbreviation (\rf{C825}) is $⨆_{n≥0}W^n$.  A similar argument holds for (b). 

\yl{D405} {\em $⎨r^0⎬ = \dW⧷\dY$.} For the forward inclusion $⎨r^0⎬⋅⊆⋅\dW⧷\dY$, it suffices to show $r^0⋅∈⋅\dW$ and $r^0⋅∉⋅\dY$.  First, $r^0$'s definition (\rf{C982}) implies $r^0⋅∈⋅W^0$, which by Claim~\rf{D406}(a) implies $r^0⋅∈⋅\dW$.  Second, to show $r^0⋅∉⋅\dY$, suppose $r^0⋅∈⋅\dY$.  Then Claim~\rf{D406}(b) implies there is $n⋅≥⋅0$ such that $r^0⋅∈⋅Y^n$, which implies $⎨r^0⎬⋅≠⋅W^n⧷Y^n$, which by $r^n$'s definition (\rf{C982}) implies $r^0⋅≠⋅r^n$, which contradicts the proposition's assumption that $r^n = r^0$.  

For the reverse inclusion $\dW⧷\dY⋅⊆⋅⎨r^0⎬$, it suffices to show that $(∀w∈\dW⧷⎨r^0⎬)$ $w⋅∈⋅\dY$.  In other words, it suffices to show that every decision node other than $r^0$ is a successor node.  Toward that end, take such a decision node $w⋅∈⋅\dW⧷⎨r^0⎬$.  Then Claim~\rf{D406}(a) implies there is $n⋅≥⋅0$ such that \ilc{D447} $w⋅∈⋅W^n⧷⎨r^0⎬$.  Note $r^n$'s definition (\rf{C982}) implies $W^n⧷Y^n = ⎨r^n⎬$,\nichts{} which implies \il{E455} $W^n⧷⎨r^n⎬⋅⊆⋅Y^n$.  Then in steps, \rf{D447} implies $w⋅∈⋅W^n⧷⎨r^0⎬$, which by the assumption $r^n = r^0$ implies $w⋅∈⋅W^n⧷⎨r^n⎬$, which by \rf{E455} implies $w⋅∈⋅Y^n$, which by Claim~\rf{D406}(b) implies $w⋅∈⋅\dY$.

\yl{D404} {\em $\dQ$ satisfies \rf{Py}.}  For any $n$, the assumption that $Q^n$ is a pentaform implies that $Q^n$ satisfies \rf{Pwy}, which implies that the $p^n = \pj{YW}(Q)$ defined in \rf{Pwy} is a function.  Meanwhile, Claim~\rf{D397} implies that $\dQ$ satisfies \rf{Pwy}, which implies that the $\dpp = \pj{YW}(\dQ)$ defined in \rf{Pwy} is a function.  Further, for each $n$, $\dQ$'s definition (\rf{D471}) implies $Q^n⋅⊆⋅\dQ$, which implies $\pj{YW}(Q^n)⋅⊆⋅\pj{YW}(\dQ)$, which by the preceding definitions implies \ilc{E433} $p^n⋅⊆⋅\dpp$ (in other words, $p^n$ is a restriction of $\dpp$).

For \rf{Py}, it suffices to show that $(∀y∈\dY)(∃ℓ≥1)$ $\dpp^ℓ(y)⋅∉⋅\dY$.  Toward that end, take a successor node $y⋅∈⋅\dY$.  By Claim~\rf{D406}(b), there is $n⋅≥⋅0$ such that $y⋅∈⋅\dY^n$.  Since $Q^n$ is a pentaform by assumption, Lemma~\rf{D478}(\rf{E532}) implies there is $ℓ⋅≥⋅0$ such that $(p^n)^ℓ(y) = r^n$, which by the assumption $r^n = r^0$ implies $(p^n)^ℓ(y) = r^0$, which by \rf{E433} implies $\dpp^ℓ(y) = r^0$, which by Claim~\rf{D405} implies $\dpp^ℓ(y)⋅∉⋅\dY$. \end{cllist}

{\em Conclusion}. Claim~\rf{D405} implies that $\dQ$ satisfies \rf{Pr}.  Claims \rf{D396}--\rf{D401} and \rf{D404} show that $\dQ$ satisfies the other seven axioms.  Thus $\dQ$ is a pentaform.  Finally, definition (\rf{C982}) states that $\dQ$'s root is the sole member of $\dW⧷\dY$, which by Claim~\rf{D405} is $r^0$. \end{npf}

\renewcommand{\sectitle}{For Equivalence (Section~\rf{C826})}
\section{\sectitle}\label{B567}\showit
\markb{\sc Appendix \rf{B567}. \sectitle}

\begin{lemma}[{{\bf implies Theorem~\rf{C689}}}]\label{B405} Suppose $(X,E,\HH,α,ι,u)$ is a traditional game.  Let $(\hp{Q},\hp{u}) = \PB(X,E,\HH,α,ι,u)$.  Then \ttr{a}{B406}~$(\hp{Q},\hp{u})$ is a pentaform game with information-set situations (\rf{D319}).  Further,
\ttr{b}{B4h}~$\hp{J} = \HH$,
\ttr{c}{B4x}~$\hp{W}⋃\hp{Y} = X$,
\ttr{d}{B4e}~$\pj{WY}(\hp{Q}) = E$,
\ttr{e}{B4L}~$⎨⁅⁅w˛y⁆˛a⁆|⁅w˛y˛a⁆∈\pj{WYA}(\hp{Q})⎬ = α$, and
\ttr{f}{B4M}~$\pj{WI}(\hp{Q}) = ι$.
\end{lemma}\nichts{}

\begin{pfb} Part (\rf{B406}) follows from Claim~\rf{E456}; part~(\rf{B4h}) from Claim~\rf{B420}; parts (\rf{B4x})--(\rf{B4e}) from Claims~\rf{C731}--\rf{C730}; and parts~(\rf{B4L})--(\rf{B4M}) from Claims \rf{B4Lp}--\rf{B4Mp}.\nichts{} \begin{cllist}

\yl{B424} {\em $E⋅∋⋅⁅w˛y⁆ \mapsto ι(w)$ is a surjection to~$I$.} This is the composition of two surjections.  First, $E⋅∋⋅⁅w˛y⁆ \mapsto w$ is a surjection to $π_1E$ by inspection.  Second, $π_1E⋅∋⋅w \mapsto ι(w)$ is a surjection to $I$ because of \rf{T5} (Definition~\rf{C843}) and because $I$ is the range of $ι$ by definition (\rf{E377}).

\yl{B428} {\em $E⋅∋⋅⁅w˛y⁆ \mapsto H_w$ is a surjection to $\HH$.} This is the composition of two surjections.  First, $E⋅∋⋅⁅w˛y⁆ \mapsto w$ is a surjection to $π_1E$ by inspection.  Second, $π_1E⋅∋⋅w \mapsto H_w$ is a surjection to $\HH$ by \rf{T2} (Definition~\rf{C843}) and by the definition of $⁅H_w⁆_{w∈π_1E}$ (after definition (\rf{C678})). 

\yl{B427} {\em $E⋅∋⋅⁅w˛y⁆ \mapsto α(w˛y)$ is a surjection to~$A$.} This holds because of \rf{T3} and because $A$ is the range of $α$ by definition (\rf{E378}).

\yl{B418} {\em $\hp{Q}$ is well-defined.} This follows from definition (\rf{C679}) and Claims \rf{B424}--\rf{B427}. 

\yl{B419} $\hp{I} = I$.  By abbreviation (\rf{C825}), $\hp{I}$ is the projection $π_I(\hp{Q})$, which by definition (\rf{C679}) and the surjection of Claim~\rf{B424} is equal to~$I$.  

\yl{B420} $\hp{J} = \HH$.  By abbreviation (\rf{C825}), $\hp{J}$ is the projection $π_J(\hp{Q})$, which by definition (\rf{C679}) and the surjection of Claim~\rf{B428} is equal to $\HH$.  

\yl{B421} $\hp{W} = π_1E$.  By abbreviation (\rf{C825}), $\hp{W}$ is the projection $π_W(\hp{Q})$, which by definition (\rf{C679}) is equal to $π_1E$.

\yl{B423} $\hp{Y} = π_2E$.  By abbreviation (\rf{C825}), $\hp{W}$ is the projection $π_Y(\hp{Q})$, which by definition (\rf{C679}) is equal to $π_2E$.

\yl{B407} {\em $\hp{Q}$ satisfies \rf{Pij}.}  Note that $\pj{JI}(\hp{Q}) = ⎨˙⁅H_w˛ι(w)⁆˙|˙⁅w˛y⁆∈E˙⎬$ by definition (\rf{C679}).  Thus $\pj{JI}(\hp{Q}) = ⎨˙⁅H_w˛ι(w)⁆˙|˙w∈π_1E˙⎬$ by inspection.  To show that this is a function, it suffices to show that $(∀w_1∈π_1E,w_2∈π_1E)$ the condition $H_{w_1} = H_{w_2}$ implies $ι(w_1) = ι(w_2)$.  Toward that end, suppose $w_1⋅∈⋅π_1E$ and $w_2⋅∈⋅π_1E$ are two decision nodes such that $H_{w_1} = H_{w_2}$.  Then the definition of $⁅H_w⁆_{w∈π_1E}$ (after definition (\rf{C678})) implies that $w_1$ and $w_2$ belong to the same cell of the partition $\HH$, which by the measurability (\rf{D351}) of \rf{T5} implies $ι(w_1) = ι(w_2)$.

\yl{B408} {\em $\hp{Q}$ satisfies \rf{Pjw}.}  Note that $\pj{WJ}(\hp{Q}) = ⎨˙⁅w˛H_w⁆˙|˙⁅w˛y⁆∈E˙⎬$ by definition (\rf{C679}).  Thus $\pj{WJ}(\hp{Q}) = ⎨˙⁅w˛H_w⁆˙|˙w∈π_1E˙⎬$ by inspection.  This is the function $⁅H_w⁆_{w∈π_1E}$ (defined after definition (\rf{C678})). 

\yl{C725} {\em $\hp{F} = F$.} The correspondence $F{:}π_1E⇉A$ is defined in equation (\rf{D169}) by $F(w) = ⎨˙a˙|˙(∃y)˙α(w˛y){=}a˙⎬$.  Since a correspondence is a set of couples (footnote \rf{B359} on \pgrf{B359}), this implies that \begin{gather}
\zz
F =  ⎨˙⁅w˛a⁆˙|˙w∈π_1E,˙(∃y)α(w˛y){=}a˙⎬.\notag
\zz
\end{gather} Note \rf{T3} implies that the domain of $α$ is~$E$.  Hence $F$ is $⎨˙⁅w˛α(w˛y)⁆˙|˙⁅w˛y⁆∈E˙⎬$, which by definition (\rf{C679}) is $\pj{WA}(\hp{Q})$, which by definition (\rf{C981}) is $\hp{F}$.

\yl{C728} {\em $(∀H∈\HH)⋅⎨˙w∈π_1E˙|˙H_w{=}H˙⎬ = H$. (For intuition, regard this as an identity which holds in every traditional game.)}  To prove this equality, note that $\HH$ partitions $π_1E$ (by \rf{T2}), and that (for each $w⋅∈⋅π_1E$) $H_w$ is the cell that contains~$w$ (by the definition of $H_w$ after (\rf{C678})).  Now consider a cell $H⋅∈⋅\HH$.  Then $⎨˙w∈π_1E˙|˙H_w{=}H˙⎬$ consists of the $w$ in the cell $H$, which is $H$ itself.

\yl{C727} {\em $\hp{Q}$ has information-set situations (\rf{D319}), that is, $(∀j∈\hp{J})$ $\hp{W}_j = j$.}  Take a situation $j⋅∈⋅\hp{J}$.  Definition~(\rf{C679}) implies that the slice $\hp{Q}_j$ satisfies\begin{gather}
\zz
\hp{Q}_j = ⎨⋅⁅ι(w)˛H_w˛w˛α(w˛y)˛y⁆⋅|⋅⁅w˛y⁆∈E,˙H_w{=}j⋅⎬, \notag
\zz
\end{gather} which by projection implies $π_W(\hp{Q}_j) = ⎨˙w˙|˙(∃y)⁅w˛y⁆∈E,˙H_w{=}j˙⎬$, which by abbreviation (\rf{C825}) implies $\hp{W}_j = ⎨˙w˙|˙(∃y)⁅w˛y⁆∈E,˙H_w{=}j˙⎬$, which by inspection implies $\hp{W}_j = ⎨˙w˙|˙w∈π_1E,˙H_w{=}j˙⎬$, which by manipulation implies  \begin{gather}
\zz
\hp{W}_j =  ⎨˙w∈π_1E˙|˙H_w{=}j˙⎬. \label{E434}
\zz
\end{gather} Meanwhile, notice that $j$ is a cell in the partition $\HH$ because $j⋅∈⋅\hp{J}$ by assumption and because $\hp{J} = \HH$ by Claim~\rf{B420}.  Thus Claim~\rf{C728} with $j$ replacing $H$ implies $⎨˙w∈π_1E˙|˙H_w{=}j˙⎬ = j$.  This and (\rf{E434}) imply the result.

\yl{C726} {\em $\hp{Q}$ satisfies \rf{Pwa}.}  By Proposition~\rf{C601}(\rf{C602}$⇐$\rf{C604}) and Claim~\rf{B408}, it suffices to show that\begin{gather}
\zz
(∀j∈\hp{J},w_1∈\hp{W}_j,w_2∈\hp{W}_j)⋅\hp{F}(w_1) = \hp{F}(w_2).\notag
\zz
\end{gather} By replacing $\hp{J}$ with $\HH$ via Claim~\rf{B420}, by replacing the two appearances of $\hp{W}_j$ with $j$ via Claim~\rf{C727}, and by replacing $\hp{F}$ with $F$ via Claim~\rf{C725}, this is equivalent to \begin{gather}
\zz
(∀j∈\HH,w_1∈j,w_2∈j)⋅F(w_1) = F(w_2). \notag
\zz
\end{gather} By a change of variables, this is equivalent to $(∀H∈\HH,w_1∈H,w_2∈H)$ $F(w_1) = F(w_2)$, which holds by the measurability (\rf{D350}) of \rf{T4}.  

\yl{C729} {\em $\hp{Q}$ satisfies \rf{Pway}.} Note $\pj{WAY}(\hp{Q}) = ⎨˙⁅w˛α(w˛y)˛y⁆˙|˙⁅w˛y⁆∈E˙⎬$ by Definition (\rf{C679}).  Thus it suffices to show, for all $⁅w_1˛y_1⁆⋅∈⋅E$ and $⁅w_2˛y_2⁆⋅∈⋅E$, that \begin{gather}
\zz
⁅w_1˛α(w_1˛y_1)⁆ = ⁅w_2˛α(w_2˛y_2)⁆⋅\text{implies}⋅y_1 = y_2. \notag
\zz
\end{gather}  Toward that end, take $⁅w_1˛y_1⁆⋅∈⋅E$ and $⁅w_2˛y_2⁆⋅∈⋅E$ such that $⁅w_1˛α(w_1˛y_1)⁆ = ⁅w_2˛α(w_2˛y_2)⁆$.  Since the equality's first coordinate implies $w_1 = w_2$, the rest of the previous sentence implies that we have two edges of the form $⁅w,y_1⁆$ and $⁅w,y_2⁆$ such that $α(w˛y_1) = α(w,y_2)$.  Thus the local injectivity (\rf{D347}) of \rf{T3} implies $y_1 = y_2$.

\yl{C731} {\em $\hp{W}⋃\hp{Y} = X$.}  In steps, $\hp{W}⋃\hp{Y}$ by Claims \rf{B421} and \rf{B423} is equal to $π_1E⋃π_2E$, which by Lemma~\rf{E320}(\rf{E347}) and \rf{T1} is equal to~$X$.

\yl{C730} {\em $\pj{WY}(\hp{Q}) = E$.}  In steps, $\pj{WY}(\hp{Q})$ by (\rf{C679}) is equal to $⎨˙⁅w˛y⁆˙|˙⁅w˛y⁆∈E˙⎬$, which by inspection is equal to~$E$.

\yl{C732} {\em $\hp{Q}$ satisfies \rf{Pwy}, \rf{Py}, and \rf{Pr}.}  Since \rf{T1} implies that $(X,E)$ is a nontrivial out-tree, Claims \rf{C731} and \rf{C730} imply that $(\hp{W}⋃\hp{Y},\pj{WY}(\hp{Q}))$ is a nontrivial out-tree.  Thus the reverse direction of Proposition~\rf{C632}(a) implies that $\hp{Q}$ satisfies \rf{Pwy}, \rf{Py}, and \rf{Pr}.

\yl{C733} {\em $\hp{Q}$ satisfies \rf{Pay}.}  Note $\pj{YA}(\hp{Q}) = ⎨˙⁅y˛α(w˛y)⁆˙|˙⁅w˛y⁆∈E˙⎬$ by definition (\rf{C679}).  Thus it suffices to show, for all $⁅w_1˛y_1⁆⋅∈⋅E$ and $⁅w_2˛y_2⁆⋅∈⋅E$, that \begin{gather}
\zz
y_1 = y_2⋅\text{implies}⋅α(w_1˛y_1) = α(w_2˛y_2). \notag
\zz
\end{gather} Toward that end, take $⁅w_1˛y_1⁆⋅∈⋅E$ and $⁅w_2˛y_2⁆⋅∈⋅E$ such that \lic{E417} $y_1 = y_2$.  Claim~\rf{C730} implies $⁅w_1˛y_1⁆⋅∈⋅\pj{WY}(\hp{Q})$ and $⁅w_2˛y_2⁆⋅∈⋅\pj{WY}(\hp{Q})$.  Further, Claim~\rf{C732} implies \rf{Pwy}, which implies $\pj{YW}(\hp{Q})$ is a function.  Thus \rf{E417} implies \li{E418} $w_1 = w_2$.  Together, \rf{E417} and \rf{E418} imply $⁅w_1˛y_1⁆ = ⁅w_2˛y_2⁆$, which implies $α(w_1˛y_1) = α(w_2,y_2)$. 

\yl{C734} {\em $\hp{Q}$ is a pentaform.} This follows from Claims \rf{B407}, \rf{B408}, \rf{C726}, \rf{C729}, \rf{C732}, and \rf{C733}.

\yl{C736} {\em $\hp{\ZZ} = \ZZ$.} Claims \rf{C731} and \rf{C730} imply that $(\hp{W}⋃\hp{Y},\pj{WY}(\hp{Q}))$ equals the out-tree $(X,E)$ from \rf{T1}.  This suffices since both pentaform games (end of Section~\rf{C657}) and traditional games (Section~\rf{C665} paragraph~\rf{D321}) derive their run collections from their out-trees (via equation (\rf{E355})).  

\yl{E456} {\em $(\hp{Q},\hp{u})$ is a pentaform game with information-set situations}.  Claims~\rf{C727} and \rf{C734} imply $\hp{Q}$ is a pentaform with information-set situations.  Thus by Definition~\rf{C668} it suffices to show that $\hp{u}$ is a utility-function profile for~$\hp{Q}$.  Definition (\rf{C680}) implies $\hp{u}$ is equal to $u$, which by \rf{T6} has the form $⁅u_i{:}\ZZ→\eR⁆_{i∈I}$, which is identical to the form $⁅u_i{:}\hp{\ZZ}→\eR⁆_{i∈\hp{I}}$ because $I = \hp{I}$ by Claim~\rf{B419} and because $\ZZ = \hp{\ZZ}$ by Claim~\rf{C736}.

\yl{B4Lp} {\em $⎨⁅⁅w˛y⁆˛a⁆|⁅w˛y˛a⁆∈\pj{WYA}(\hp{Q})⎬ = α$.} By definition (\rf{C679}), the left-hand side  is\begin{gather}
\zz
⎨⋅⁅⁅w˛y⁆˛a⁆⋅|⋅⁅w˛y˛a⁆⋅∈⋅⎨⁅w˛y˛α(w˛y)⁆|⁅w˛y⁆∈E⎬⋅⎬, \notag
\zz
\end{gather} which by inspection is $⎨˙⁅⁅w˛y⁆˛a⁆˙|˙⁅w˛y⁆∈E,˙a{=}α(w˛y)˙⎬$, which by inspection is  $⎨˙⁅⁅w˛y⁆˛α(w˛y)⁆˙|˙⁅w˛y⁆∈E˙⎬$, which by \rf{T3} is~$α$.

\yl{B4Mp} {\em $\pj{WI}(\hp{Q}) = ι$.} In steps, $\pj{WI}(\hp{Q})$ by (\rf{C679}) is $⎨˙⁅w˛ι(w)⁆˙|˙(∃y)⁅w˛y⁆∈E˙⎬$, which by inspection is $⎨˙⁅w˛ι(w)⁆˙|˙w∈π_1E˙⎬$, which by \rf{T5} is~$ι$. \hfill $\Box$ \end{cllist} \end{pfb}

\begin{lemma}[{{\bf implies Theorem~\rf{C691}}}]\label{B331}  Suppose $(Q,u)$ is a pentaform game.  Let $(\hs{X},\hs{E},\hs{\HH},\hs{α},\hs{ι},\hs{u}) = \TB(Q,u)$.  Then, \ttr{a}{E437} $(\hs{X},\hs{E},\hs{\HH},\hs{α},\hs{ι},\hs{u})$ is a traditional game.  Further, \ttr{b}{E438} $\hs{r} = r$, \ttr{c}{E443} $π_1\hs{E} = W$, \ttr{d}{E444} $π_2\hs{E} = Y$, \ttr{e}{E445} $\hs{E}^{-1} = p$. \ttr{f}{E446} $\hs{≼} = ≼$, \ttr{g}{E447} $\hs{≺} = ≺$, \ttr{h}{E448} $\hs{\ZZ} = \ZZ$, \ttr{i}{E449} $\hs{A} = A$, \ttr{j}{E451} $\hs{F} = F$, and \ttr{k}{E450} $\hs{I} = I$. \end{lemma}

\newcommand{\noteextraclaims}{\nichts{}}

\begin{pf} Part (\rf{E437}) follows from Claim~\rf{E452}; parts (\rf{E438})--(\rf{E448}) from Claims \rf{E441}--\rf{C763}; and parts (\rf{E449}), (\rf{E451}), and (\rf{E450}) from Claims \rf{E442}, \rf{C753}, and \rf{C757}.{\noteextraclaims}\begin{cllist}

\yl{E379} {\em $(\hs{X},\hs{E})$ is equal to $(W⋃Y,\pj{WY}(Q))$ and is a nontrivial out-tree.} Definitions (\rf{C693})--(\rf{C694}) imply $(\hs{X},\hs{E})$ is equal to $(W⋃Y,\pj{WY}(Q))$. Proposition~\rf{C632}(a)'s forward direction implies $(W⋃Y,\pj{WY}(Q))$ is a nontrivial out-tree.    

\yl{C744} {\em \rf{T1} holds.} $(\hs{X},\hs{E})$ is a nontrivial out-tree by Claim~\rf{E379}.

\yl{E441} {\em $\hs{r} = r$.} Since $(\hs{X},\hs{E})$ is an out-tree by Claim~\rf{E379}, Section~\rf{C665} paragraph~\rf{D321} defines $\hs{r}$ to be the root of $(\hs{X},\hs{E})$.  By Claim~\rf{E379} again, this $\hs{r}$ is the root of the out-tree $(W⋃Y,\pj{WY}(Q))$, which by Proposition~\rf{C632}(b) is the unique element of $W⧷Y$, which by definition (\rf{C982}) is $r$.  
 
\yl{C745} {\em $π_1\hs{E} = W$.}  In steps, $π_1\hs{E}$ by $\hs{E}$'s definition (\rf{C694}) equals $π_1(π_{WY}(Q))$, which by inspection equals $π_W(Q)$, which by abbreviation (\rf{C825}) equals~$W$.

\yl{E439} {\em $π_2\hs{E} = Y$.}  In steps, $π_2\hs{E}$ by $\hs{E}$'s definition (\rf{C694}) equals $π_2(π_{WY}(Q))$, which by inspection equals $π_Y(Q)$, which by abbreviation (\rf{C825}) equals~$Y$.

\yl{E440} {\em $\hs{E}^{-1} = p$.}  In steps, $\hs{E}^{-1}$ by definition (\rf{C694}) is $(\pj{WY}(Q))^{-1}$, which by inspection is $\pj{YW}(Q)$, which by the definition of $p$ within \rf{Pwy} is $p$.

\yl{C763} {\em (a) $\hs{≼} = ≼$. (b) $\hs{≺} = ≼$.  (c) $\hs{\ZZ} = \ZZ$.}  Claim~\rf{E379} suffices because both traditional games (Section~\rf{C665}, paragraph~\rf{D321}) and pentaform games (end of Section~\rf{C657}) derive their precedence relations and run collections from their out-trees (via (\rf{E351}) and (\rf{E355}) in Section~\rf{E349}).

\yl{C750} {\em \rf{T2} holds.} Proposition~\rf{D328}(\rf{D329}$⇒$\rf{D331}) and axiom \rf{Pjw} imply $⎨W_j|j∈J⎬$ partitions~$W$, which by $\hs{\HH}$'s definition (\rf{C695}) implies $\hs{\HH}$ partitions $W$, which by Claim~\rf{C745} implies $\hs{\HH}$ partitions $π_1\hs{E}$.

\yl{C751} {\em $\hs{α}$ is a function from $\hs{E}$.}  By $\hs{α}$'s definition~(\rf{C696}), \begin{gather}
\zz
\hs{α} = ⎨˙⁅⁅w˛y⁆˛a⁆˙|˙⁅w˛y˛a⁆∈\pj{WYA}(Q)˙⎬. \label{E384}
\zz
\end{gather} First, for functionhood, it suffices (by footnotes~\rf{C273} and \rf{E401} on \pgrf{C273}) to show \begin{gather}
\zz
⁅⁅w˛y⁆˛a^1⁆⋅∈⋅\hs{α}⋅\text{and}⋅⁅⁅w˛y⁆˛a^2⁆⋅∈⋅\hs{α}⋅\text{imply}⋅a^1 = a^2. \notag 
\zz
\end{gather} Toward that end, suppose $⁅⁅w˛y⁆,a^1⁆$ and $⁅⁅w˛y⁆˛a^2⁆$ belong to $\hs{α}$.  Then (\rf{E384}) implies $⁅w˛y˛a^1⁆$ and $⁅w˛y˛a^2⁆$ belong to $\pj{WYA}(Q)$, which implies $⁅y˛a^1⁆$ and $⁅y˛a^2⁆$ belong to $\pj{YA}(Q)$, which by \rf{Pay} implies $a^1 = a^2$.  Second, the domain of the function $\hs{α}$ is $π_1\hs{α}$, which by (\rf{E384}) equals $\pj{WY}(Q)$, which by $\hs{E}$'s definition (\rf{C694}) is $\hs{E}$.
 
\yl{C752} {\em \rf{T3} holds.} Because of Claim~\rf{C751}, it suffices to show that $\hs{α}$ is locally injective (\rf{D347}).  We will prove the contrapositive.  Toward that end, consider two edges $⁅w˛y_1⁆⋅∈⋅E$ and $⁅w˛y_2⁆⋅∈⋅E$ from decision node $w$ which are both assigned the action $\hs{α}(w˛y_1) = \hs{α}(w˛y_2)$.  It suffices to show $y_1 = y_2$. 

Let $a$ denote the common action $\hs{α}(w˛y_1) = \hs{α}(w˛y_2)$.  Then $\hs{α}$'s definition (\rf{C696}) implies that both $⁅w˛y_1˛a⁆$ and $⁅w˛y_2˛a⁆$ are in $\pj{WYA}(Q)$.  Thus by rearrangement, both $⁅w˛a˛y_1⁆$ and $⁅w˛a˛y_2⁆$ are in $\pj{WAY}(Q)$.  Hence axiom \rf{Pway} implies $y_1 = y_2$.

\yl{E442} {\em $\hs{A} = A$.}  In steps, the action set $\hs{A}$ by its definition (\rf{E378}) is the range of $\hs{α}$, which is $⎨˙\hs{α}(w˛y)⎬˙|˙⁅w˛y⁆∈\hs{E}˙⎬$, which by Claim~\rf{C751} and $\hs{α}$'s definition (\rf{C696}) is $π_2(⎨˙⁅⁅w˛y⁆˛a⁆˙|˙⁅w˛y˛a⁆∈\pj{WYA}(Q)˙⎬)$, which by inspection is $π_3(\pj{WYA}(Q))$, which by inspection is $π_A(Q)$, which by abbreviation (\rf{C825}) is $A$. 

\yl{C753} {\em $\hs{F} = F$.}  The correspondence $\hs{F}{:}π_1\hs{E}⇉\hs{A}$ is defined in (\rf{D169}) by $\hs{F}(w) = ⎨˙a˙|˙(∃y)˙\hs{α}(w˛y){=}a˙⎬$.  Thus  $\hs{F}$ (by footnote~\rf{B359} on \pgrf{B359}) is equal to \begin{gather}
\zz
⎨˙⁅w˛a⁆˙|˙w∈π_1\hs{E},˙(∃y)\hs{α}(w˛y){=}a˙⎬,\notag
\zz
\end{gather} which by $\hs{E}$'s definition (\rf{C694}) and $\hs{α}$'s definition (\rf{C696}) is equal to \begin{gather}
\zz
⎨˙⁅w˛a⁆˙|˙w∈π_1(\pj{WY}(Q)),˙(∃y)⁅w˛y˛a⁆∈\pj{WYA}(Q)˙⎬, \notag
\zz
\end{gather} which by inspection is equal to $⎨˙⁅w˛a⁆˙|˙w∈π_W(Q),˙(∃y)⁅w˛a˛y⁆∈\pj{WAY}(Q)˙⎬$, which by inspection is equal to $\pj{WA}(Q)$, which by $F$'s definition (\rf{C981}) is equal to~$F$.

\yl{C755} {\em \rf{T4} holds.} Proposition~\rf{C601}(\rf{C602}$⇒$\rf{C604}) and axioms \rf{Pjw} and \rf{Pwa} imply that $(∀j∈J,w_1∈W_j,w_2∈W_j)$ $F(w_1) = F(w_2)$.  Since Proposition~\rf{D328}(\rf{D329}$⇒$\rf{D331}) and axiom \rf{Pjw} imply that $⁅W_j⁆_{j∈J}$ is an injectively indexed partition, this is equivalent to \begin{gather}
\zz
(∀H∈⎨W_{j}|j∈J⎬,w_1∈H,w_2∈H)⋅F(w_1) = F(w_2), \notag
\zz
\end{gather} which by $\hs{\HH}$'s definition (\rf{C695}) and Claim~\rf{C753} implies \begin{gather}
\zz
(∀H∈\hs{\HH},w_1∈H,w_2∈H)⋅\hs{F}(w_1) = \hs{F}(w_2), \notag
\zz
\end{gather} which implies the measurability (\rf{D350}) of $\dF$.

\yl{D338} {\em (a) $\pj{WJ}(Q)$ is a surjection from $W$ to~$J$.  (b) $\pj{JI}(Q)$ is a surjection from $J$ to~$I$.  }  Part (a) holds by axiom \rf{Pjw}, and because $W = π_W(Q)$ and $J = π_J(Q)$ by abbreviation (\rf{C825}).  Similarly, (b) holds by axiom \rf{Pij}, and because $J = π_J(Q)$ and $I = π_I(Q)$ by abbreviation (\rf{C825}).  

\yl{E381} {\em $\pj{JI}(Q)○\pj{WJ}(Q) = \pj{WI}(Q)$.} By definition, $\pj{JI}(Q)○\pj{WJ}(Q)$ is equal to $⎨˙⁅w˛i⁆˙|˙(∃j)˙⁅w˛j⁆∈\pj{WJ}(Q), ⁅j˛i⁆∈\pj{JI}(Q)˙⎬$, which by inspection is equal to \begin{gather}
\zz
⎨˙⁅w˛i⁆˙|˙(∃j,i\up)˙⁅w˛j˛i\up⁆∈\pj{WJI}(Q), ⁅j˛i⁆∈\pj{JI}(Q)˙⎬. \notag 
\zz
\end{gather} Consider the conditions $⎨w˛y˛i\up⎬∈\pj{WJI}(Q)$ and $⁅j˛i⁆∈\pj{JI}(Q)$.  With axiom \rf{Pij}, they imply $i\up = i$.  Thus the set is equal to \begin{gather}
\zz
⎨˙⁅w˛i⁆˙|˙(∃j)˙⁅w˛j˛i⁆∈\pj{WJI}(Q), ⁅j˛i⁆∈\pj{JI}(Q)˙⎬, \notag 
\zz
\end{gather} which by inspection is equal to $⎨˙⁅w˛i⁆˙|˙(∃j)˙⁅w˛j˛i⁆∈\pj{WJI}(Q)˙⎬$, which by inspection is equal to $\pj{WI}(Q)$.

\yl{E383} {\em $\pj{WI}(Q)$ is a surjection from $W$ to~$I$.} This holds by Claims \rf{D338} and \rf{E381}.

\yl{C758} {\em $\hs{ι}$ is a function from $π_1\hs{E}$.}  Claim~\rf{E383} implies that $\pj{WI}(Q)$ is a function from $W$, which by Claim~\rf{C745} implies that $\pj{WI}(Q)$ is a function from $π_1\hs{E}$, which by $\hs{ι}$'s definition (\rf{C697}) implies that $\hs{ι}$ is a function from $π_1\hs{E}$.  

\yl{E380} {\em \rf{T5} holds.} Because of Claim~\rf{C758}, it suffices to show $\hs{ι}$ is measurable (\rf{D351}).  In other words, it suffices to show that \begin{gather}
\zz
(∀H∈\hs{\HH},w_1∈H,w_2∈H)⋅\hs{ι}(w_1) = \hs{ι}(w_2).\notag
\zz
\end{gather} By $\hs{\HH}$'s definition (\rf{C695}) and $\hs{ι}$'s definition (\rf{C697}), this is equivalent to \begin{gather}
\zz
(∀H∈⎨W_j|j∈J⎬,w_2∈H,w_2∈H)⋅\pj{WI}(Q)(w_1) = \pj{WI}(Q)(w_2),\notag
\zz
\end{gather} which, because $⁅W_j⁆_{j∈J}$ is an injectively indexed partition by Proposition~\rf{D328}(\rf{D329}$⇒$\rf{D331}) and axiom \rf{Pjw}, is equivalent to \begin{gather}
\zz
(∀j∈J,w_1∈W_j,w_2∈W_j)⋅\pj{WI}(Q)(w_1) = \pj{WI}(Q)(w_2),\notag
\zz
\end{gather} which by Claim~\rf{E381} is equivalent to \begin{gather}
\zz
(∀j∈J,w_1∈W_j,w_2∈W_j)⋅\pj{JI}(Q)˙○˙\pj{WJ}(Q)(w_1) = \pj{JI}(Q)˙○\pj{WJ}(Q)(w_2).\notag
\zz
\end{gather} 

Now, to show this, take a situation $j⋅∈⋅J$ and two decision nodes $w_1$ and $w_2$ in its information set $W_j$.  Then $w_1⋅∈⋅W_j$ and Lemma~\rf{D333}(\rf{D366}) imply $⁅w_1˛j⁆⋅∈⋅\pj{WJ}(Q)$, which by axiom~\rf{Pjw}\nichts{} implies $\pj{WJ}(Q)(w_1) = j$.  By the same reasoning, $\pj{WJ}(Q)(w_2) = j$.  Hence $\pj{WJ}(Q)(w_1) = \pj{WJ}(Q)(w_2)$, which by axiom~\rf{Pij}\nichts{} implies the desired equality.

\yl{C757} {\em $\hs{I} = I$.} In steps, $\hs{I}$ by its definition (\rf{E377}) is equal to the range of $\hs{ι}$, which by $\hs{ι}$'s definition (\rf{C697}) is equal to the range of $\pj{WI}(Q)$, which by Claim~\rf{E383} is equal to~$I$.

\yl{C760} {\em \rf{T6} holds.} Definition (\rf{C698}) sets $\hs{u}$ equal to $u$, which by the definition of a pentaform game (Definition~\rf{C668}) has the form $⁅u_i{:}\ZZ→\eR⁆_{i∈I}$, which is identical to the form $⁅u_i{:}\hs{\ZZ}→\eR⁆_{i∈\smash{\hs{I}}}$ because $\hs{\ZZ} = \ZZ$ by Claim~\rf{C763}(c) and because $\hs{I} = I$ by Claim~\rf{C757}. 

\yl{E452} {\em $(\hs{X},\hs{E},\hs{\HH},\hs{α},\hs{ι},\hs{u})$ is a traditional game.}  This follows from Definition~\rf{C843} and Claims \rf{C744}, \rf{C750}, \rf{C752}, \rf{C755}, \rf{E380}, and \rf{C760}.  \end{cllist} \unskipcl \end{pf}

\begin{lemma}\label{B463} Let $(X,E,\HH,α,ι,u)$ be a traditional game.  Then $\TB\PB(X,E,\HH,α,ι,u)$ $=$ $(X,E,\HH,α,ι,u)$. \end{lemma}

\begin{pf} Let $(\hp{Q},\hp{u}) = \PB (X,E,\HH,α,ι,u)$, which by Theorem~\rf{C689} is a pentaform game with information-set situations.  Next let $(\hs{X},\hs{E},\hs{\HH},\hs{α},\hs{ι},\hs{u}) = \TB(\hp{Q},\hp{u})$.  Since $(\hs{X},\hs{E},\hs{\HH},\hs{α},\hs{ι},\hs{U}) = \TB\PB(X,E,\HH,α,ι,u)$ by inspection, it suffices to show $(\hs{X},\hs{E},\hs{\HH},\hs{α},\hs{ι},\hs{U}) = (X,E,\HH,α,ι,U)$.  This is done, one component at a time, by Claims \rf{B489}--\rf{B494}. \begin{cllist}

\yl{B489} $\hs{X} = X$.  Note $\hs{X}$ by its definition (\rf{C693}) is $\hp{W}⋃\hp{Y}$, which by Lemma~\rf{B405}(\rf{B4x}) is~$X$.

\yl{B493} $\hs{E} = E$.  Note $\hs{E}$ by its definition (\rf{C694}) is $\pj{WY}(\hp{Q})$, which by Lemma~\rf{B405}(\rf{B4e}) is~$E$.

\yl{B503} $\hs{\HH} = \HH$.  By the proof's first sentence, $(\hp{Q},\hp{u})$ has information-set situations (\rf{D319}).  In other words, $(∀j∈\hp{J})$ $\hp{W}_j = j$.  Then in steps, $\hs{\HH}$ by its definition (\rf{C695}) is $⎨\hp{W}_j|j∈\hp{J}⎬$, which by the previous sentence is $⎨j|j∈\hp{J}⎬$, which reduces to $\hp{J}$, which by Lemma~\rf{B405}(\rf{B4h}) is $\HH$.

\yl{B495} $\hs{α} = α$.  Note $\hs{α}$ by its definition (\rf{C696}) is $⎨˙⁅⁅w˛y⁆˛a⁆˙|˙⁅w˛y˛a⁆∈\pj{WYA}(\hp{Q})˙⎬$, which by Lemma \rf{B405}(\rf{B4L}) is~$α$.

\yl{B496} $\hs{ι} = ι$.  Note $\hs{ι}$ by its definition (\rf{C697}) is $\pj{WI}(\hp{Q})$, which by Lemma \rf{B405}(\rf{B4M}) is~$ι$. 

\yl{B494} $\hs{u} = u$.  Note $\hs{u}$ by its definition (\rf{C698}) is $\hp{u}$, which by its definition (\rf{C680}) is~$u$. \end{cllist} \unskipcl \end{pf}

\begin{lemma}\label{B462} Suppose $(Q,u)$ is a pentaform game with information-set situations.  Then $\PB\TB(Q,u) = (Q,u)$. \end{lemma}

\begin{pf}  Let $(\hs{X},\hs{E},\hs{\HH},\hs{α},\hs{ι},\hs{u}) = \TB(Q,u)$, which by Theorem~\rf{C691} is a traditional game.  Next let $(\hp{Q},\hp{u}) = \PB(\hs{X},\hs{E},\hs{\HH},\hs{α},\hs{ι},\hs{u})$.  Since $(\hp{Q},\hp{u}) = \PB\TB(Q,u)$ by inspection, it suffices to show that $(\hp{Q},\hp{u}) = (Q,u)$.  Definitions (\rf{C680}) and (\rf{C698}) imply $\hp{u} = \hs{u} = u$.  Thus it suffices to show $\hp{Q} = Q$.  This is Claim~\rf{C774} below.  (Looking ahead, claims \rf{D141}--\rf{B476} concern the starting point $Q$ and the midpoint $(\hs{X},\hs{E},\hs{\HH},\hs{α},\hs{ι})$ but not end point $\hp{Q}$.) \begin{cllist}

\yl{D141} {\em $(∀⁅j˛w⁆∈\pj{JW}(Q))$ $j = \hs{H}_w$ (where $\hs{H}_w$ is the information set in $\hs{\HH}$ that contains w, as defined below (\rf{C678})).} Take an original situation/decision-node couple $⁅j˛w⁆⋅∈⋅\pj{JW}(Q)$.  Easily, projection implies $j⋅∈⋅π_J(Q)$, which by abbreviation (\rf{C825}) implies \ilc{D144} $j⋅∈⋅J$.  Further, the assumption $⁅j˛w⁆⋅∈⋅\pj{JW}(Q)$ and Lemma~\rf{D333}(\rf{D366}) imply \il{C771} $w⋅∈⋅W_j$.  

Note that \rf{D144} and $\hs{\HH}$'s definition (\rf{C695}) imply $W_j⋅∈⋅\hs{\HH}$, that is, that the original pentaform information set $W_j$ is a traditional information set in $\hs{\HH}$.  Thus since $\hs{\HH}$ is a partition by \rf{T2}, \rf{C771} implies that $W_j$ is the member of $\hs{\HH}$ that contains~$w$.  In other words, \il{E457} $W_j = \hs{H}_{w}$.  Meanwhile, since $(Q,u)$ has information-set situations (\rf{D319}) by assumption, $W_j = j$.  This and \rf{E457} imply $j = \hs{H}_w$. 

\yl{B475} {\em $Q⋅⊆⋅⎨˙⁅\hs{ι}(w),\hs{H}_w,w,\hs{α}(w˛y),y⁆˙|˙⁅w˛y⁆∈\hs{E}˙⎬$.}  (The following argument will letter its observations in an unusual way.)  Take an original quintuple \tts{q}{C770} $⁅i˛j˛w˛a˛y⁆$ $∈$~$Q$.  We begin with four preliminary observations. 
First, \rf{C770} implies $⁅w˛i⁆⋅∈⋅\pj{WI}(Q)$, which by $\hs{ι}$'s definition (\rf{C697}) implies \tts{i}{B482} $i = \hs{ι}(w)$.  
Second, \rf{C770} implies $⁅w˛j⁆⋅∈⋅\pj{WJ}(Q)$, which by Claim~\rf{D141} implies \tts{j}{D145} $j = \hs{H}_w$.  
Third, \rf{C770} implies $⁅w˛y˛a⁆⋅∈⋅\pj{WYA}(Q)$, which by $\hs{α}$'s definition (\rf{C696}) implies \tts{a}{B481} $a = \hs{α}(w˛y)$.  
Fourth, \rf{C770} implies $⁅w˛y⁆⋅∈⋅\pj{WY}(Q)$, which by $\hs{E}$'s definition (\rf{C694}) implies \tts{e}{B552} $⁅w˛y⁆⋅∈⋅\hs{E}$.  
In conclusion, \rf{B482}, \rf{D145}, and \rf{B481} imply $⁅i˛j˛w˛a˛y⁆ = ⁅\hs{ι}(w),\hs{H}_w,w,\hs{α}(w˛y),y⁆$, which with \rf{B552} completes the argument.  

\yl{B476} {\em $Q = ⎨˙⁅\hs{ι}(w),\hs{H}_w,w,\hs{α}(w˛y),y⁆˙|˙⁅w˛y⁆∈\hs{E}˙⎬$.}  Claim~\rf{B475} shows the forward inclusion.  From another perspective, Claim~\rf{B475} shows that the set $Q$ is a subset of a function from $\hs{E}$ (to be clear, the function's argument $⁅w˛y⁆⋅∈⋅\hs{E}$ appears in the third and fifth coordinates, and the function takes each $⁅w˛y⁆⋅∈⋅\hs{E}$ to the triple $⁅\hs{ι}(w),\hs{H}_w,\hs{α}(w˛y)⁆$ which appears in the first, second, and fourth coordinates).  Thus the set $Q$ and the function are equal if the projection $\pj{WY}(Q)$ is equal to the domain~$\hs{E}$ (as opposed to merely being a subset of $\hs{E}$).  This holds by $\hs{E}$'s definition~(\rf{C694}).

\yl{C774} {\em $Q = \hp{Q}$.}  The right-hand side of Claim~\rf{B476}'s equality is equal to $\hp{Q}$ by $\hp{Q}$'s definition (\rf{C679}).  \end{cllist} \unskipcl 
\end{pf}

\begin{npf}[{{\bf for Theorem~\rf{B560}}}]\label{B560p} Theorem~\rf{C689} shows that $\fP$ maps traditional games to pentaform games with information-set situations.  Hence the theorem follows from Lemmas \rf{B463} and \rf{B462}. \end{npf}

\begin{npf}[{{\bf for Propositions \rf{C699} and \rf{C777}}}]\label{C777p} As suggested in the text, Proposition~\rf{C777} is more general than Proposition~\rf{C699}.

\smallskip{\em Proposition~\rf{C777}}.  By assumption, $(X,E,\HH,α,ι,u) = \TB(\hp{Q},\hp{u})$.  Parts \rf{C7x}, \rf{C7e}, \rf{E453}, \rf{C7L}, \rf{C7M}, and \rf{C7u} are implied by definitions (\rf{C693})--(\rf{C698}), respectively.  Parts \rf{C7r}--\rf{C7z}, \rf{C7a}--\rf{C7f}, and \rf{C7i} are implied by Lemma~\rf{B331}(\rf{E438})--(\rf{E450}), respectively.  (Note that $(X,E,\HH,α,ι,u)$ and $(\hp{Q},\hp{u})$ here and in Proposition~\rf{C777} correspond to $(\dX,\dE,\dHH,\bar{α},\bar{ι},\du)$ and $(Q,u)$ in definition (\rf{C677}) and Lemma~\rf{B331}.)

\smallskip{\em Proposition~\rf{C699}}.  By assumption, $\PB(X,E,\HH,α,ι,u) = (\hp{Q},\hp{u})$.  Thus Theorem~\rf{B560} implies that $(\hp{Q},\hp{u})$ is a pentaform game with information-set situations and that $(X,E,\HH,α,ι,u) = \TB(\hp{Q},\hp{u})$.  Hence the previous paragraph (that is, Proposition~\rf{C777}) implies conclusions \rf{C7x}--\rf{C7z}, \rf{E453}, and \rf{C7L}--\rf{C7u}.  Since $(\hp{Q},\hp{u})$ has information-set situations (\rf{D319}), conclusion \rf{E453} implies conclusion \rf{C7h}.  \end{npf}

\eput